\documentclass{aa}
\usepackage[usenames,dvipsnames]{xcolor}
\usepackage[T1]{fontenc}
\usepackage{natbib}
\usepackage{csquotes}
\usepackage{xspace}
\usepackage{amssymb}
\usepackage{amsmath}
\usepackage{graphicx}
\usepackage{hyperref}
\usepackage{siunitx}
\usepackage{tikz}
\usepackage{subcaption}
\usepackage{aas_macros}
\usepackage{threeparttable}

\usetikzlibrary{arrows.meta}
\tikzset{%
  >={Latex[width=2mm,length=2mm]},
            base/.style = {rectangle, rounded corners, draw=black,
                           minimum width=2cm, minimum height=0.5cm,
                           text centered, font=\sffamily},
  activityStarts/.style = {base, fill=blue!30},
       startstop/.style = {base, fill=red!30},
    activityRuns/.style = {base, fill=green!30},
         process/.style = {base, minimum width=2.5cm, fill=orange!15,
                           font=\ttfamily},
}

\DeclareSIUnit\erg{erg}
\DeclareSIUnit\Myr{Myr}
\DeclareSIUnit\AU{AU}

\def\max{\mathrm{max}}
\def\erg{\hbox{erg}}

\def\cm{\hbox{cm}}

\def\yr{\hbox{yr}}
\def\AU{\hbox{AU}}

\begin{document}

\title{
How drifting and evaporating pebbles shape giant planets I: Heavy element content and atmospheric C/O
}  \titlerunning{Drifting and evaporating pebbles shape giant planets I: Heavy element content and C/O}
\authorrunning{Schneider \& Bitsch}
\author{Aaron David~Schneider$^{1,2,3}$ \& Bertram~Bitsch$^{1}$}
\institute{
  (1) Max-Planck-Institut f\"ur Astronomie, K\"onigstuhl 17, 69117 Heidelberg, Germany\\
  (2) Centre for ExoLife Sciences, Niels Bohr Institute, Øster Voldgade 5, 1350 Copenhagen, Denmark\\
  (3) Instituut voor Sterrenkunde, KU Leuven, Celestijnenlaan 200D, B-3001 Leuven, Belgium
} \date{\today}
\offprints{B. Bitsch,\\ \email{bitsch@mpia.de}}
\abstract{Recent observations of extrasolar gas giants suggest super-stellar C/O ratios in planetary atmospheres, while interior models of observed extrasolar giant planets additionally suggest high heavy element contents. Furthermore, recent observations of protoplanetary disks revealed super-solar C/H ratios, which are explained by inward drifting and evaporating pebbles enhancing the volatile content of the disk. We investigate in this work how the inward drift and evaporation of volatile-rich pebbles influences the atmospheric C/O ratio and heavy element content of giant planets growing by pebble and gas accretion.
To achieve this goal, we perform semi-analytical 1D models of protoplanetary disks, including the treatment of viscous evolution and heating, pebble drift, and simple chemistry to simulate the growth of planets from planetary embryos to Jupiter-mass objects by the accretion of pebbles and gas while they migrate through the disk. 
Our simulations show that the composition of the planetary gas atmosphere is dominated by the accretion of vapor that originates from inward drifting evaporating pebbles at evaporation fronts.
This process allows the giant planets to harbor large heavy element contents, in contrast to models that do not take pebble evaporation into account. In addition, our model reveals that giant planets originating farther away from the central star have a higher C/O ratio on average due to the evaporation of methane-rich pebbles in the outer disk. These planets can then also harbor super-solar C/O ratios, in line with exoplanet observations. However, planets formed in the outer disk harbor a smaller heavy element content due to a smaller vapor enrichment of the outer disk compared to the inner disk, where the very abundant water ice also evaporates. Our model predicts that giant planets with low or large atmospheric C/O should harbor a large or low total heavy element content. We further conclude that the inclusion of pebble evaporation at evaporation lines is a key ingredient for determining the heavy element content and composition of giant planets.
  }

\maketitle


\section{Introduction}
\begin{figure*}
        \begin{subfigure}{\textwidth}
                \includegraphics[width=\textwidth]{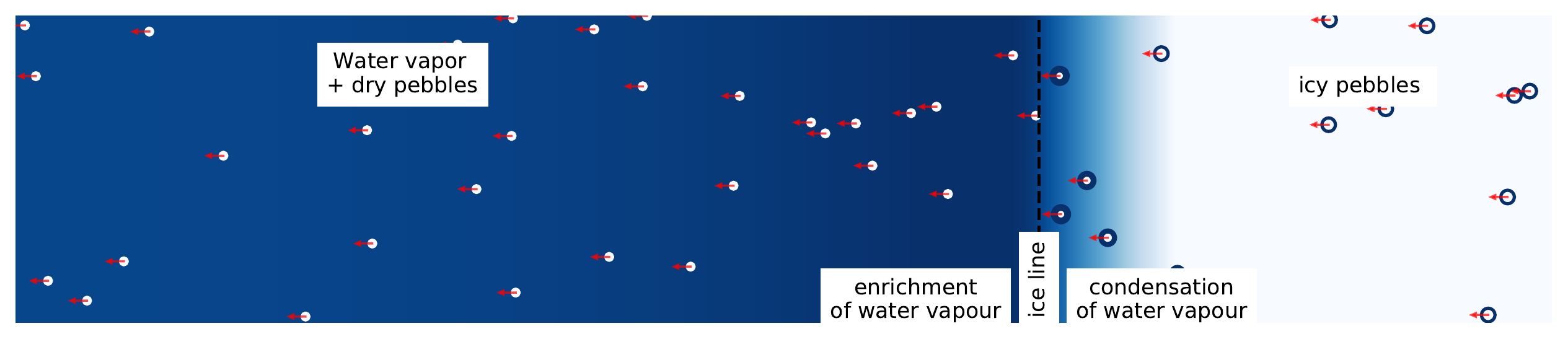}   
        \end{subfigure}
        \begin{subfigure}{\textwidth}
                \includegraphics[width=\textwidth]{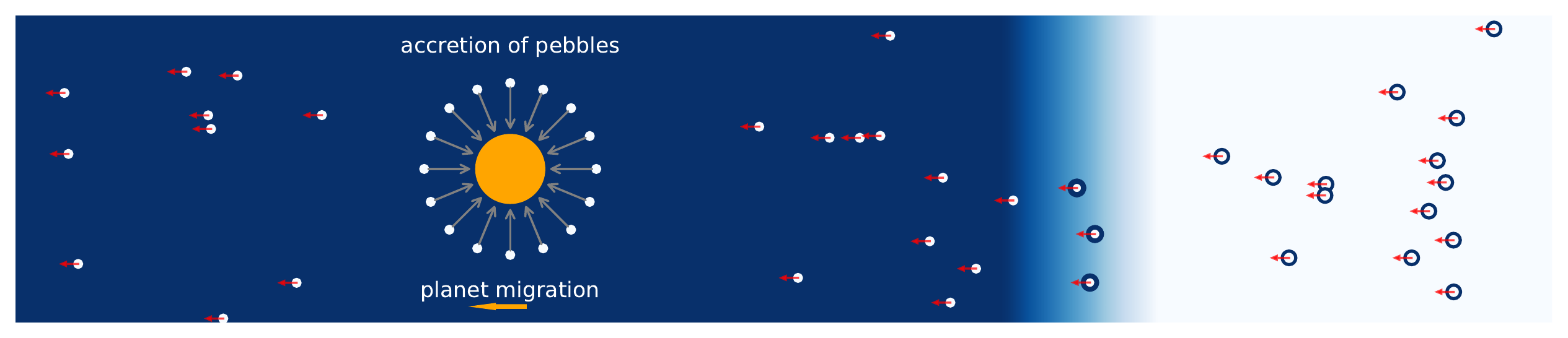}   
        \end{subfigure}
        \begin{subfigure}{\textwidth}
                \includegraphics[width=\textwidth]{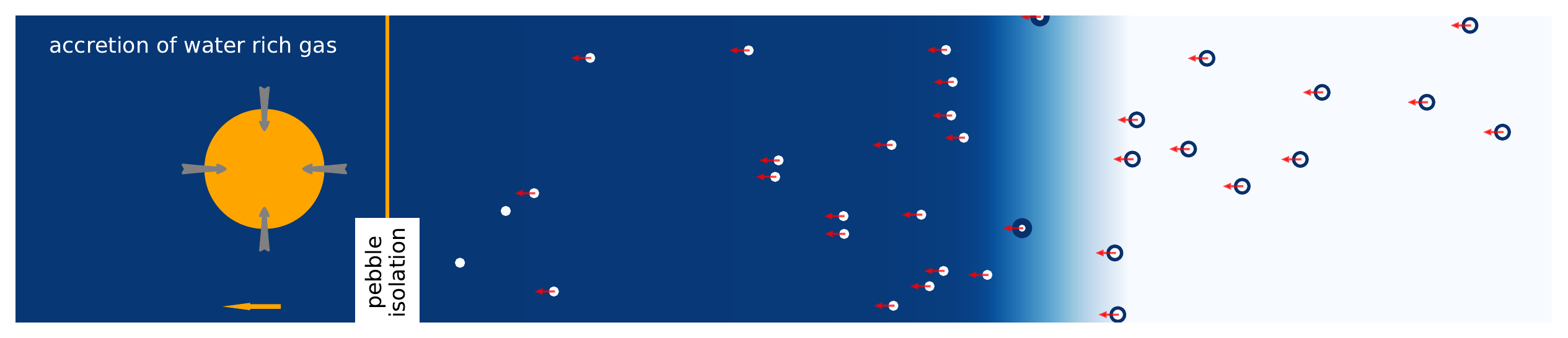}    
        \end{subfigure} 
        \caption{Phases of planetary growth. \\
        Top: Dust particles grow to pebbles (small dots) and drift toward the star. Icy pebbles that cross the water ice line (dashed line) evaporate their water content and enrich the gas with water vapor. Water vapor that crosses the ice line condenses onto pebbles, increasing their water content.\\
        Middle: The core of the planet is formed by pebble accretion while the planet migrates. Depending on the formation path, the core composition can be icy or dry. In the cartoon shown here, the core would be water poor.\\
        Bottom: Once the planet is heavy enough to reach pebble isolation and form a pressure bump, pebbles are stopped and cannot be accreted by the planet. The planet will then start to accrete gas that is enriched with water vapor.\\
        The water content of the disk (in solid or gaseous form) is color coded, where a darker color indicates a higher water content. We restrict ourselves in this cartoon to the water evaporation front, but the same applies for the evaporation of all solids in our model.}
        \label{fig:phases}
\end{figure*}

The number of observed exoplanets is increasing every day, with now more than 4000 observed exoplanets \citep{NASA_exoplanet_archive}. However, how exactly the planets form and how some of their properties can be explained  (e.g., the C/O ratio and their heavy element content) is still unclear. Observed atmospheric abundances of exoplanets could thus be used to constrain the planet formation pathway (e.g., \citealt{Mordasini2016, Madhusudhan2017, Cridland2019}).

One quantity to constrain the planet formation pathway could be the atmospheric C/O ratios, which can be measured to high precision \citep{Molliere20151Dmodel}. It has been clear since the pioneering works of \citet{Oeberg2011} that atmospheric C/O ratios strongly depend on the formation environment of the planet. Although atmospheric measurements are still imprecise, \citet{Brewer2017} constrain atmospheric C/O ratios for different observed hot gas giants and find that C/O ratios differ from the host star's C/O ratio. Some might even be enhanced by more than a factor of two compared to the host star's value. However, these data will hopefully be improved with the {\it James Webb Space Telescope} (JWST) and the {\it Atmospheric Remote-sensing Infrared Exoplanet Large-survey} (ARIEL) by the end of the decade.

If the mass and radius of an observed planet are known, one can also apply interior models to constrain the amount of heavy elements (elements with atomic numbers higher than two) that is needed to match those parameters (e.g., \citealt{Miller2011}). \citet{Thorngren2016} uses a sample of non-inflated hot gas giants to relate the amount of heavy elements to the planetary mass and radius. They find that the heavy element mass is approximately proportional to the square root of the planetary mass and that an average Jupiter-mass planet should harbor \SI{57.9}{M_\oplus} of heavy elements. However, recent work by \citet{Mueller2020} questions whether some of the planets from the catalog of \citet{Thorngren2016} are inflated, thus making the analysis of \citet{Thorngren2016} less applicable. In any case, the origin of this large heavy element content is still unknown.

From the formation side, models either include the growth of planetary cores via planetesimals \citep{IdaLinI,Alibert2005,IdaLinIV,IdaLinV,IdaLinVI,Mordasini2012,Alibert2013,Cridland2019, Emsenhuber2020} or pebbles \citep{Bitsch2015, Bitsch2019,Lambrechts2012, Lambrechts2014, Lambrechts2019, Ndugu2018, Ali-Dib2017,Bruegger2018}. Even though there are plenty of models that deal with the formation aspect, only a few models include the chemical composition that allows a more detailed comparison to observations. For example, \citet{Venturini_2020_pla_chem} investigates the heavy element content of planets built by pebbles or planetesimals but do not treat multiple chemical species. On the other hand, \citet{Booth2017_pebble_evap} include the physics of radial dust drift and pebble evaporation at evaporation lines to infer the C/O ratios of planets built by pebble and gas accretion but do not include condensation and do not investigate the heavy element content of gas giants.

Other approaches focus solely on the disk chemistry and handle the chemical reactions on the surface of dust grains \citep{Semenov2010, Eistrup2016, Notsu2020}. It is, however, difficult to include the physical processes of grain growth and radial dust drift in these models, though first attempts have been made \citep[e.g.,][]{Booth2019}. A recent work by \citet{Krijt2020} also expands this in much more detail for a 2D disk model. However, it seems that chemical reactions operate on longer timescales compared to the radial drift of millimeter- and centimeter-sized pebbles \citep[e.g.,][]{Booth2019}.

In this work we focus thus on a disk model that includes the growth and drift of pebbles \citep{Birnstiel2012} as well as evaporation and condensation at evaporation fronts. We then model the growth of planets via the accretion of pebbles \citep{Johansen2017peb} and gas \citep{Ndugu2021} inside these disks, while tracing the chemical composition of the accreted material to derive the atmospheric C/O ratio as well as the heavy element content. Our model approach is outlined in the cartoon shown in Fig.~\ref{fig:phases}.

Our newly developed code \texttt{chemcomp} simulates the formation of planets in viscously evolving protoplanetary disks by the accretion of pebbles and gas. The chemical composition of planetary building blocks (pebbles and gas) is traced by including a physical approach of the evaporation and condensation of volatiles at evaporation lines.

This paper is structured as follows: The first part includes the outline of the disk model and the description of the planetary formation model. We then explain the numerical methods (Sect.~\ref{sec:model}) used in \texttt{chemcomp} and show results that have been obtained (Sects.~\ref{sec:results} and \ref{sec:res_heavy_elements}). We then discuss the shortcomings and implications of our simulations in Sect.~\ref{sec:discussion} before concluding in Sect.~\ref{sec:conclusion}.

\section{Model}\label{sec:model}

In this section we discuss the different ingredients and parameters of our model. We list all variables and their meaning in Tables~\ref{tab:variables1} and~\ref{tab:variables2}. We start by discussing first the parts of our model that are relevant for the disk structure and evolution, namely the dust growth and the chemical composition of the disk followed by the viscous evolution, including pebble evaporation and condensation as well as the disk dispersal. Our model regarding planetary growth includes planet migration, pebble and gas accretion, and gap opening in our disk model. We then close this section by discussing the operating principle of our newly developed code \texttt{chemcomp}. 

We follow here a classic viscous disk evolution model \citep[e.g.,][]{Lynden-Bell1974,Bell1997}, utilizing the alpha-viscosity prescription \citep{Shakura1973}. The viscosity is then given by
 \begin{equation}
        \nu = \alpha\frac{c_s^2}{\Omega_K},
 \end{equation}
 where $\alpha$ is a dimensionless factor that describes the turbulent strength and $\Omega_K=\sqrt{\frac{GM_\star}{r^3}}$ is the Keplerian angular frequency.
 The isothermal sound speed $c_s$ can be linked to the midplane temperature  $T_\mathrm{mid}$ (see Appendix~\ref{sec:num_T_it} for details of our disk temperature calculation) by
 \begin{equation}
        c_s     = \sqrt{\frac{k_\mathrm{B}T_\mathrm{mid}}{\mu m_\mathrm{p}}},
 \end{equation}
 where $\mu$ is the mean molecular weight, which can change due to the enrichment of the disk's gas with vapor, as we discuss below, and $m_\mathrm{p}$ is the proton mass. For simplicity, we keep the disk's temperature fixed in time.

\subsection{Dust growth}\label{sec:dustgrowth}
We followed the two populations approach from \citet{Birnstiel2012} to model the growth of dust to pebbles limited by fragmentation, drift, and drift-induced fragmentation. We compare the gas and solid evolution of our code to the code of \citet{Birnstiel2012} in Appendix~\ref{sec:codecomp}. Since solid particles try to move Keplerian, the gas asserts an aerodynamic headwind on the solid particles due to the sub-Keplerian azimuthal speed $\Delta v$ of the gas given by
\begin{equation}
        \Delta v = v_\mathrm{K} - v_\varphi = - \frac{1}{2}\frac{\mathrm{d}\ln P}{\mathrm{d}\ln r} \left(\frac{H_\mathrm{gas}}{r}\right)^2v_\mathrm{K},        
\end{equation}
where $v_\mathrm{K}=\Omega_\mathrm{K} r$ is the Keplerian velocity, $H_\mathrm{gas}=\frac{c_s}{\Omega_\mathrm{K}}$ is the scale height of the disk, and $P$ is the gas pressure. 
This friction can be quantified in the Epstein drag regime \citep{Brauer2008, Birnstiel2010} by quantifying the friction time, $\tau_\mathrm{f}$, and the Stokes number, $\mathrm{St}$, of a particle
\begin{equation}\label{eq:Stokes}
        \mathrm{St} = \Omega_\mathrm{K} \tau_\mathrm{f} = \frac{\pi}{2}\frac{a\rho_\bullet}{\Sigma_\mathrm{gas}},
\end{equation}
where $a$ and $\rho_\bullet$ denote the radius of the solid particle and its density.\footnote{We follow the approach outlined in Eq. 12 of \citet{Drazkowska2017} and calculate $\rho_\bullet$ dynamically according to the abundance of ices in solids, which changes due to condensation, where silicates have a density of 3g/cm$^3$ and ices a density of 1g/cm$^3$.} $\Sigma_\mathrm{gas}$ corresponds to the gas surface density. Small particles have low friction times and are therefore strongly coupled to the gas, whereas large particles have large friction times and therefore only couple weakly to the gas.
Due to this headwind, dust grains will spiral inward with the radial velocity $u_\mathrm{Z}$ \citep{Weidenschilling1977dyn, Brauer2008, Birnstiel2012}
\begin{equation}\label{eq:vel_dust}
        u_\mathrm{Z} = \frac{1}{1+\mathrm{St}^2}u_\mathrm{gas} - \frac{2}{\mathrm{St}^{-1}+\mathrm{St}}\Delta v.
\end{equation} 
Thus, growing from small grains to large grains implies an increase in the radial dust velocity.
Growth can therefore be limited by radial drift (larger grains \enquote{drift away}), especially in the outer disk regions.
When the velocity of the dust grains exceeds a certain velocity boundary \citep{Birnstiel2009}, the dust will fragment (and therefore decrease its size) upon collision. This fragmentation velocity is normally measured in laboratory experiments that yield fragmentation velocities of \SIrange{1}{10}{\m\per\s} for silicate and ice particles, respectively \citep{Gundlach2015}. In our model we always use a fixed fragmentation velocity of 5m/s.

The solid to gas ratio $\epsilon$ is defined as
\begin{equation}
        \epsilon=\frac{\Sigma_\mathrm{Z}}{\Sigma_\mathrm{gas}},
\end{equation}
where $\Sigma_\mathrm{Z}$ denotes the total solid surface density (composed of dust and pebbles). The pebble surface density can be constrained from the solid surface density by multiplication with a numerical factor $f_m$ taken from the two population model \citep{Birnstiel2012}:
\begin{equation}
        \Sigma_\mathrm{peb} = f_m \times \Sigma_\mathrm{Z},
\end{equation}
where $f_m=0.97$ in the drift limited case and $f_m=0.75$ in the fragmentation limited case.

We can now calculate the mass averaged dust velocity (important for dust transport; see Sect.~\ref{sec:evo}):
\begin{equation}\label{eq:peb_velocity}
        \bar u_\mathrm{Z} = (1-f_\mathrm{m})u_\mathrm{dust} - f_\mathrm{m}u_\mathrm{peb} \ .
\end{equation}
However, other growth limiting mechanisms such as the bouncing barrier \citep{Guettler2010} or the charging barrier \citep{Okuzumi2009} exist but are, for the sake of simplicity, not considered in this work. In principle, bouncing would lead to smaller particle sizes compared to a fragmentation or drift-limited particle size \citep[e.g.,][]{Lorek2018}.

This approach to the dust evolution is within a few percent compared to a "full" model, which includes individual velocities for all grains within the grain size distribution \citep{Birnstiel2012}. Previous planetesimal \citep{Drazkowska2016, Drazkowska2017} and planet formation simulations \citep{Guliera2020} have used this full model. While the full model allows for a slightly larger accuracy for the dust evolution, the here used approach of \citet{Birnstiel2012} results in a shorter computational run time, needed to probe all the different disk and planetary parameters (Table~\ref{tab:parameters}) in our work. We further note that our approach does not take into account the reduction of the dust and pebble velocities due to increases in the local dust-to-gas ratio, as considered in other works (e.g., \citealt{Nakagawa1986, Drazkowska2016}).

\subsection{Compositions}\label{sec:composition}
\begin{table}
 \centering
 \begin{tabular}{c c}
 \hline
 \hline
 Species (X) & Abundance \\ \hline
 He/H & 0.085\\
 O/H & $4.90\times 10^{-4}$ \\ 
 C/H & $2.69\times 10^{-4}$ \\
 N/H & $6.76 \times 10^{-5}$ \\
 Mg/H & $3.98\times 10^{-5}$ \\
 Si/H & $3.24\times 10^{-5}$ \\
 Fe/H & $3.16\times 10^{-5}$ \\
 S/H & $1.32\times 10^{-5}$ \\
 Al/H & $2.82\times 10^{-6}$ \\
 Na/H & $1.74\times 10^{-6}$ \\
 K/H & $1.07\times 10^{-7}$ \\
 Ti/H & $8.91\times 10^{-8}$ \\
 V/H & $8.59\times 10^{-9}$\\
 \hline
 \end{tabular}
 \caption{Elemental number ratios used in our model, corresponding to the abundance of element X compared to hydrogen in the solar photosphere \citep{Asplund2009}.}
 \label{tab:solar_values}
\end{table}
\begin{table*}
\centering
\begin{tabular}{c c c c}
\hline\hline
Species (Y) & $T_{\text{cond}}$ {[}K{]} & $v_{\text{Y, no C}}$ & $v_{\text{Y, with C}}$\\ \hline 
CO & 20  & 0.45 $\times$ C/H &  0.45 $\times$ C/H  \\
N$_2$ & 20  & 0.9 $\times$ N/H &  see $v_{\text{Y, no C}}$   \\
CH$_4$ & 30 & 0.45 $\times$ C/H  & 0.25 $\times$ C/H \\
CO$_2$ & 70 & 0.1 $\times$ C/H & 0.1 $\times$ C/H \\
NH$_3$ & 90 & 0.1 $\times$ N/H & see $v_{\text{Y, no C}}$ \\
H$_2$S & 150 & 0.1 $\times$ S/H & see $v_{\text{Y, no C}}$ \\
H$_2$O & 150 & O/H - (3 $\times$ MgSiO$_3$/H + 4 $\times$ Mg$_2$SiO$_4$/H + CO/H \\
& & + 2 $\times$ CO$_2$/H + 3 $\times$ Fe$_2$O$_3$/H + 4 $\times$ Fe$_3$O$_4$/H + VO/H \\ 
& & + TiO/H + 3$\times$Al$_2$O$_3$ + 8$\times$NaAlSi$_3$O$_8$ + 8$\times$KAlSi$_3$O$_8$) & see $v_{\text{Y, no C}}$ \\
Fe$_3$O$_4$ & 371 & (1/6) $\times$ (Fe/H - 0.9 $\times$ S/H) & see $v_{\text{Y, no C}}$ \\
C (carbon grains) & 631 & 0 & 0.2 $\times$ C/H\\
FeS & 704 & 0.9 $\times$ S/H & see $v_{\text{Y, no C}}$ \\
NaAlSi$_3$O$_8$ & 958 & Na/H & see $v_{\text{Y, no C}}$ \\
KAlSi$_3$O$_8$ & 1006 & K/H & see $v_{\text{Y, no C}}$ \\
Mg$_2$SiO$_4$ & 1354 & Mg/H - (Si/H - 3$\times$K/H - 3$\times$Na/H) & see $v_{\text{Y, no C}}$  \\ 
Fe$_2$O$_3$ & 1357 & 0.25 $\times$ (Fe/H - 0.9 $\times$ S/H) & see $v_{\text{Y, no C}}$\\ 
VO & 1423 & V/H & see $v_{\text{Y, no C}}$ \\ 
MgSiO$_3$ & 1500 & Mg/H - 2$\times$(Mg/H - (Si/H - 3$\times$K/H - 3$\times$Na/H)) & see $v_{\text{Y, no C}}$ \\ 
Al$_2$O$_3$ & 1653 & 0.5$\times$(Al/H - (K/H + Na/H)) & see $v_{\text{Y, no C}}$ \\
TiO & 2000 & Ti/H & see $v_{\text{Y, no C}}$ \\  \hline
\end{tabular}
\caption[Condensation temperatures]{Condensation temperatures and volume mixing ratios of the chemical species.}
    \begin{tablenotes}
          \item \textbf{Notes:} Condensation temperatures for molecules are taken from \citet{Lodders2003}. For Fe$_2$O$_3$ the condensation temperature for pure iron is adopted \citep{Lodders2003}. Volume mixing ratios $v_{\rm Y}$ (i.e., by number) are adopted for the species as a function of disk elemental abundances (see, e.g., \citealt{Madhusudhan2014}). We expand here on the different mixing ratios from \citet{Bitsch2020chem}. Results in this work apply the model without pure carbon grains. A comparison with a model that includes carbon grains can be found in Appendix~\ref{sec:with_C}.
    \end{tablenotes}
\label{tab:comp}
\end{table*}

We make the assumption that the original chemical composition of the disk is similar to the host star composition. We use here the solar abundances [X/H] from \citet{Asplund2009} (see also Table \ref{tab:solar_values}), whereas our chemical model is based on the model presented in \citet{Bitsch2020chem}. 
The disk temperature is dependent on the orbital distance (see Appendix~\ref{sec:num_T_it} and Fig. \ref{fig:T-profil}) and therefore the composition of dust and gas is likewise dependent on the orbital distance. We used a simple chemical partitioning model to distribute the elements into molecules, Y (see Table \ref{tab:comp}; extended from \citealt{Bitsch2020chem}).

Based on the condensation temperature, a molecule of species Y will generally either be available in gas form (evaporated) when the disk temperature is above the condensation temperature or in solid form (condensated) when the disk temperature is below the condensation temperature. The transition point, where the midplane temperature equals the condensation temperature of species Y, is referred to as the evaporation line of species Y.

Sulfur is mostly available in refractory form in protoplanetary disks \citep{Kama2019}, leaving only a small component in volatile form. For nitrogen we use N$_2$ and NH$_3$, where most of the nitrogen should be in the form of N$_2$ \citep[e.g.,][]{Bosman2019}. Even though Ti, Al, K, Na, and V are not very abundant and thus do not contribute significantly to the planetary accretion rates, we include these elements in our model because they can be observed in the atmospheres of hot Jupiters and could also play a crucial role for the chemical evolution inside the atmospheres \citep{Ramirez2020}.

\begin{figure}
        \centering
        \includegraphics[width=.45\textwidth]{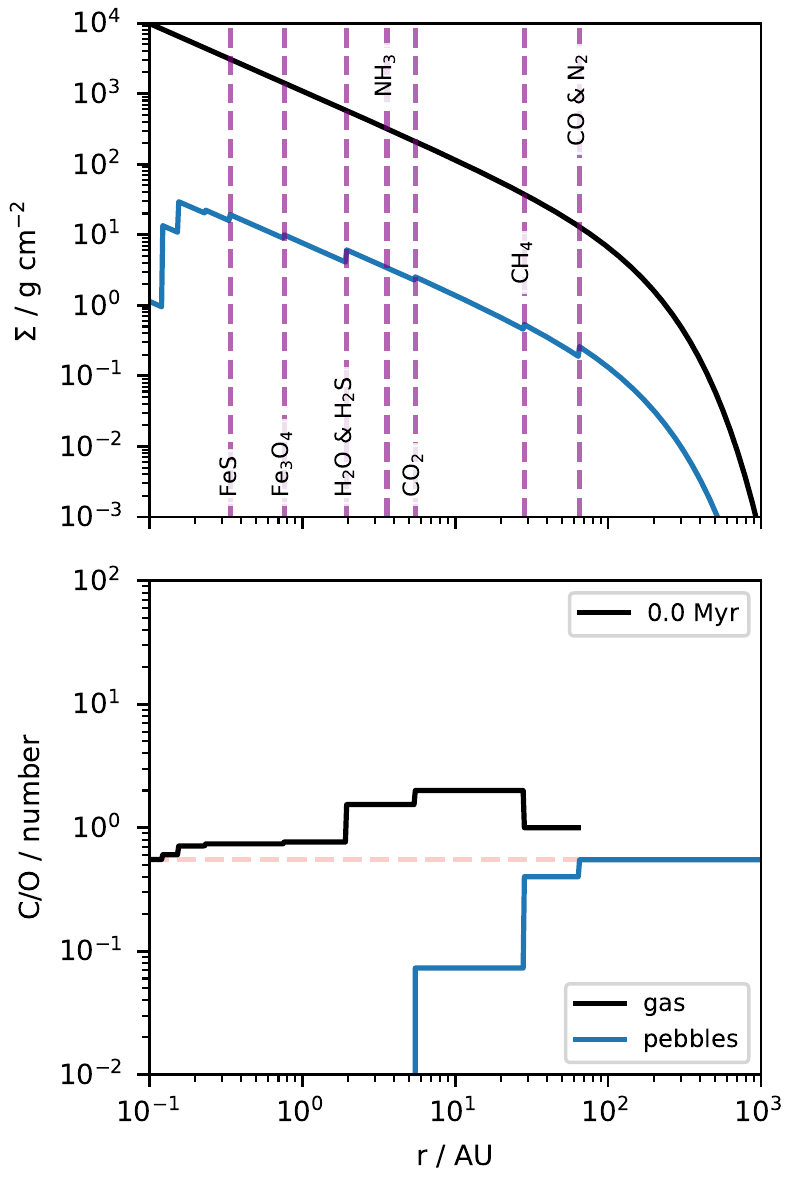}
        \caption{Initial surface density and C/O ratio. Top: Initial surface density of pebbles (blue line) and gas (black line). Bottom: C/O number ratio in the disk in pebbles and gas. Evaporation lines are labeled and indicated as dashed purple lines. Evaporation lines for molecular species with condensation temperatures higher than \SI{704}{K} are not shown for simplicity (see Table~\ref{tab:comp}). The solar C/O value of $~0.54$ is indicated as a red horizontal dashed line. We use our standard disk parameters with $\alpha=5\times 10^{-4}$ for this simulation, as stated in Table \ref{tab:parameters}.}
        \label{fig:sigma_0}
\end{figure}

The above-described partitioning model is used for the initial chemical composition in our disk (e.g., Fig. \ref{fig:sigma_0}) as well as the chemical composition at all times for our simplest model, which is basically an extension of the step-function-like compositions in the work of \citet{Oeberg2011}. In the course of this work we also include the evaporation of drifting grains at evaporation lines, which will change the chemical composition of the disk (see below).

With this we can define the elemental number ratio
\begin{equation}
\label{eq:def_CO}
    \rm {X1/X2} = \frac{m_{\rm X1}}{m_{\rm X2}}\frac{\mu_{\rm X2}}{\mu_{\rm X1}}, 
\end{equation}
where $m_{\rm X1}$ and $m_{\rm X2}$ is the mass fraction of the elements X1 and X2, respectively, in a medium of mass or density $m$ (e.g., $m=\Sigma_{\rm gas}$) and $\mu_{\rm X1}$ and $\mu_{\rm X2}$ are the atomic masses of the specific element. In our work we mainly use this definition to calculate C/O.

In our model, we include the change of the mean molecular weight $\mu$ due to evaporation of inward drifting pebbles that increase the vapor content in the gas phase in time. This can lead to an increase in $\mu$ from 2.3 (standard hydrogen-helium mixture) up to $4$, if the disk is heavily enriched in vapor. We show the derivation of how we calculate $\mu$ in our model in Appendix~\ref{sec:mucompo}.

For all icy species (H$_2$O, H$_2$S, NH$_3$, CO$_2$, CH$_4$, N$_2,$ and CO) we assume a pebble density of $\rho_{\bullet, \mathrm{ice}} =\SI{1}{\g\per\cm\cubed}$, while refractory species (Fe$_3$O$_4$ and higher condensation temperatures) have a pebble density of $\rho_{\bullet, \mathrm{ref}} =\SI{3}{\g\per\cm\cubed}$. The exact pebble density is then computed dynamically during the simulation via their composition
\begin{equation}
    \rho_\bullet = (m_{\rm ref} + m_{\rm ice}) \cdot \left( \frac{m_{\rm ref}}{\rho_{\bullet, \mathrm{ref}}} + \frac{m_{\rm ice}}{\rho_{\bullet, \mathrm{ice}}} \right)^{-1} \ .
\end{equation}
This process accounts automatically for a change in the pebble density due to condensation or evaporation of volatile species.

\subsection{Viscous evolution}\label{sec:evo}
\begin{figure*}
        \centering
        \includegraphics[width=1.0\textwidth]{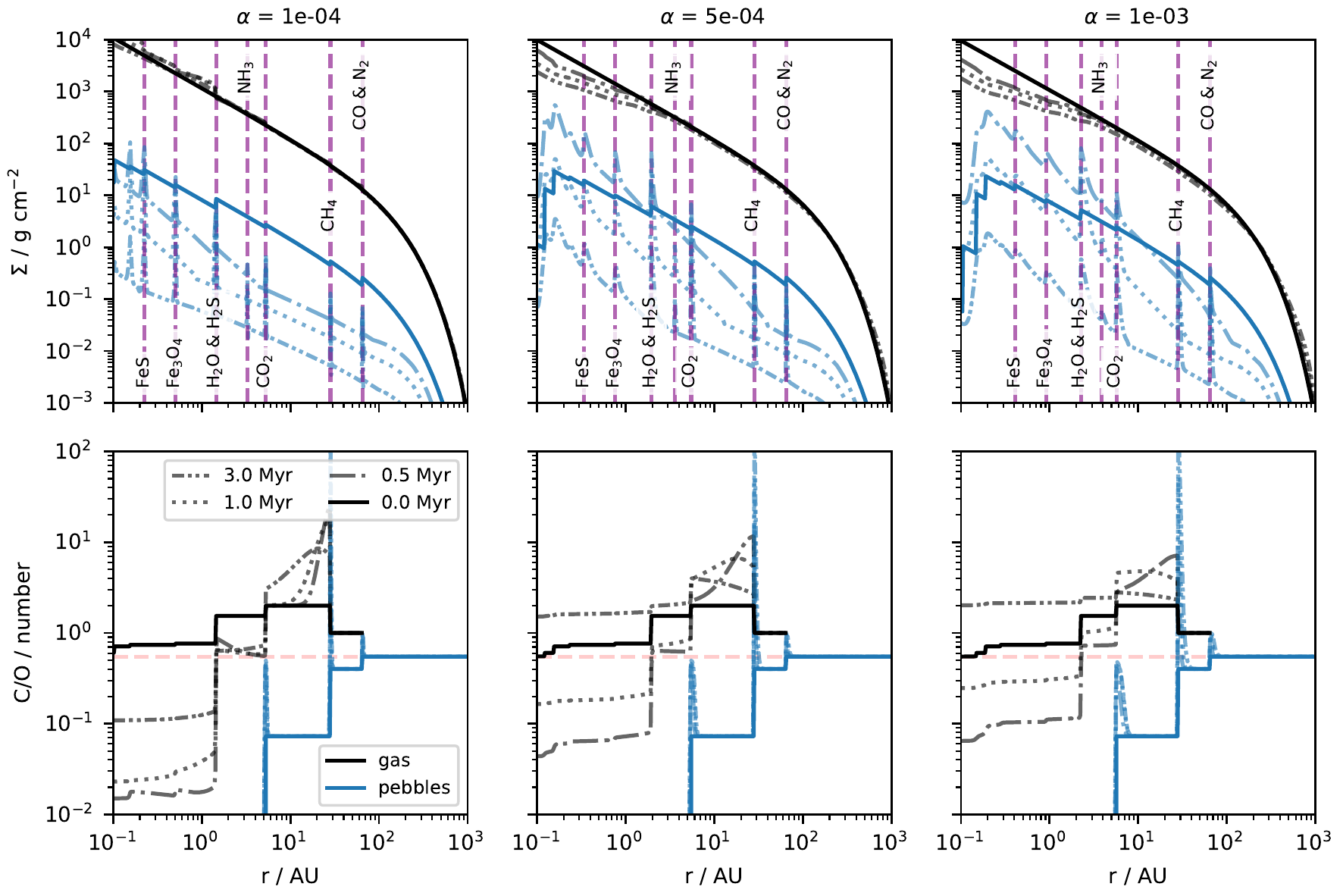}
        \caption{Like Fig. \ref{fig:sigma_0} but now including the time evolution of the disk, also for $\alpha=10^{-4}$ (left) and $\alpha=10^{-3}$ (right). Evaporation line positions are shown with vertical lines, which are different for the different simulations due to an increase in viscous heating with increasing $\alpha$ (Appendix~\ref{sec:num_T_it}). We use the same disk parameters (except $\alpha$) as in Fig.~\ref{fig:sigma_0}, corresponding to the standard parameters shown in Table~\ref{tab:parameters}.}
        
        \label{fig:sigma_evap}
\end{figure*}

Given mass conservation and conservation of angular momentum, one can derive \citep{Pringle1981,Armitage} the viscous disk equation
\begin{equation}\label{eq:evo_evo_gas}
        \frac{\partial\Sigma_{\mathrm{gas,Y}}}{\partial t} - \frac3r\frac{\partial}{\partial r}\left[\sqrt{r}\frac{\partial}{\partial r}\left(\sqrt{r}\nu\Sigma_{\mathrm{gas,Y}}\right)\right] = \dot\Sigma_\mathrm{Y},
\end{equation}
where $\dot\Sigma_\mathrm{Y}$ is a source term for a given species Y described later on. The viscous disk equation describes how the gas surface density evolves in time. With this equation we yield the radial gas velocity, given by
\begin{equation}\label{eq_the:radial_velocity_general}
        u_\mathrm{gas} = -\frac{3}{\Sigma_\mathrm{gas}\sqrt r}\frac{\partial}{\partial r}\left(\nu \Sigma_\mathrm{gas}\sqrt r\right).
\end{equation}
We compare the radial velocities of the gas, dust and planets in Appendix~\ref{sec:velocities}.

For the initialization of simulations we used the analytical solution (without the source term $\dot \Sigma_Y$) found by \citet{Lynden-Bell1974}, which is dependent on a scaling radius $R_0$ and the initial disk mass $M_0$ and can be expressed as \citep{Lodato2017}
\begin{multline}\label{eq:gas_density_analytically}
        \Sigma_\mathrm{gas}(r,t) = \frac{M_0}{2\pi R_0^2}(2-\psi)\left(\frac{r}{R_0}\right)^{-\psi}\xi^{\frac{5/2-\psi}{2-\psi}}\\ 
        \times\exp\left(-\frac{(r/R_0)^{2-\psi}}{\xi}\right),
\end{multline}
where $t$ is the time, $\psi=\left(\frac{\mathrm{d}\ln{\nu}}{\mathrm{d}\ln{r}}\right)_{r=r_\mathrm{in}}\approx 1.08$, which is evaluated at the inner edge of the disk ($r_\mathrm{in}$) and $\xi=1+\frac{t}{t_\nu}$ with the viscous time $t_\nu$
\begin{equation}
        t_\nu = \frac{R_0^2}{3(2-\psi)^2\nu(R_0)}.
\end{equation}
We used the numerical approach described in \citet{Birnstiel2010} to solve the evolution of the gas surface density.

Considering the inward drift of pebbles (Eq. \ref{eq:peb_velocity}), the evolution of the solid surface density can be described by \citep{Birnstiel2010,Birnstiel2012,Birnstiel2015}
\begin{multline}\label{eq:evo_evo_dust}
    \frac{\partial\Sigma_{\mathrm{Z,Y}}}{\partial t} +\frac1r\frac{\partial}{\partial r}\left\{ r\left[\Sigma_{\mathrm{Z,Y}}\cdot \bar u_{\mathrm{Z}}-\frac{\partial}{\partial r}\left(\frac{\Sigma_{\mathrm{Z,Y}}}{\Sigma_{\mathrm{gas}}}\right)\cdot\Sigma_{\mathrm{gas}} \nu \right]\right\}\\
    = - \dot\Sigma_\mathrm{Y} - \dot\Sigma_{\rm Y}^\mathrm{acc,peb},
\end{multline}
where $\dot\Sigma_{\rm Y}^\mathrm{acc,peb}$ is the source term that accounts for the discount of accreted pebbles from the dust surface density in the grid cell of the planet.
The source term $\dot\Sigma_\mathrm{Y}$ originates from pebble evaporation and condensation and is given by
\begin{equation}\label{eq:sink_ice}
        \dot\Sigma_\mathrm{Y} = 
        \begin{cases}
                \dot\Sigma^{\mathrm{evap}}_\mathrm{Y} & r<r_{\mathrm{ice,Y}}\\ 
                \dot\Sigma^{\mathrm{cond}}_\mathrm{Y}    & r\geq r_{\mathrm{ice,Y}}.\\ 
        \end{cases}
\end{equation}
Here $\dot\Sigma^{\mathrm{evap}}_Y$ and $\dot\Sigma^{\mathrm{cond}}_Y$ are the evaporation and condensation source terms of species Y for the two transport Eqs. \ref{eq:evo_evo_gas} and \ref{eq:evo_evo_dust}.

In order to allow for mass conservation we require that no more than 90\% of the local surface density is evaporated or condensed in one time step $\Delta t$:
\begin{equation}
    |\dot\Sigma_{\rm Y}| = \min\left[|\dot\Sigma_{\rm Y}|, 0.9\frac{\min\left(\Sigma_\mathrm{gas, Y},\Sigma_\mathrm{Z, Y}\right)}{\Delta t}\right]
.\end{equation}
We used a fixed timestep of $\Delta t = \SI{10}{\yr}$ for all of our models.

For the condensation term we assume that gas can only condensate by sticking on the surface (with efficiency $\epsilon_p=0.5$) of existent solids. The condensation term is then given by
\begin{equation}
        \dot\Sigma^{\mathrm{cond}}_\mathrm{Y} = 
\frac{3\epsilon_p}{2\pi\rho_\bullet}\Sigma_{\mathrm{gas,Y}}\left(\frac{\Sigma_\mathrm{Z}}{a_\mathrm{dust}}+\frac{\Sigma_\mathrm{peb}}{a_\mathrm{peb}}\right)\Omega_\mathrm{k}\sqrt{\frac{\mu}{\mu_\mathrm{Y}}},
\end{equation}
where $\mu_\mathrm{Y}$ is the mass (in proton masses) of a molecule of species Y. Here $a_\mathrm{dust}$ and $a_\mathrm{peb}$ are the particle sizes of the small and large dust distribution, respectively (see Sect.~\ref{sec:dustgrowth}). A derivation of the above equation can be found in Appendix~\ref{sec:cond}.

For the evaporation term we assumed that the flux of solids that drifts through the evaporation line is evaporated into the gas within \SI{1e-3}{AU}
\begin{equation}\label{eq:sink_ice_evap}
        \dot\Sigma^{\mathrm{evap}}_Y = \frac{\Sigma_\mathrm{Z,Y}\cdot \bar u_{\mathrm{Z}}}{\SI{1e-3}{AU}} \ .
\end{equation}
We show how the water ice fraction evolves in time for disks with different viscosities in Appendix~\ref{sec:watercontent}.

The time evolution of the gas and pebble surface density as shown in Fig. \ref{fig:sigma_evap} reveals that due to the inward drift of pebbles, the gas surface density changes on longer timescales compared to the solid surface density. This effect is enhanced for low viscosities, which allow larger grains and thus faster drift as well as slower viscous gas evolution.

The low viscosity has two main effects regarding the C/O ratio of the protoplanetary disk. At low viscosity, the pebbles grow larger (e.g., \citealt{Birnstiel2012}), which allows faster radial drift of the pebbles, enriching the gas with vapor when the pebbles evaporate at evaporation lines. As a consequence, the C/O ratio in the very inner disk is very low initially, due to the evaporation of water-rich pebbles. The low viscosity is then inefficient in diffusing the vapor inward, so that it takes several megayears to diffuse the gaseous methane inward, resulting in an increase in C/O only at late times. This effect is further aided by the inward diffusion of water vapor, which also increases the C/O in the gas phase. We discuss more about the water content in the protoplanetary disk below. This effect clearly depends on the disk's viscosities, where larger viscosities allow an increase to super-solar C/O values, while disks with low viscosities ($\alpha=10^{-4}$) always have sub-solar C/O values interior of the water ice line (Fig.~\ref{fig:sigma_evap}). Furthermore, the evaporation of inward drifting pebbles does not only influence the C/O ratio in the disk, but also the heavy element content in the gas phase (Fig.~\ref{fig:heavy}), which is larger in the initial phase, but then decreases in time, as the pebble supply originating from the outer disk diminishes in time.

The heavy element enrichment of the gas phase of the disk interior to the evaporation lines is key to understand the heavy element enrichment of gas accreting planets. Hints for this icy pebble migration across evaporation lines could also explain features of observed protoplanetary disks \citep{Banzatti2020, Zhang2020}.

Gas diffuses and spreads outward, crossing the evaporation lines where it can condense, leading to pebble pileups that can be seen especially around the water ice line in the pebble surface density (Fig~\ref{fig:sigma_evap}). This process is also invoked by \citet{Aguichine2020} to explain the composition of refractory materials in the solar systems at so-called rock lines. \citet{Mousis2021} uses the same processes, but around volatile lines, to explain the chemical composition of the comet C/2016 R2. The results of our simulations are in line with theirs.

\subsection{Disk lifetime}\label{sec:lifetime}

Observations of disks indicate that disks live for a few million years \citep{Mamajek2009}. The final stages of the disk is determined by photoevaporation \citep[e.g.,][]{Ercolano2009,Pascucci2009,Ercolano2010, Owen2012,Owen2013}. We did not include an exact photoevaporation model here but instead began to drastically decay the gas surface density once it reached a critical time. We used a sink term in the viscous evolution (Eq. \ref{eq:evo_evo_gas}) given by 
\begin{equation}\label{eq:sink_evap}
        \dot\Sigma_\mathrm{gas,Y}^{W} = 
                \begin{cases}
                        0 & t<t_\mathrm{evap}\\\frac{\Sigma_\mathrm{gas,Y}}{\tau_\mathrm{decay}} & t\geq t_\mathrm{evap}
                \end{cases}.
\end{equation}
The starting time of the disk clearance $t_\mathrm{evap}$ was generally set to $\SI{3}{Myr}$ and the decay timescale $\tau_\mathrm{decay}$ to $\SI{10}{kyr}$. A similar approach is used in the N-body simulations of \citet{Izidoro2019} and \citet{Bitsch2019}. We discuss more about how photoevaporation would influence our results in Sect.~\ref{sec:discussion}.

\subsection{Migration}\label{sec:migration}

Growing planets with mass $M$ naturally interact gravitationally with the ambient disk, which causes the planet to change its orbital elements (for a review, see \citealt{Kley2012} and \citealt{Baruteau2014}). This migration process depends on the angular momentum of the planet
\begin{equation}
        J_p = M a_p^2 \Omega_\mathrm{K}.
\end{equation}
The change of orbital position $a_p$ is then given as \citep{Armitage} 
\begin{equation}\label{eq:migrate}
        \dot{a_p}=\frac{\mathrm{d} a_p}{\mathrm{d}t} = \frac{a_p}{\tau_M} = 2 a_p \frac{\Gamma}{J_p},
\end{equation}
where $\tau_M$ is the migration timescale and $\Gamma$ the torque that acts on the planet. Low mass planets only disturb the disk slightly and migrate in the type-I fashion, where the torque acting on the planet is a combination of the Lindblad and corotation torques. We used here the torque formula by \citet{Paardekooper2011}, which includes the Lindblad as well as as the barotropic and entropy related corotation torques.

Besides these classical torque, new studies reveal that the thermal torque \citep{Lega2014, Benitez-Llambay2015, Masset2017}, originating from density perturbations close to the planet due to thermal heat exchange between the planet and the disk, as well as the dynamical torque \citep{Paardekooper2014,Pierens2015}, originating from feedback processes of the migration rate of the planet on the torque, can play a vital role in the orbital evolution of the planet.

The thermal torque consists of a cooling torque and a heating torque due to the bombardment and ablation of solids. We followed the description of \citet{Masset2017} to include the thermal torque using the accretion luminosity as defined by \citet{Chrenko2017}
\begin{equation}
        L = \frac{G M}{a_p}\dot M_\mathrm{peb},
\end{equation}
where $\dot M_\mathrm{peb}$ is the accretion rate of pebbles onto the planet (see below). Depending on the accretion efficiency of the planet, this thermal torque can lead to outward migration \citep{Guilera2019, Baumann2020}.

For the dynamical torque we followed \citet{Paardekooper2014} and replaced Eq. \ref{eq:migrate} by
\begin{equation}\label{eq:migrate_dyn}
        \frac{\mathrm{d} a_p}{\mathrm{d}t} = \frac{a_p}{\tau_M} = 2 a_p \Theta \frac{\Gamma}{J_p},
\end{equation}
where $\Theta$ is the numerical parameter that determines the effects of migration onto the migration rate (Eqs. 31 and 32 from \citet{Paardekooper2014}). The dynamical torque can also help to significantly slow down inward migration of low mass planets, preventing large-scale migration of planets \citep{Ndugu2021}.

Planets that start to accrete gas efficiently, open gaps in the protoplanetary disk (\citealt{Crida2006, CridaMorbidelli2007}). The gap opening is caused by gravitational interactions between the disk and the planet, but can also be aided by gas accretion itself \citep{CridaBitsch2017, Bergez2020}. A combined approach of gap opening via gravity and gas accretion is implemented in \citet{Ndugu2021} and we follow their approach. We first describe here the gap opening by gravity and include later on the gap opening via gas accretion.

A gap by gravity can be opened (with $\Sigma_{\rm Gap} < 0.1 \Sigma_{\rm gas}$), when
\begin{equation}
\label{eq:gapopen}
 \mathcal{P} = \frac{3}{4} \frac{H_{\rm gas}}{R_{\rm H}} + \frac{50}{q \mathcal{R}} \leq 1 \ ,
\end{equation}
where $R_\mathrm{H}=a_p\left(\frac{M}{3M_\star}\right)^{1/3}$ is the planetary Hill radius, $q=M / M_\star$, and $\mathcal{R}$ the Reynolds number given by $\mathcal{R} = a_p^2 \Omega_{\rm K} / \nu$ \citep{Crida2006}. The depth of the gap caused by gravity is given in \citet{CridaMorbidelli2007} as
\begin{equation}
 f(\mathcal{P}) = \left\{
  \begin{array}{cc}
   \frac{\mathcal{P}-0.541}{4} &\quad \text{if} \quad \mathcal{P}<2.4646 \\
   1.0-\exp\left(-\frac{\mathcal{P}^{3/4}}{3}\right) &\quad \text{otherwise}
  \end{array}
  \right. . 
  \end{equation}
We note that the gap depth can also be influenced by gas accretion, so we used in our code
\begin{equation}
    f_{\rm gap} = f(\mathcal{P}) f_{\rm A} \ 
\end{equation}
to calculate the gap depth relevant to switch from type-I to type-II migration. Here $f_{\rm A}$ corresponds to the contribution of accretion to the gap depth (see below in Sect.~\ref{sec:gas_accretion}), as discussed in \citet{Ndugu2021}.

If the planet becomes massive enough to achieve a gap depth of 10\% of the unperturbed gas surface density, it opens up a gap in the disk, and it migrates in the pure type-II regime, where the migration timescale is given as $\tau_{\rm visc} = a_p^2 / \nu$. However, if the planet is much more massive than the gas outside the gap, it will slow down. This happens if $M > 4\pi \Sigma_{\rm gas} a_p^2$, which leads to the migration timescale of
\begin{equation}
\label{eq:typeII}
 \tau_{\rm II} = \tau_{\nu} \times \max \left(1 , \frac{M}{4\pi \Sigma_{\rm gas} a_p^2} \right) \ ,
\end{equation}
resulting in slower inward migration for more massive planets \citep{Baruteau2014}.

In addition, we used a linear smoothing function for the transition between planets that open partial gaps inside the disk (that migrate with the reduced type-I speed by the factor $f_{\rm gap}$) and planets that migrate with type-II, because even the reduced type-I migration rate (if the gap is fully opened with $\mathcal{P}<1$) is different from the nominal type-II rate. This approach is also used in \citet{Bitsch2015}.

\subsection{Accretion}\label{sec:accretion}

We illustrate the planetary growth in our model in Fig.~\ref{fig:phases}. Planetary embryos are planted into the disk with masses, where pebble accretion starts to become efficient \citep[e.g.,][]{Lambrechts2012,Johansen2017peb}:
\begin{equation}
        M_t = \sqrt\frac{1}{3}\frac{\Delta v^3}{G\Omega_\mathrm{K}},
\end{equation}
with a typical value of $M_t \approx \num{5e-3} M_\odot$ at \SI{10}{AU}. Our planets then start to accrete pebbles (see Sect.~\ref{sec:pebble_accretion}). When the planet is massive enough to form a pressure bump in the gas surface density, pebbles are hindered from reaching the planet. This transition occurs at the pebble isolation mass $M_\mathrm{iso}$ \citep[see][]{Morbidelli2012,Lambrechts2014_iso,Ataiee2018,Bitsch2018_peb_iso}, where we follow the pebble isolation mass prescription of \citet{Bitsch2018_peb_iso}.

At $M=M_\mathrm{iso}$ the accretion of pebbles ends and the core is completely formed. Core accretion is then followed by envelope contraction and envelope accretion (see Sect.~\ref{sec:gas_accretion}).

During the buildup of the core, we set the fraction of matter that is accreted to the core (with mass $M_c$) at $M<M_\mathrm{iso}$ to $90\%$ of the total accreted matter ($10\%$ of accreted pebbles contribute to the primary envelope). This approach is supposed to account for pebble evaporation into the planetary envelope during core buildup in a simplified way compared to more sophisticated models (e.g., \citealt{Brouwers2020, Ormel2021}), who actually show that even less than 50\% of the solids accreted via pebbles make up the core and the evaporated pebbles make up most of the heavy element content of these forming planets \citep{Ormel2021}. The initial growth during the core buildup in our model is thus simply described via
\begin{equation}
    \dot{M}_{\rm core} = 0.9 \dot{M}_{\rm peb}; \quad
    \dot{M}_{\rm gas} = 0.1 \dot{M}_{\rm peb} \ ,
\end{equation}
where $\dot{M}_{\rm peb}$ describes the pebble accretion rate onto the planetary core (see below). We discuss this assumption, which mainly influences the atmospheric C/O, but not the total heavy element content, in more detail in Appendix~\ref{sec:50:50}. After the pebble isolation mass is reached, all material is accreted into the planetary envelope $M_a$.

\subsubsection{Pebble accretion}\label{sec:pebble_accretion}

The accretion of small millimeter to centimeter sized objects, so-called pebbles, is thought to significantly accelerate the growth process of planetary cores \citep{Ormel2010, Johansen2010, Lambrechts2012}. Here, we follow the pebble accretion rates derived in \citet{Johansen2017peb}. The accretion rate of pebbles on a growing protoplanet is determined by the azimuthal flux of pebbles $\rho_\mathrm{peb}\delta v$ through the cross section of the accretion sphere $\pi R_\mathrm{acc}^2$ of the planet:
\begin{equation}\label{eq:m_dot_peb}
        \dot M_\mathrm{peb} = \begin{cases}
        \pi R_\mathrm{acc}^2\rho_\mathrm{peb}\delta v & \textrm{3D accretion}\\
         2 R_\mathrm{acc}\Sigma_\mathrm{peb}\delta v & \textrm{2D accretion}
         \end{cases}.
\end{equation}
The radius of the accretion sphere $R_\mathrm{acc}$ strongly depends on the Stokes number of the pebbles:
\begin{equation}\label{eq:peb_acc_radius}
        R_\mathrm{acc} \propto \left(\frac{\Omega_\mathrm{K}\mathrm{St}}{0.1}\right)^{1/3}R_\mathrm{H},
\end{equation} 
where $R_\mathrm{H}$ is the planetary hill radius. The optimal Stokes number for pebble accretion is approximately 0.1. 

The transition criterion for the transition from 3D to 2D accretion is given by \citep{Morbidelli2015}
\begin{equation}
        H_\mathrm{peb} < \frac{2\sqrt{2\pi}     R_\mathrm{acc}}{\pi},
\end{equation}
where we use the pebble scale height $H_\mathrm{peb} = H_\mathrm{gas}\sqrt{\alpha_z/\mathrm{St}}$ and the relation $\rho_\mathrm{peb}=\Sigma_\mathrm{peb}/(\sqrt{2\pi}H_\mathrm{peb})$. In the case of 2D pebble accretion, the planetary accretion radius is larger than the midplane pebble scale height of the disk, so that the planet can accrete from the full pebble flux passing the planetary orbit. This is not the case in the 3D pebble accretion regime, where the planetary accretion radius is smaller than the pebble scale height and only a fraction of the pebble flux passing the planet can contribute to the accretion. We use here $\alpha_z=\num{1e-4}$ for all our simulations (Table \ref{tab:parameters}), motivated by the constraints from protoplanetary disk observations, which show low level of vertical stirring \citep{Dullemond2018DSHARP}. Motivated by simulations \citep{Nelson2013, Turner2014,Flock2015} and observations \citep{Dullemond2018DSHARP, Flaherty2018}, \citet{Pinilla2021} study how different sources and values of turbulence for vertical stirring, radial diffusion, and gas viscous evolution influence grain growth and drift, finding that indeed different values for these parameters allow a better match to the observations. For simplicity we thus keep $\alpha_{\rm z}$ fixed in our simulations and only vary $\alpha$, responsible for the disk evolution, gas accretion and migration.

The pebble surface density and the Stokes number are a natural outcome of our disk evolution model (see Sect.~\ref{sec:dustgrowth}) and depend on the initial solid to gas ratio ($\epsilon_0$). This approach is an improvement compared to previous planet formation simulations via pebble accretion, where mostly a simplified pebble growth and drift model is used (e.g., \citealt{Lambrechts2014, Bitsch2015, Ndugu2018}), but approaches using a model with accretion rates depending on a full pebble size distribution have also been implemented in other works \citep{Guliera2020, Venturini2020a, Savvidou2021, Drazkowska2021}.

\subsubsection{Gas accretion}\label{sec:gas_accretion}

In reality, gas contraction can already happen during the buildup of the core, where the efficiency increases once the heat released by infalling solids stops. We follow here a simple two-step process, where the planetary envelope can quickly contract and runaway gas accretion can start once the planet has reached its pebble isolation mass, where the accretion of solids stop.

In our model, the gas accretion onto the planet is given by the minimum between the accretion rates given by \citet{Ikoma2000}, \citet{Machida2010} and by the gas the disk can viscously provide into the horseshoe region after the planet has emptied the horseshoe region \citep{Ndugu2021}. We follow here the approach outlined in \citet{Ndugu2021}, which is derived for H-He gas. We discuss how vapor-enriched gas accretion (e.g., \citealt{Hori2011, Venturini2015}) could influence our results in Sect.~\ref{sec:discussion}.

The gas accretion rates of \cite{Ikoma2000} are given by 
\begin{equation}  
\label{Ikoma2000}
{\dot{M}_{\rm gas, Ikoma}} =  \frac{M}{\tau_{\rm KH}} ,
\end{equation} 
where $\tau_{\rm KH}$ is the Kelvin-Helmholtz contraction rate and  scales as
\begin{equation}
{\tau_{\rm KH} }= 10^{3}\left(\frac{M_{\rm c}}{30 M_{\rm E}}\right)^{-2.5}\left(\frac{\kappa_{\rm env}}{0.05 \rm cm^{2}g^{-1}}\right){\rm year}.
\end{equation}
Here $M_{\rm c}$ is the mass of the planet's core; this is in contrast with $M$, which is the full planet mass (core plus envelope). We set the envelope opacity for simplicity to 0.05 cm$^2$/g for all our planets. 

\cite{Machida2010} give the gas accretion rate $\dot{M}_{\rm gas, Machida}$ as the minimum of:
\begin{equation}
 \dot{M}_{\rm gas,\rm low} = 0.83\Omega_{\rm k}\Sigma_{\rm gas}H_{\rm gas}^{2}\left(\frac{R_{\rm H}}{H_{\rm gas}}\right)^{\frac{9}{2}}
 \label{Machida1}
\end{equation}
and 
\begin{equation}
 \dot{M}_{\rm gas,\rm high} = 0.14 \Omega_{\rm k}\Sigma_{\rm gas} H_{\rm gas}^{2} \ .
 \label{Machida2}
\end{equation}
The \cite{Machida2010} rate is derived from shearing box simulations, where gap formation is not taken fully into account. However, once a gap is opened, obviously the planet cannot accrete more gas than the disk can supply. Throughout our simulations, we modeled the disk supply rate
\begin{equation}
\label{disk}
\dot{M}_{\rm disk}= - 2\pi r \Sigma_{\rm gas} u_{\rm gas},
\end{equation} 
where $u_{\rm gas}$ is the radial gas velocity (Eq. \ref{eq_the:radial_velocity_general}). The radial gas velocity depends linearly on $\alpha$, which therefore sets the accretion flow in the disk, so it provides the gas accretion rate to the planet. Therefore, a change in $\alpha$ would imply a different accretion flow through the disk and thus sets a different limit on the planet accretion rate. In summary, our gas accretion rate, ${\dot{M}_{\rm gas}}$ onto the planet is taken as
\begin{equation}
\label{eq:gasaccrete}
 \dot{M}_{\rm gas} = \min{\left( \dot{M}_{\rm gas, Ikoma}, \dot{M}_{\rm gas, Machida},\dot{M}_{\rm disk}+\dot{M}_{\rm HS}\right)},
\end{equation} where $ \dot{M}_{\rm HS}$ is the horseshoe depletion rate, given by 
\begin{equation}
 \dot{M}_{\rm HS}=M_{\rm HS}/(2T_{\rm HS}),
\end{equation} 
where $M_{\rm HS}$ is the mass of the horseshoe region, $T_{\rm HS}=2\pi / \Omega_{\rm HS}$ is the synodic period at its border with $\Omega_{\rm HS} = 1.5 \pi \Omega r_{\rm HS}/a_p$, $r_{\rm HS} = x_s a_p$ is its half-width \citep[with $x_s$ from ][Eq. 48]{Paardekooper2011}. At each time step $\Delta t$ we compute the mass accretion rate $\dot{M}_{\rm HS}$ that could be provided by the horseshoe region.

Following the same philosophy as for gravitational gap opening, we introduce an additional parameter $f_{\rm A}$, initially equal to 1, which is computed every time step. $f_{\rm A}$ scales as
\begin{equation} 
\label{eq:gapacc}
f_{\rm A} = 1-\frac{{\dot{M}_{\rm gas}}\delta{t}}{f({\cal{P}})\hat{M}_{\rm HS}}.
\end{equation}
Here $\hat{M}_{\rm HS}$ is the mass inside the horseshoe region in the absence of gas accretion onto the planet and in absence of gravitational gap opening. $\hat{M}_{\rm HS}$ is given by
\begin{equation}
 \hat{M}_{\rm HS} =2\pi a_p r_{\rm HS} \hat{\Sigma}_{\rm HS}. 
\end{equation} 
The full depth of the gap is therefore
 \begin{equation}
 \label{gapeff}
 f_{\rm gap}={f(\cal{P})} f_{\rm A}.
\end{equation}
The formulae above (Eq.~\ref{eq:gapacc}) requires us to monitor the mass of the horseshoe region 
$M_{\rm HS}$ as a function of time. By definition of the $f(\cal{P})$ and $f_{\rm A}$ 
factors, the mass of the horseshoe region scales as
\begin{equation}
\label{HSmass}
M_{\rm HS}={f(\cal{P})}f_{\rm A}\hat{M}_{\rm HS}.
\end{equation}
The quantity $\hat{M}_{\rm HS}$ evolves over time because the width of the horseshoe region $r_{\rm HS}$ changes with the planet mass and location. For simplicity, we assume that the vortensity in the horseshoe region is conserved (strictly speaking this is true only in the limit of vanishing viscosity), so that if the location of the planet changes from $a_p$ to $a_p'$, the horseshoe gas density changes from 
\begin{equation}
\hat{\Sigma}_{\rm HS}=\hat{M}_{\rm HS}/(4\pi a_p r_{\rm HS})
\end{equation} 
to 
\begin{equation}
\hat{\Sigma}'_{\rm HS}=\hat{\Sigma}_{\rm HS}(a_p/a_p')^{3/2}.
\end{equation}

Thus, when $r_{\rm HS}$ changes to $r'_{\rm HS}$, we compute the quantity
\begin{equation} 
\hat{M}'_{\rm HS}=4\pi a_p'r'_{\rm HS} \hat{\Sigma}'_{\rm HS}. 
\end{equation}
If $\hat{M}'_{\rm HS}<\hat{M}_{\rm HS}$, we refill the horseshoe region at the disk's viscous spreading rate and recompute $\hat{M}'_{\rm HS}$ as
\begin{equation}
 \hat{M}'_{\rm HS} =  \hat{M}_{\rm HS} + (\dot{M}_{\rm disk}-\dot{M}_{\rm gas})\Delta t, 
\end{equation} where $\dot{M}_{\rm disk}$ and $\dot{M}_{\rm gas}$ are defined in Eqs.~\ref{disk} and \ref{eq:gasaccrete}, respectively. If the opposite is true, it means that the horseshoe region has expanded and must have captured new gas from the disk, with a density $\Sigma_{\rm gas}$. Thus, we compute the new value of $\hat{M}_{\rm HS}$ as:
\begin{equation}
\label{eq:newMH}
\hat{M}'_{\rm HS}=\hat{M}_{\rm HS} + \left(4\pi a_p' r'_{\rm HS} -\frac{\hat{M}_{\rm HS}}{\hat{\Sigma}'_{\rm HS}}\right) {\Sigma_{\rm gas}}\ .
\end{equation} 
Once $\hat{M}'_{\rm HS}$ is computed, the new value of $\hat{\Sigma}'_{\rm HS}$ is recomputed as $\hat{\Sigma}'_{\rm HS}=\hat{M}'_{\rm HS}/(4\pi a_p' r'_{\rm HS})$. This procedure is then repeated at every time step. This procedure automatically captures the gas surface density decay during the disk's evolution because $\Sigma_{\rm gas}$ is evaluated at each time step. This approach is outlined, as described, in \citet{Ndugu2021}.

\subsection{Gap profile}

\begin{figure*}
        \includegraphics[width=\textwidth]{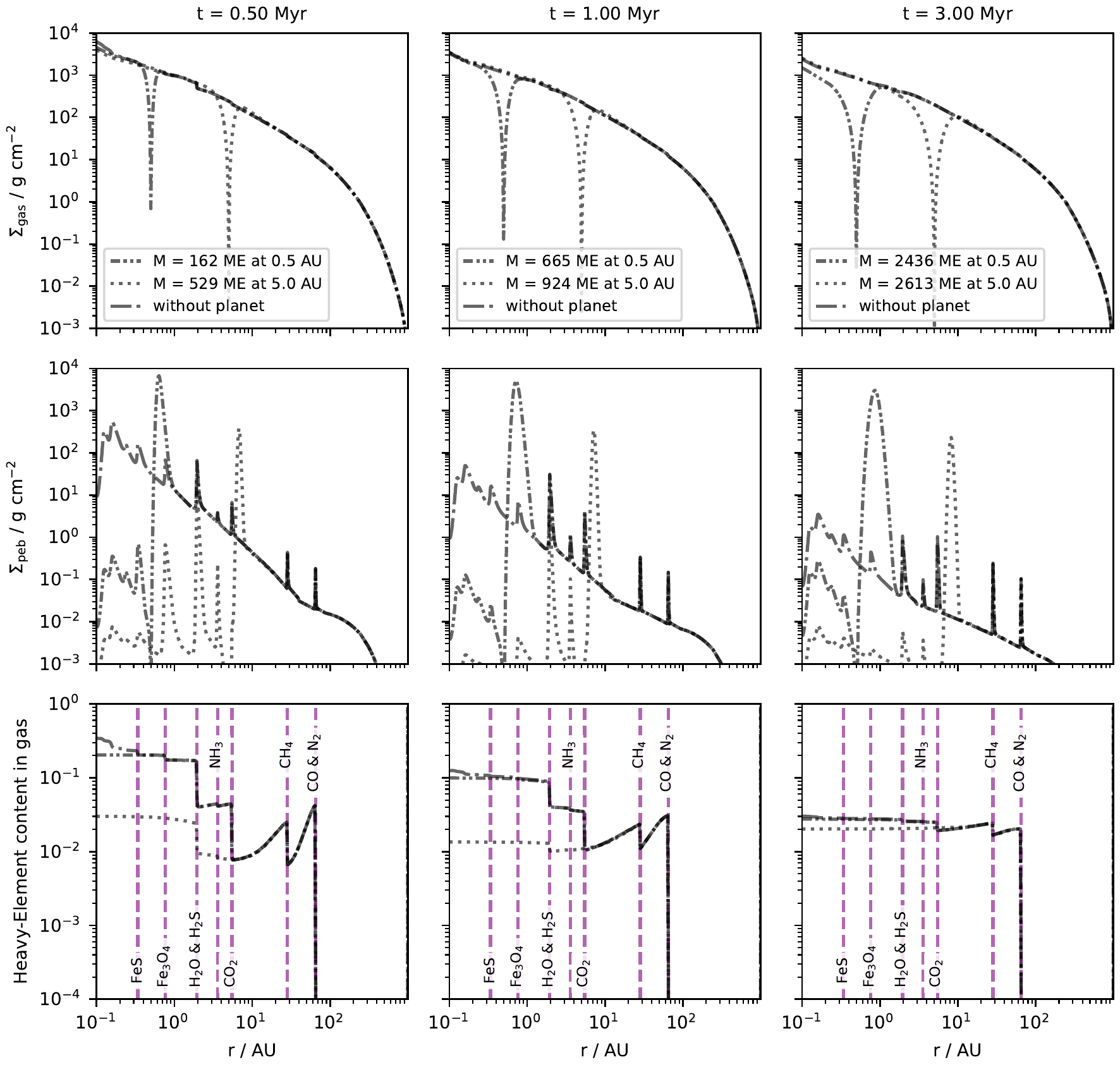}
        \caption{Gas surface density (top), pebble surface density (middle), and heavy element content $\Sigma_\mathrm{gas,Y}/\Sigma_\mathrm{gas}$ (bottom) of the gas surface density as a function of radius for different times using the evaporation and condensation treatment described in Sect.~\ref{sec:evo} in the case of either no planet or planets fixed at \num{0.5} or \SI{5}{AU}. These planets are either located interior (\SI{0.5}{AU}) or exterior (\SI{5}{AU}) to the water ice line. Disk parameters can be found in Table \ref{tab:parameters}; here we use $\alpha=\num{0.0005}$.}
        \label{fig:heavy}
\end{figure*}

\begin{figure*}
    \centering
    \includegraphics[width=\textwidth]{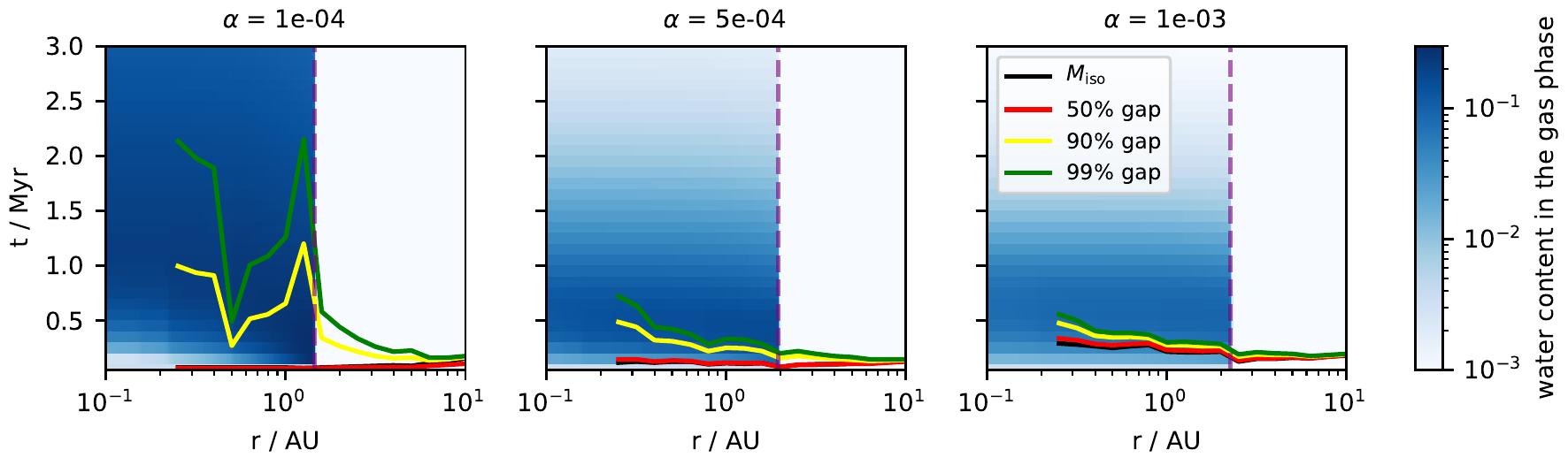}
    \caption{Water content in the gas phase as a function of time for different disk viscosities. The black line marks the time a non-migrating growing planet starting at $t=\SI{0.05}{Myr}$ needs to reach the pebble isolation mass, while the red, yellow, and green lines indicate the time the same growing planet needs to reach a certain gap depth.}
    \label{fig:watertimenoplanet}
\end{figure*}
\begin{figure*}
    \centering
    \includegraphics[width=\textwidth]{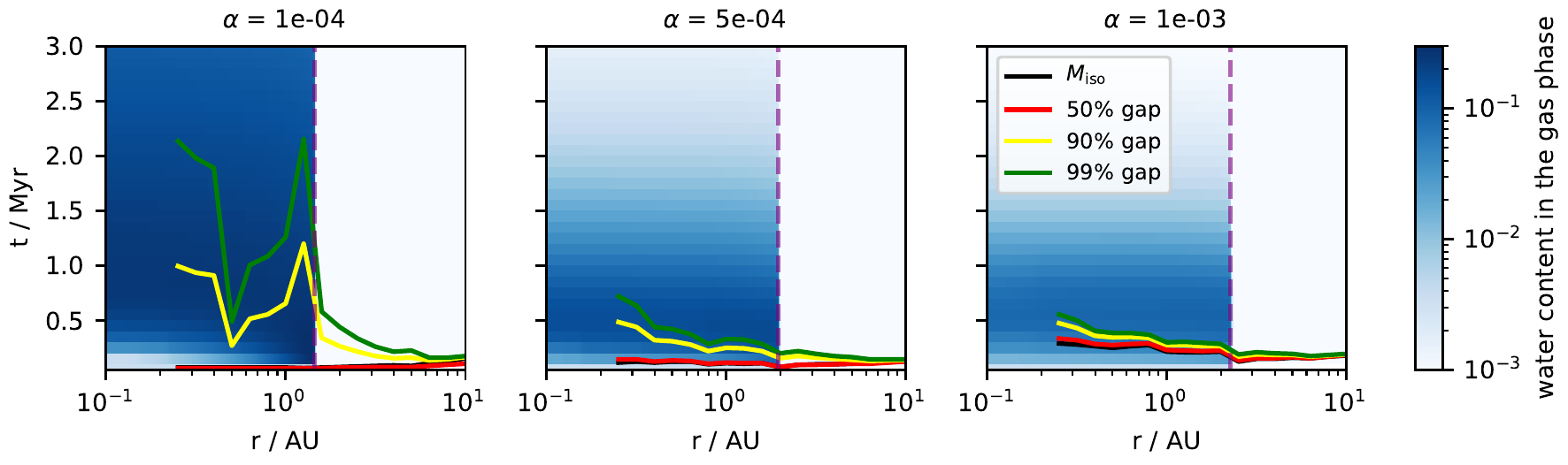}
    \caption{Same as Fig.~\ref{fig:watertimenoplanet} but with a growing non-migrating planet placed at 0.5 AU. The water content of the growing planet is displayed in Fig.~\ref{fig:watertime}. The planet growing interior to the water ice line has only a little influence on the disk's water vapor content.}
    \label{fig:watertime05}
\end{figure*}

\begin{figure*}
    \centering
    \includegraphics[width=\textwidth]{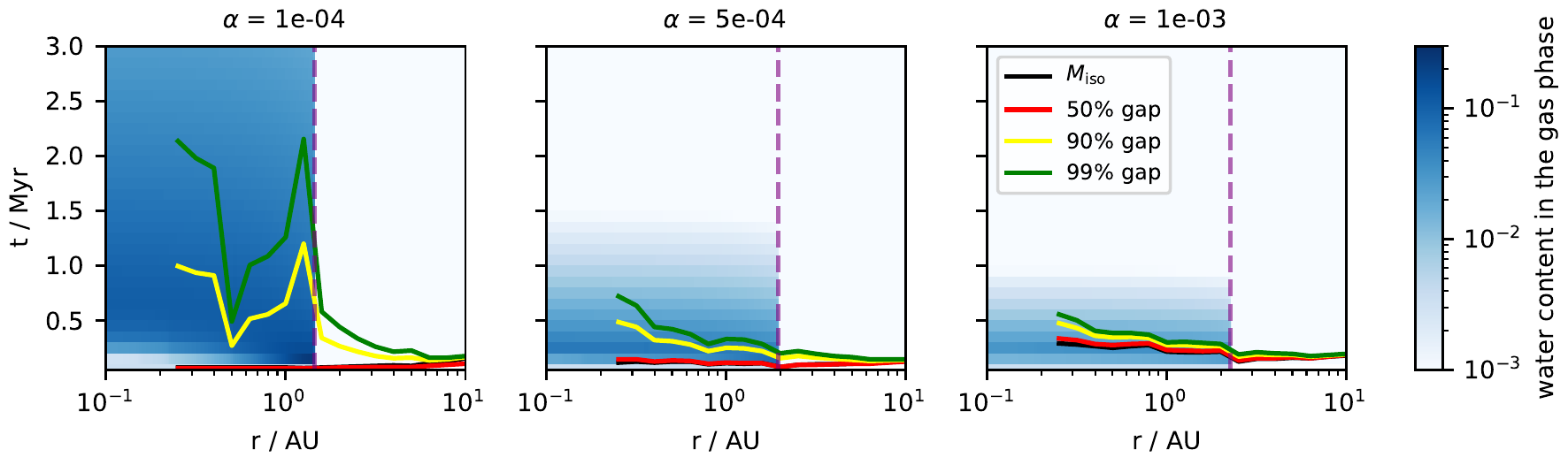}
    \caption{Same as Fig.~\ref{fig:watertimenoplanet} but with a growing non-migrating planet placed at 5.0 AU. The water content of the growing planet is displayed in Fig.~\ref{fig:watertime}. The planet exterior to the water ice line has a strong influence on the water vapor content in the inner disk, compared to planets interior to the water ice line (Fig.~\ref{fig:watertime05}) or if there are no planets (Fig.~\ref{fig:watertimenoplanet}).}
    \label{fig:watertime50}
\end{figure*}

Pebbles will stop drifting inward when the planet is massive enough to open a small gap in the protoplanetary disk. In reality, the torque from the planet acting on the disk is responsible for the opening of the gap (e.g., \citealt{Crida2006}). For simplicity, we chose an approach with varying viscosity to mimic the effect of a pressure bump caused by a growing planet (e.g., \citealt{Pinilla2021}). We did so by applying a gap profile inversely to the viscosity once the planet has reached the pebble isolation mass in order to keep the mass accretion rate of the gas phase of the protoplanetary disk constant. The viscous $\alpha$ parameter in Eq. \ref{eq:evo_evo_gas} is then given at the planet's location by
\begin{equation}
        \alpha = \alpha / \aleph(r),
\end{equation}
where the numerical factor $\aleph(r)$ describes the gap profile, which is approximated by a Gaussian distribution
\begin{equation}
        \aleph(r) = 1 - [1-f_\mathrm{gap}]\exp\left(0.5\left[(r-a_p)/\sigma\right]^2\right)
\end{equation}
around the planetary position with a width given by the horseshoe width
\begin{equation}
        \sigma = 2 r_\mathrm{HS} / [2 \sqrt{2 \log(2)}].        
\end{equation}

This increased $\alpha$ parameter is then only used to calculate the gap in the gas surface density profile. All other calculations within the code use the default $\alpha$ parameter. We interpolate quantities to the planetary position by excluding the gap (i.e., use only those grid cells that satisfy $\aleph < 1\%$), since our descriptions of the migration and accretion rates depend on the unperturbed disk.

The effects of the planetary gap on the gas surface density, the heavy element content and the pebble distribution can be seen in Fig.~\ref{fig:heavy}, where we compare simulations of a protoplanetary disk that include non-migrating planets at 0.5 or 5 AU with a protoplanetary disk that does not host a planet. The gap generated by the growing planet results in deep deletions in the gas surface density, where consequently pressure bumps exterior to the planetary orbit are formed. The gap widens in time, as the planet grows. We additionally show in Figs.~\ref{fig:watertimenoplanet}, \ref{fig:watertime05}, and \ref{fig:watertime50} the water vapor content in disks without and with planets (at 0.5 and 5.0 AU) as a function of time for different disk viscosities.

It is clear from Fig.~\ref{fig:heavy} that the growing planets cause pressure bumps exterior to their orbits in the protoplanetary disk, where pebbles can accumulate. These pile-ups in the pebble surface density are much larger than the pile-ups caused by condensation in the outer disk regions. Exterior to the pressure bump exerted by the planet, the disk structure is not affected and the pile-ups in the pebble surface density are caused by condensation and evaporation. Furthermore, the planets are very efficient at blocking the pebbles, resulting in low pebble surface densities interior to the planetary orbits, because our pebble evolution model does not contain multiple pebble species during the pebble drift step, which prevents a detailed filtering mechanism at pressure bumps (e.g., \citealt{Weber2018}). Interior to the pressure bump generated by the giant planet, spikes in the pebble distribution at the evaporation lines are still present, especially around the water ice line\footnote{ At this point in the disk, the water ice makes up 33\% of the total solid mass, clearly overpowering the contribution of H$_2$S.}. This spike in the pebble distribution is fuelled by outward diffusing condensing vapor. However, in time, the vapor has diffused inward, hindering efficient condensation, diminishing the pebble pile-up at the evaporation fronts interior to the giant planet, resulting in a very low pebble density. This effect can be seen clearly in the middle panel with $\alpha=5 \times 10^{-4}$ of Fig.~\ref{fig:watertime50}, where the water vapor content decreases sharply after \SI{1}{Myr}.

The heavy element content in the gas phase in the inner disk increases toward the star because pebbles drift inward very efficiently, where they evaporate once they cross the evaporation lines aided by the fact that gas diffusion is initially inefficient at low viscosities. This can be seen in Fig.~\ref{fig:watertimenoplanet} where the water vapor content is initially larger close to the water ice line, but then slowly increases in the whole inner disk as water vapor diffusion becomes efficient after a few \SI{100}{kyr}. The same effect is shown in the simulations of \citet{Garate2020}.

Especially at the water ice line, a jump in the heavy element content of the gas phase is observed, due the large water ice abundance. This is not only the case for the simulations without planets, but is also the case if the planet is interior to the water ice line, which only blocks the inward drifting pebbles once they have evaporated at the water ice line. Especially the water vapor content in the inner disk is very similar in both cases (Figs.~\ref{fig:watertimenoplanet} and~\ref{fig:watertime05}). In contrast, the heavy element content in the gas is much lower, if a planet starts to block pebbles exterior to the water ice line. Nevertheless, our simulations still show initially a small jump in the heavy element content of the gas phase for this case because the inward drifting pebbles enrich the gas before the planet generates a pressure bump to block the inward drifting pebbles. The water vapor then diffuses inward, reducing the heavy element content (Fig.~\ref{fig:heavy}). At very late times, however, the heavy element content in the gas phase rises again due to the inward diffusion of methane and CO, while the water vapor content is depleted (Fig.~\ref{fig:watertime50}). Consequently, the C/O ratio in the gas phase also increases (see Fig.~\ref{fig:sigma_evap} for the case without planet). It is clear that the positions of planets relative to the evaporation fronts influences the heavy element content in the gas phase (see also \citealt{Bitsch+2021}).

In time the heavy element content of the gas in the inner regions diminishes because the volatile vapor is slowly accreted onto the star and the disk has transported most of its pebbles into the inner disk (Fig.~\ref{fig:sigma_evap}), cutting the supply of new pebbles that could evaporate and contribute to the heavy element content of the gas.

The enrichment in heavy elements in the inner disk regions is a strong function of the disk's viscosity. At low viscosities, the pebbles grow larger and thus drift faster inward where they evaporate, increasing the heavy element content in the gas phase. In addition, the low viscosity is very inefficient in removing the volatile-rich gas onto the star. At high viscosity, pebbles are smaller and the disk is more efficient in diffusing the vapor-rich gas inward. As a consequence, the heavy element enrichment in the inner disk decreases with increasing viscosity. This effect can be seen very clearly for the evolution of the water vapor content in time (Figs.~\ref{fig:watertimenoplanet}, \ref{fig:watertime05}, and \ref{fig:watertime50}). This clearly indicates that the enrichment of the inner disk, and consequently planetary atmospheres, as discussed below, is a strong function of the disk's viscosity, where lower viscosities will allow a larger enrichment of the disk with vapor and consequently of planetary atmospheres with vapor. Observations of the water vapor content in the inner disk, could thus be helpful to constrain the disk's viscosity.

\subsection{Operating principle}
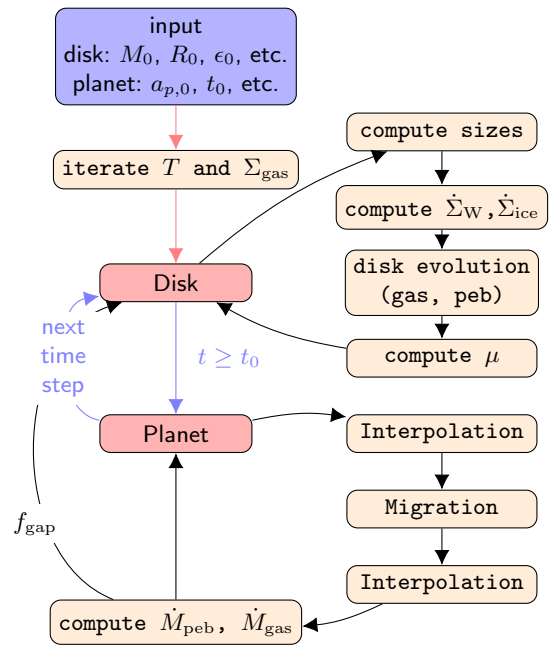
\begin{figure}
\centering
\begin{tikzpicture}[node distance=1cm,
    every node/.style={fill=white, font=\sffamily}, align=center]
  \node (input)    [activityStarts]                              {input\\disk: $M_0$, $R_0$, $\epsilon_0$, etc.\\planet: $a_{p,0}$, $t_0$, etc.};
  \node (input_T) [process, below of=input, yshift=-0.5cm] {iterate $T$ and $\Sigma_\mathrm{gas}$};
  \node (disk)     [startstop, below of=input_T, yshift=-0.5cm]   {Disk};
  
  \node (twopop)   [process, right of=disk, xshift=2.5cm, yshift=+2cm] {compute sizes};
  \node (sigdot)   [process, below of=twopop] {compute $\dot\Sigma_\mathrm{W}$,$\dot\Sigma_\mathrm{ice}$};
  \node (visc)     [process, below of=sigdot] {disk evolution\\(gas, peb)};
  \node (mu)     [process, below of=visc] {compute $\mu$};
  
  \node (planet)   [startstop, below of=disk, yshift=-1cm]   {Planet};
  \node (interp1)[process, right of=planet, xshift=2.5cm]{Interpolation};
  \node (migration)[process, below of=interp1]{Migration};  
  \node (interp2)[process, below of=migration]{Interpolation};
  
  \node (mdot)     [process, below of=planet, yshift=-1.5cm] {compute $\dot M_\mathrm{peb}$, $\dot M_\mathrm{gas}$};

  \draw[->, red!50]     (input) to (input_T);
  \draw[->, red!50]     (input_T) to (disk);
  
  \draw[->]                (disk) to[bend left=10] (twopop);
  \draw[->]                (twopop) to (sigdot);
  \draw[->]                (sigdot) to (visc);
  \draw[->]                (visc) to (mu);
  \draw[->]                (mu) to[bend left=10] (disk);
  
  \draw[->, blue!50]     (disk) to node[text width=0.9cm, xshift= 0.7cm]
                                   {$t \geq t_0$} (planet); 
  \draw[->]                (planet) to[bend left=10] (interp1);
  \draw[->]                (interp1) to (migration);
  \draw[->]                (migration) to (interp2);
  \draw[->]                (interp2) to[bend left=10] (mdot);
  
  \draw[->]     (mdot) to (planet);

  \draw[->]                (mdot) to[bend left=70] node[text width=3cm, yshift= -1cm] {$f_\mathrm{gap}$} (disk);
  
  \draw[->,blue!50]     (planet) to[bend left=80] node[text width=1cm]
                                   {next time step} (disk);

  \end{tikzpicture}

  \caption{Operating principle of \texttt{chemcomp}. The main loop is shown in blue. Black arrows connect the individual steps (beige nodes) that are performed in each time step. Red arrows indicate initialization steps.}
  \label{fig:op_principle}
        
\end{figure}

\begin{table}
        \begin{subtable}{.45\textwidth}
        \centering
        \begin{tabular}{c c c}
                \hline\hline
                Quantity & Value\\
                \hline
                $a_{p,0}$                        & $\SI{3}, \SI{10}, \SI{30}{AU}^{\star}$ & initial position\\
                $t_0$                        & $\SI{0.05}{Myr}^{\star}$ & starting time\\
                $\kappa_\mathrm{env}$& $\SI{0.05}{\cm\squared\per\g}$ & envelope opacity\\
                \hline
        \end{tabular}
        \caption{Planet}
        \end{subtable}  
        \begin{subtable}{.45\textwidth}
                \centering
        \begin{tabular}{c c c}
                \hline\hline
                Quantity & Value\\
                \hline
                $r_\mathrm{in}$       & $\SI{0.1}{AU}$ & inner edge\\
                $r_\mathrm{out}$      & $\SI{1000}{AU}$ & outer edge\\
                $N_\mathrm{Grid}$     & $\num{500}$ & Number of gridcells\\
                \hline
        \end{tabular}
        \caption{Grid}
        \end{subtable}
        \begin{subtable}{.45\textwidth}
                \centering
                \begin{tabular}{c c c}
                \hline\hline
                Quantity & Value\\
                \hline
                $\alpha$           & $(1, 5, 10) \times 10^{-4}$ & viscous alpha parameter\\
                $\alpha_z$         & $\num{1e-4}$ & vertical mixing\\
                $M_0$              & $\SI{0.128}{M_\odot}$ & initial disk mass\\
                $R_0$              & $\SI{137}{AU}$ & initial disk radius\\
                $[\mathrm{Fe}/\mathrm{H}]$ & 0 & host star metallicity\\
                $t_\mathrm{evap}$  & $\SI{3}{Myr}$ & disk lifetime\\
                $\epsilon_0$       & $\SI{2}{\%}^\star$ & solid to gas ratio\\
                $u_\mathrm{frag}$  & $\SI{5}{\m\per\s}$ & fragmentation velocity\\
                \hline
        \end{tabular}
        \caption{Disk}
        \end{subtable}
        \caption{Parameters used throughout this paper.}
        \begin{tablenotes}
        	\item \textbf{Notes:} The star symbol indicates those parameters whose values are varied in the figures associated with Table \ref{tab:thorngren_param}.
        \end{tablenotes}

        \label{tab:parameters}
\end{table}

Calculations in this paper are performed using the newly developed 1D code \texttt{chemcomp}. It provides a platform to simulate the above described physics. It includes a disk module (attributes are defined on a log-radial grid) that deals with the formation of pebbles (see Sect.~\ref{sec:dustgrowth}) as well as the dynamics of gas and pebbles (see Sect.~\ref{sec:evo}). It calculates the temperature of the disk (see Appendix~\ref{sec:num_T_it}) and the temperature-dependent compositions of gas and pebbles (see Sect.~\ref{sec:composition}) by also including effects induced by the existence of evaporation lines (see Sect.~\ref{sec:evo}).

The code also contains a planet module that handles growth (see Sect.~\ref{sec:accretion}) and migration (see Sect.~\ref{sec:migration}) of a single planet. The planet module acts as the supervisor of the disk module and \enquote{collects} the matter available in the disk. 

The operating principle of the code can also be divided into these two modules. As Fig.~\ref{fig:op_principle} shows, each time step begins ($\Delta t = \SI{10}{\yr}$) with the disk step. The disk step begins by computing the pebble growth and then computing the sink and source terms for the viscous evolution (Eqs. \ref{eq:sink_ice} and \ref{eq:sink_evap}). We then have everything in place to evolve the surface densities in time. We use a modified version of the donor-cell scheme outlined in \citet{Birnstiel2010} to solve Eqs. \ref{eq:evo_evo_dust} and \ref{eq:evo_evo_gas} for every molecular species. The realization in \texttt{chemcomp} is an adapted version from the implementation in the unpublished code \texttt{DISKLAB} \citep{disklab}. The inner disk boundary for the gas and solid evolution is treated using a Neumann boundary condition for $\Sigma_\mathrm{gas,Y}$ and $\frac{\Sigma_\mathrm{Z,Y}}{\Sigma_\mathrm{gas}}$, respectively. The outer boundary uses a fixed Dirichlet boundary condition.

We now have a disk that is advanced in time in which we calculate the torques acting on the planet by interpolating the disk quantities from the radial grid to the planetary position and then advancing its position (Eq. \ref{eq:migrate_dyn}). After a next interpolation of the disk quantities to the new position of the planet, accretion rates for pebble accretion (Eq. \ref{eq:m_dot_peb}) and gas accretion (Eq. \ref{eq:gasaccrete}), are calculated. The calculated accretion rates already include the chemical composition of the disk because the surface densities are treated as vectors, meaning that the resulting accretion rates are also given as compositional vectors. These accretion rates are now added to the planets composition 
\begin{equation}
        M_\mathrm{Y} \rightarrow M_\mathrm{Y} + \dot M_\mathrm{Y} \cdot \Delta t.
\end{equation}
The pebble and gas accretion rates are additionally converted to sink terms that are then added to the viscous evolution for the next time step. We remove the accreted pebbles only from the cell where the planet is located, since we do not numerically resolve the Hill sphere during pebble accretion.
If the planet migrates down to 0.2 AU\footnote{ This is beyond the inner edge of our disk because grid cells interior of the planet are needed to calculate gradients relevant for planet migration and gap opening.}, we stop the accretion of gas because recycling flows penetrating into the Hill sphere of the planet prevent efficient gas accretion \citep{Ormel2015, Cimerman2017, Lambrechts2017}. Finally, we also check whether the disk has disappeared (disk mass below \SI{1e-6}{M_\odot}). If both checks evaluated negative we start a new time step.

\subsection{Initialization}
All disk quantities are defined on a logarithmically spaced grid with $N_\mathrm{grid}=500$ cells between $r_\mathrm{in}=\SI{0.1}{AU}$ and $r_\mathrm{out}=\SI{1000}{AU}$. The code is initialized with the initial gas surface density. Followed by the knowledge of the solid to gas ratio ($\epsilon_0$) it computes the dust- and pebble surface densities. The code will then compute the temperature profile using the surface densities. In this paper we use the analytical solution for the gas surface density (Eq. \ref{eq:gas_density_analytically}) to initialize the code. Since this is based on the viscosity, which depends on the temperature, we iterate the above steps until the temperature has converged to 0.1\% accuracy.
When the disk has been initialized the code begins the viscous evolution. The planetary seed is then placed at $t=t_0$ into the disk. We stop the planetary integration at the end of the disk's lifetime or if the planet reaches 0.2 AU.

\section{Growth tracks}\label{sec:results}
\begin{table}
        \centering
        \begin{tabular}{c c}
                \hline\hline
                Abbreviation & Meaning for simulations\\
                \hline
                evap / evaporation & evaporation and condensation\\ & at evaporation lines included\\
                plain & evaporation and condensation\\ & is not included\\
                \hline
                
        \end{tabular}
        \caption{Meaning of used model abbreviations in Figs. \ref{fig:growth},  \ref{fig:thorngren_alpha}, \ref{fig:thorngren_DTG}, \ref{fig:thorngren_CO}, and \ref{fig:thorngren_ma}.}
    \label{tab:abbr}
\end{table}

We discuss in this section the results of our models, where we first focus on the water content of growing giant planets and then discuss the atmospheric C/O ratio of growing planets. The disk parameters are the same as for the disk simulations discussed before and we only vary the viscosity parameter.

\subsection{Planetary water content}\label{sec:water}

\begin{figure*}
    \centering
    \includegraphics[width=0.95\textwidth]{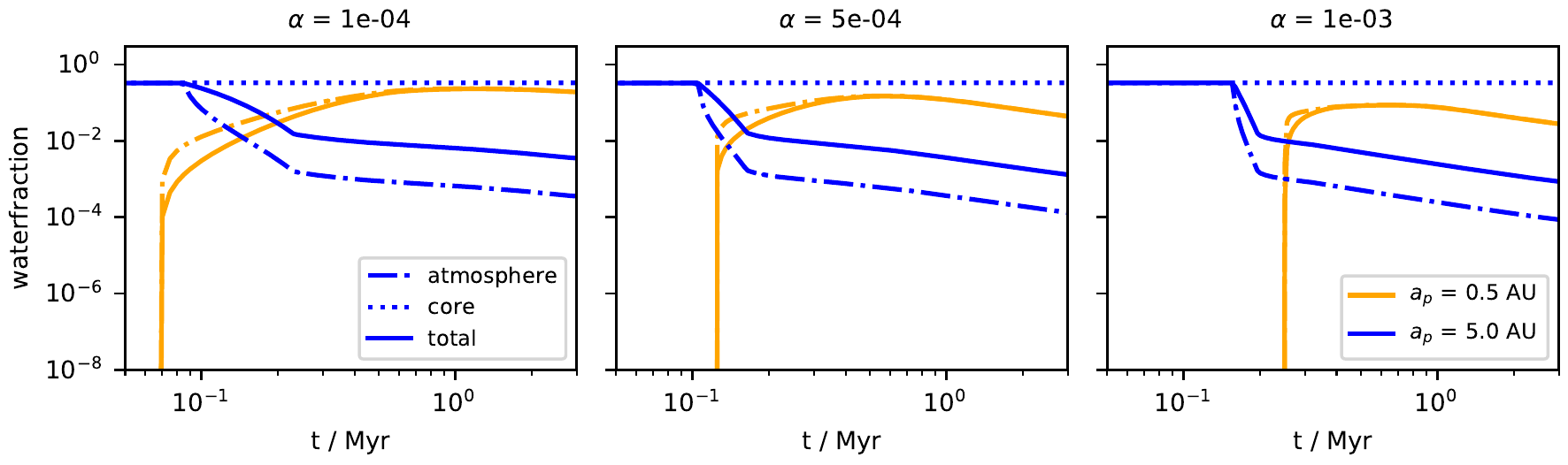}
    \caption{Water content of non-migrating planets as a function of time for different disk viscosities, displayed for the core (dotted), the atmosphere (dashed), and the whole planet (solid). The corresponding water content of the protoplanetary disk as a function of time is displayed in Figs.~\ref{fig:watertime05} and \ref{fig:watertime50}.}
    \label{fig:watertime}
\end{figure*}

In Fig.~\ref{fig:watertime} we show the water content of non-migrating planets placed at 0.5 AU and 5.0 AU for different viscosities. These planets are placed interior and exterior to the water ice line. Initially only the core is formed, so that the water content is determined only by solid accretion. Due to our assumption for the buildup of the planetary atmosphere during this phase, the water content is the same for the atmosphere and the core. During the core buildup phase, the water ice fraction of the core forming at 0.5 AU is zero because no water ice penetrates that deeply into the disk from the water ice line (see Appendix~\ref{sec:watercontent}). Once the planetary core is fully formed, gas accretion can start (indicated by the deviation of the different curves in Fig.~\ref{fig:watertime}). Due to the planet's location interior to the water ice line, the planet accretes a lot  of water vapor, increasing the planet's atmospheric and total water fraction for all different viscosities.

However, the different viscosities have an influence on the exact water content of the growing planets. At low viscosities, the planet can reach larger water contents compared to high viscosities. This is caused by the larger enrichment with vapor of the inner disk in the case of low viscosity due to the more efficient pebble drift and less efficient vapor diffusion (see also Fig.~\ref{fig:watertime05}). Toward the end of the disk's lifetime, the water content in the planetary atmosphere decreases again because the supply of pebbles that enrich the disk has run out, preventing a continuous water vapor enrichment of the accreting planet. This effect is also stronger for higher viscosities because water vapor diffuses away faster.

A planetary embryo growing at 5.0 AU, has initially a large water content in the core due to the accretion of water ice. As soon as the planet starts to accrete gas efficiently, the water content in the planetary atmosphere decreases because the gas is water poor. The kink in the atmospheric and total water abundance around \SIrange{100}{200}{kyr} is caused by a change in the planetary gas accretion rate, after the planet has emptied its horseshoe region and the planetary growth is limited by the supply rate of the disk (Eq.~\ref{eq:gasaccrete}), reducing the gas accretion rate onto the planet. Consequently the dilution of the water vapor in the atmospheres slows down in time. This effect is not visible for the planet growing at 0.5 AU, because the water content is increasing as the planet grows and the small deviation in the accretion rate is not visible within the log-scale of the plot. 

We note that the growth rate of the planetary core is reduced for larger viscosities. This is caused by the decreasing pebble sizes for increasing viscosities, resulting in lower pebble accretion rates, independent of whether the planet is interior or exterior of the water ice line.

\begin{figure*}
    \centering
    \includegraphics[width=0.95\textwidth]{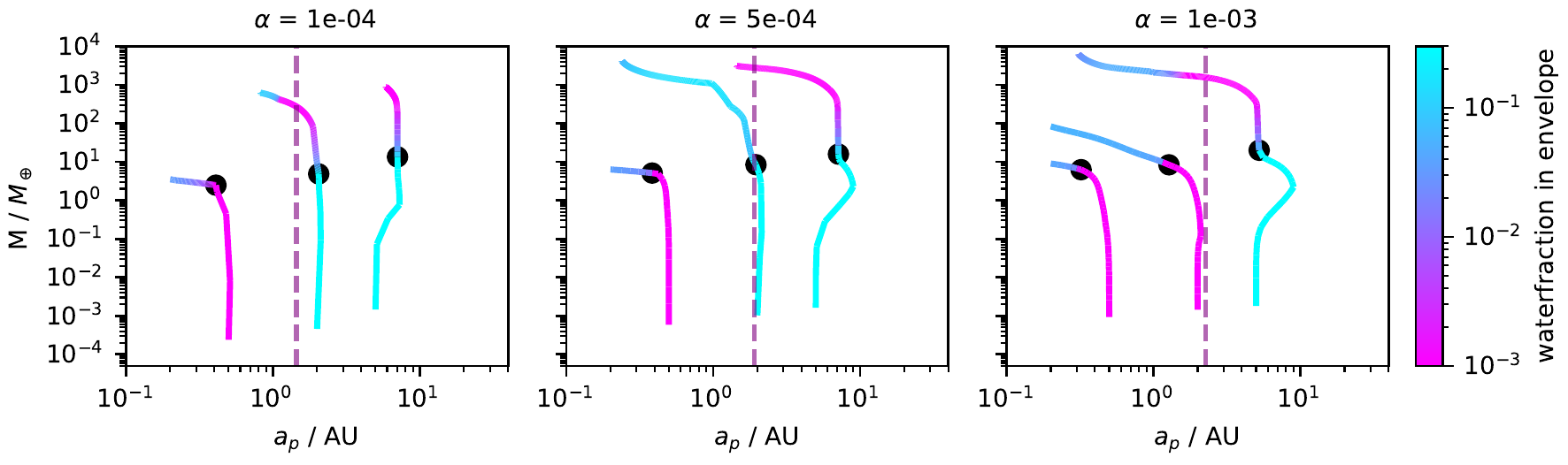}
    \caption{Atmospheric water content (in color) of migrating planets in disks with different viscosities. The dot marks the pebble isolation mass, where the planet switches to gas accretion, while the vertical line marks the water ice line.}
    \label{fig:watermig}
\end{figure*}

In Fig.~\ref{fig:watermig} we show the atmospheric water content of growing and migrating planets in disks with different viscosities. We place the planetary embryos initial interior, exterior, and close to the water ice line. The position of the water ice line is farther away from the star for disks with larger viscosities due to the increase in viscous heating.

As the planetary core grows, the water content of the planetary atmosphere is determined by the composition of the solids that evaporate in the initial atmosphere. This leads to initially water-poor planets interior to the water ice line and to initially water-rich planets exterior to the ice line, as for the non-migrating planets (Fig.~\ref{fig:watertime}). As soon as the planets start to accrete gas (marked by the dot in Fig.~\ref{fig:watermig}), the water content of the planetary atmosphere changes. Planets interior to the water ice line accrete water vapor, while planets exterior to the water ice line accrete water-poor gas. If the planets migrate across the water ice line, we observe again a change in the atmospheric water content. The final water content in the atmosphere is then an interplay between the accreted mass interior and exterior to the water ice line.

\subsection{Atmospheric C/O ratio}\label{sec:growthtracks}

\begin{figure*}
        \centering
        \includegraphics[width=1.0\textwidth]{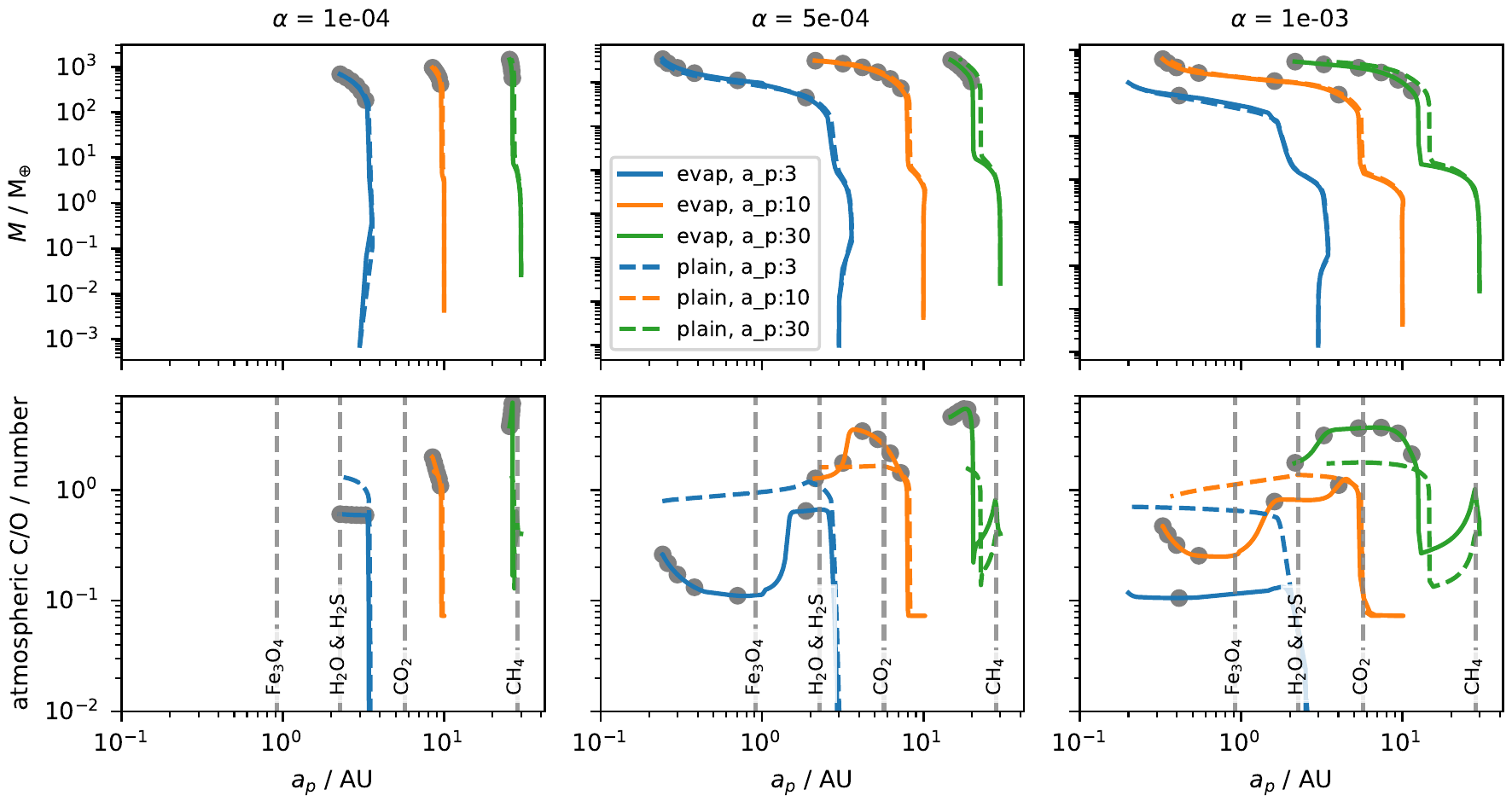}
        \caption{Evolution of planets that accrete pebbles and gas starting at different initial positions. 
        Top: Growth tracks displaying the planetary mass as a function of its position. The dashed line indicates planets that grow in disks with a static composition model (see Sect. \ref{sec:composition}), while the solid lines include the evaporation of pebbles at the evaporation lines. Bottom: Atmospheric C/O content of the same planets as a function of their position. The dashed gray lines mark the positions of evaporation lines. The gray dots mark time steps of \SI{0.5}{Myr}. We use here our standard parameters as stated in Table \ref{tab:parameters}, corresponding to the disk evolution shown in Fig.~\ref{fig:sigma_evap}, and abbreviations of the labels can be found in Table \ref{tab:abbr}.}
        \label{fig:growth}
\end{figure*}

In our model planets grow and migrate at the same time. We do not take scattering effects happening after the gas disk phase \citep[e.g.,][]{Raymond2009,Sotiriadis2017,Bitsch2020_scat} into account because our model can only include one planetary embryo at a time. We show the growth tracks of three different planets starting at different positions for three different disk viscosities in Fig. \ref{fig:growth}. The disk parameters used for this example correspond to the disk evolution shown in Figs.~\ref{fig:sigma_evap} and \ref{fig:heavy}.

The blue line depicts the evolution of planets starting at \SI{0.05}{Myr} and \SI{3}{AU}, which rapidly accrete pebbles and then gas and migrate inward to become hot Jupiters if the viscosity is large enough, allowing an efficient migration in the type-II regime. Because of the high pebble accretion rates the planet migrates only slightly or even outward during core formation (Fig.~\ref{fig:growth}) due to the heating torque. The heating torque does not contribute any more once the planet has reached a few Earth masses and the planet moves inward in type-I migration. Within $\SI{0.05}{Myr}$ (at $t=\SI{0.1}{Myr}$) the planet has already started to accrete gas and reached roughly $\SI{100}{M_\oplus}$ after 0.2 Myr, which happens slightly later in disks with higher viscosities due to the slower pebble accretion rates. During this phase, the planet accretes faster than it migrates because it feeds off the gas supply inside the planetary horseshoe region. Once the horseshoe region is emptied, the planetary accretion is slowed down and the planet migrates inward in the slower type-II migration regime, which is slower for lower viscosities, resulting in only a marginal inward migration in this case. Especially in the case of $\alpha=10^{-3}$ the inward migration is very efficient, so that the planet reaches the inner edge of the protoplanetary disk well before the end of its lifetime.

In Fig.~\ref{fig:growth} we also compare how evaporation and condensation at evaporation lines affect the formation and composition of the growing planets. The difference is minimal in the growth tracks (upper panel). However, the C/O ratio (bottom panel) shows significant differences for the two model approaches. Water ice line crossing pebbles enrich the gas in oxygen and therefore greatly reduce the C/O ratio interior to the water ice line compared to the static model approach (Fig.~\ref{fig:sigma_evap}). Furthermore, the evaporation of the methane content of drifting pebbles increases the C/O ratio in the gas significantly at late times, especially for larger viscosities, as the methane vapor diffuses inward.  

The C/O ratio of the planet forming at 3 AU in the disk with $\alpha=10^{-4}$ stays around solar, corresponding to the disk's C/O content (Fig.~\ref{fig:sigma_evap}). Its migration across the water ice line happens so late that there is no significant change in the planet's C/O. The C/O of the planet forming in the disk with $\alpha=5 \times 10^{-4}$ shows a more complicated pattern, originating from crossing various ice lines. After the planet crossed the water ice line, the planetary C/O drops (with a slight delay caused by the slower gas accretion rate, as the planetary gap is already formed) and only increases at late times again when inward diffusing carbon-rich gas reaches the planet. The planet forming in the disk with $\alpha=10^{-3}$ migrates across the water ice line very quickly and thus harbors a low C/O ratio. In contrast, the planets forming in the model without evaporation of inward drifting pebbles show C/O ratios close to solar, corresponding to the values of the inner disk regions without evaporation (Fig.~\ref{fig:sigma_evap}).

The origin of the planets forming at 10 AU (orange line) is farther away from the central star, where less solids and gas are available in the disk. This slows down the accretion rate of solids and gas as well as the migration rate. The cores of these planets form beyond the CO$_2$ evaporation line, resulting in low C/O ratios during the core accretion phase. When the planet has reached the pebble isolation mass and begins to accrete gas, the C/O ratio increases and stays approximately constant because the gas in the range between 2-10 \SI{}{AU}, features a relatively constant C/O (Fig.~\ref{fig:sigma_evap}). Only at late times does the C/O ratio increase further due to the inward diffusing methane vapor. However, the planet forming in the disk with $\alpha=10^{-3}$ migrates also significantly during its gas accretion phase, crossing several evaporation fronts. At each evaporation front, the planetary C/O ratio changes correspondingly, but also increases slightly toward the end of the disk's lifetime due the inward diffusion of carbon-rich gas.

The planets forming at 30 AU (green lines) needs more time to accrete material, resulting in further inward migration during the type-I regime, especially in disks with high viscosities, which have the lowest pebble accretion rates due to the smallest particles. However, these planets have a very efficient envelope contraction phase due the higher pebble isolation mass, which leads to core masses of $20 M_\oplus$ (twice the core mass of the inner planets) and thus shorter contraction times. This efficient contraction phase boosts the planet into rapid gas accretion, which then slows down the planet into the type-II migration phase.

The C/O ratio looks initially very similar if evaporation of pebbles is considered or not. This is related to the formation position of the planet close to the methane evaporation front, where the C/O in the gas and solid phase does not change very much (Fig.~\ref{fig:sigma_evap}). Only once the planet crosses the methane evaporation line a significant change of the planetary C/O ratio can be observed. In the cases without evaporation, the planetary C/O ratio increases slightly above 1, while including evaporation can boost the atmospheric C/O ratio up to a value of 5. However, toward later times, the C/O ratio decreases slightly because the supply of carbon-rich pebbles that fuel the carbon vapor, diminishes in time, resulting in a decrease in the C/O ratio of the disk (Fig.~\ref{fig:sigma_evap}) and consequently in the planetary atmosphere.

We note that the exact change of the atmospheric C/O ratio in the planetary atmosphere when crossing an evaporation front also depends on the gas mass the planet accretes. If the planet is already very massive and only accretes a tiny fraction of gas with a different composition, the change of the planetary C/O ratio is small.

We conclude that the inclusion of pebble evaporation effects make a huge difference in the atmospheric elemental ratios of gas giants. Our simulations show that the C/O ratio increases with orbital distance, for all disk viscosities. Observed hot Jupiter planets that have formed as cold Jupiters and were scattered to sub AU orbits might therefore have significantly larger C/O ratios compared to planets that formed by smooth inward migration from initially closer-in orbits.

\section{Heavy element content}\label{sec:res_heavy_elements}

\begin{table*}
\centering
        \begin{tabular}{c  c}
        \hline\hline
        parameter & values\\
        \hline
                $\epsilon_0$ & \numlist{0.01; 0.015; 0.02; 0.025}\\
                $\alpha$ & \numlist{1e-4; 5e-4; 1e-3}\\
                $t_\mathrm{evap}$ & \SIlist{1;2;3}{Myr}\\
                $t_0$ & \SIlist{0.05; 0.15; 0.25; 0.35; 0.45}{Myr}\\
                $a_{p,0}$ & \SIlist{1; 2; 3; 5; 10; 15; 20; 25; 30}{AU}\\
        \hline
        \end{tabular}
        \caption{Parameters different from the standard parameters in Table \ref{tab:parameters} that are used for simulations in Sect.~\ref{sec:res_heavy_elements}. Simulations are performed on a grid by simulating each of the 1620 possible combinations. Like \citet{Thorngren2016}, only planets with final masses in the range of $20 M_\oplus$ and $20 M_{J}$ and stellar insulations less than $F_\star < \SI{2e8}{\erg\per\s\per\cm\squared}$ and additionally only planets with final orbits with less than \SI{1}{\AU} are considered (except in Fig.~\ref{fig:thorngren_ma}).}
        \label{tab:thorngren_param}
\end{table*}

\begin{figure*}
        \centering
        \includegraphics[width=.95\textwidth]{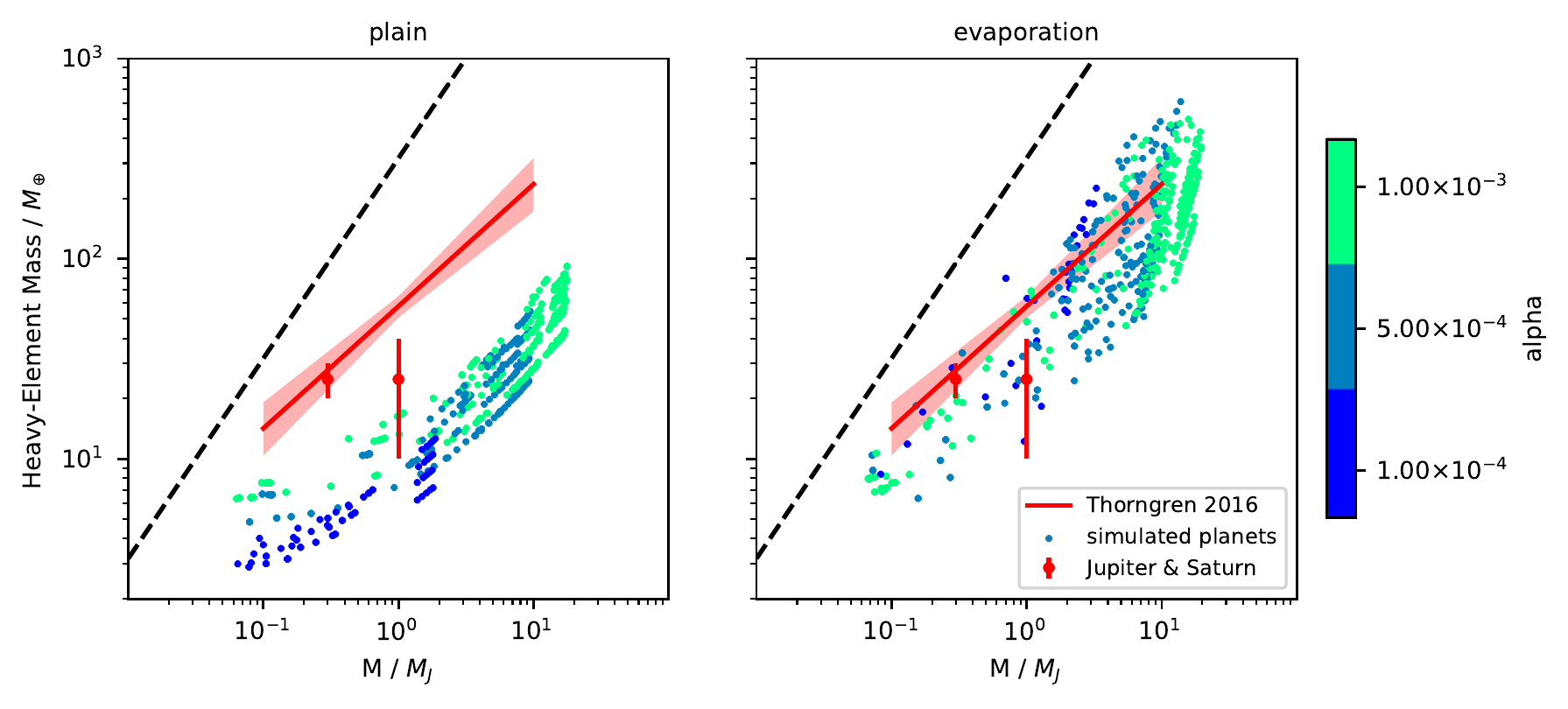}
        \caption{Total heavy element mass (core plus envelope) as a function of planetary mass. Simulated planets (see Table \ref{tab:thorngren_param}) are compared to interior models of observed hot Jupiters from \citet{Thorngren2016}, marked with the red line. The color coding in this figure shows the dependence on $\alpha$. The different panels indicate whether evaporation line effects have been taken into account for the simulations (see Table \ref{tab:abbr}).}
        \label{fig:thorngren_alpha}
\end{figure*}

\begin{figure*}
        \centering
        \includegraphics[width=.95\textwidth]{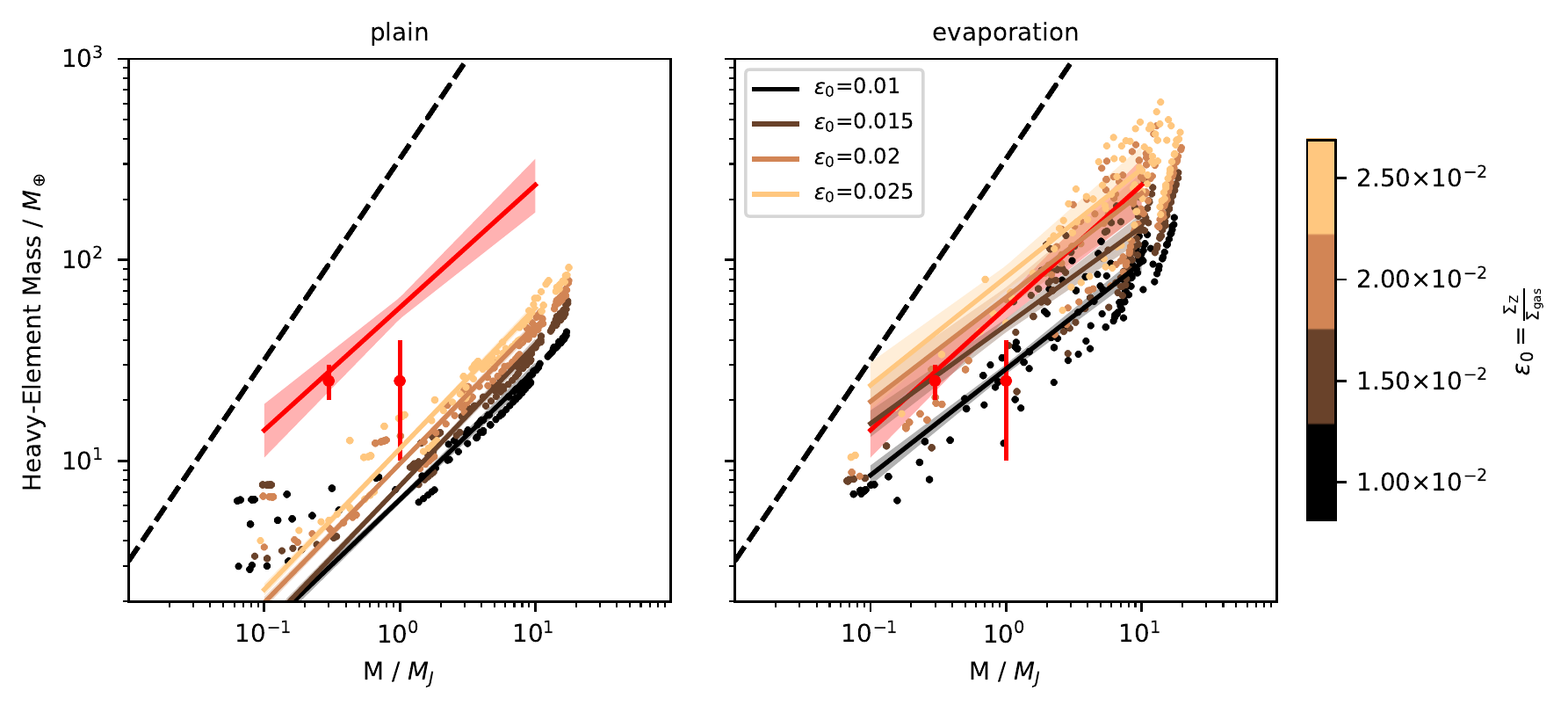}
        \caption{Like Fig.~\ref{fig:thorngren_alpha} but using the solid to gas ratio $\epsilon_0$ for the color coding. Each solid to gas ratio has been fitted for comparison to the original fit from \citet{Thorngren2016}. Fit results can be found in Table \ref{tab:thorngren_fit}.}
        \label{fig:thorngren_DTG}
\end{figure*}

\begin{figure*}
        \centering
        \includegraphics[width=.95\textwidth]{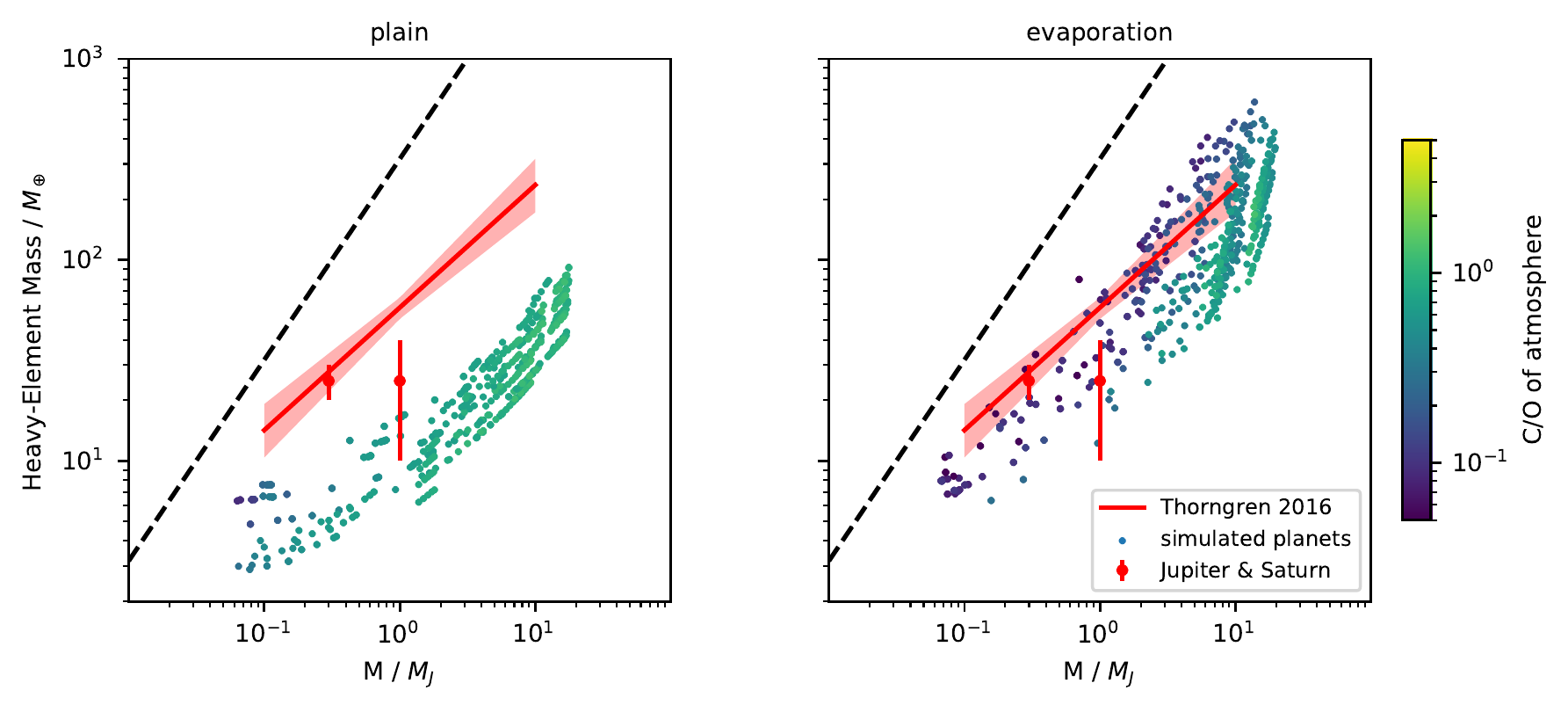}
        \caption{Like Fig.~\ref{fig:thorngren_alpha} but using the atmospheric C/O ratio for the color coding.}
        \label{fig:thorngren_CO}
\end{figure*}

\begin{figure*}
        \centering
        \includegraphics[width=\textwidth]{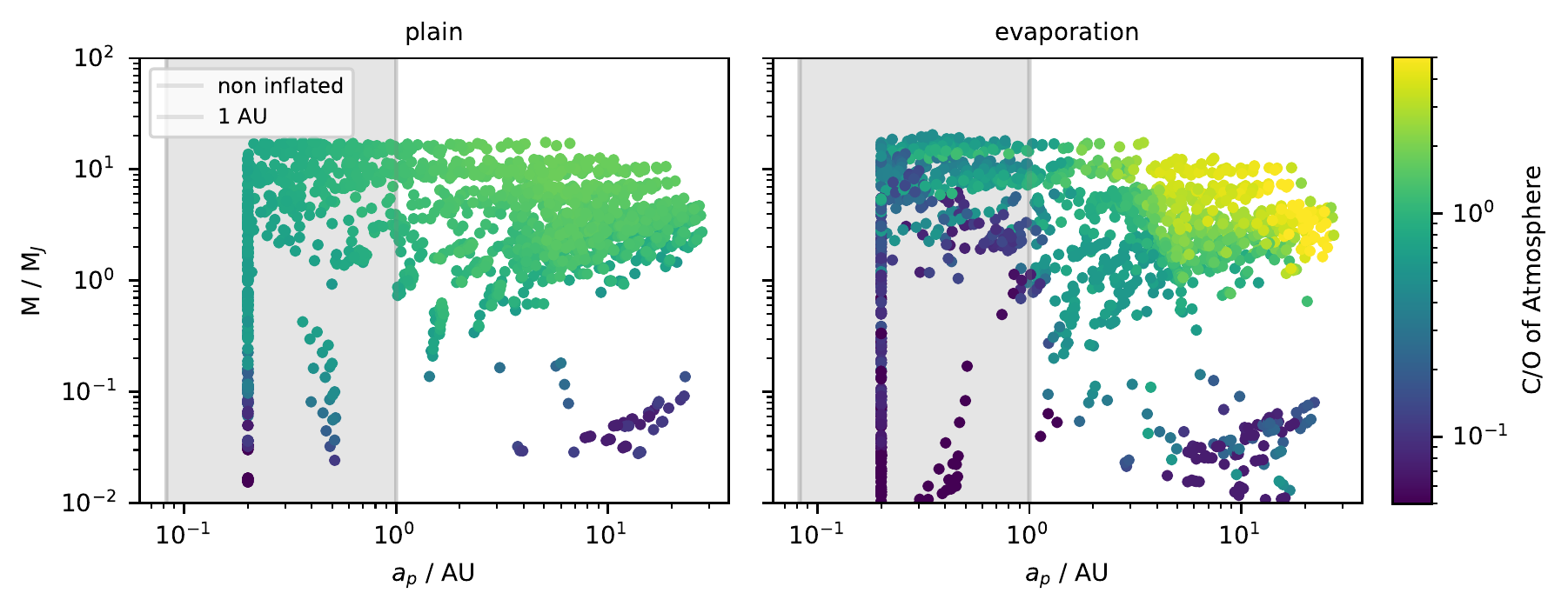}
        \includegraphics[width=\textwidth]{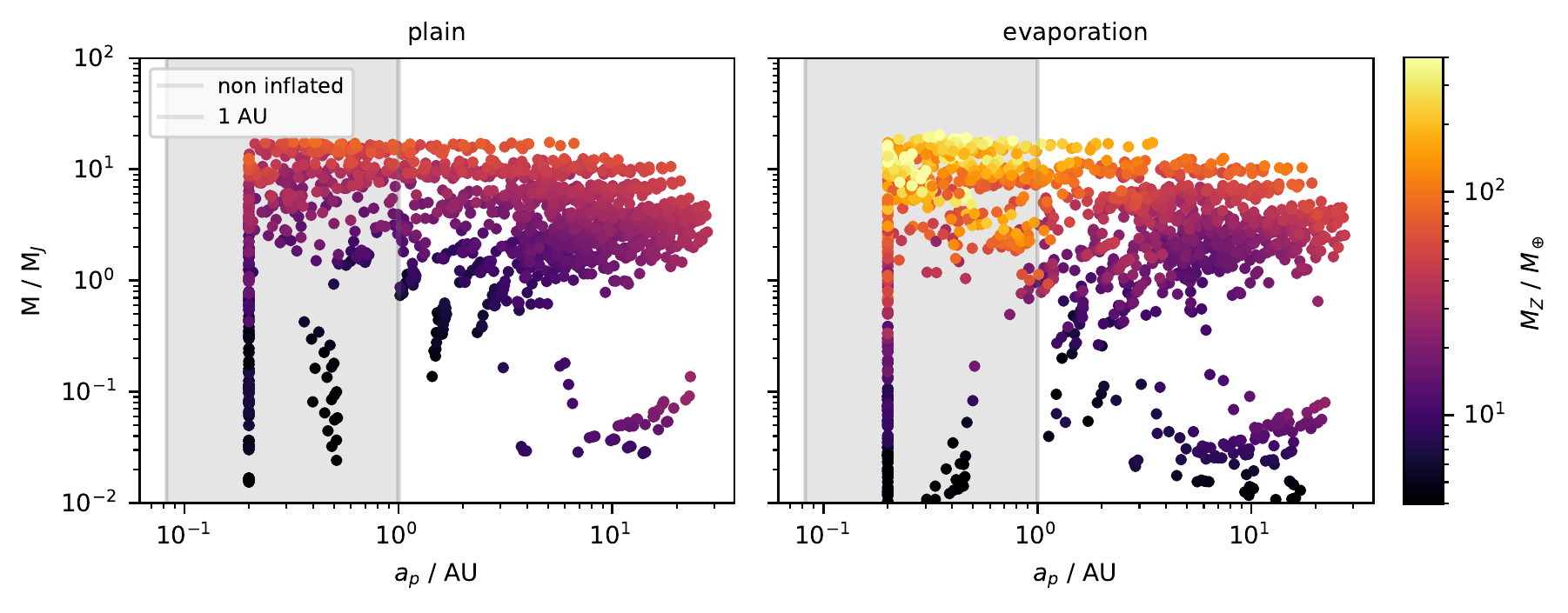}
        \caption{Final position and masses of planets formed with the initial conditions stated in Table~\ref{tab:thorngren_param}, color coded by their atmospheric C/O (top) and their heavy element content (bottom). We mark in the gray band the planets that are featured in Figs. \ref{fig:thorngren_alpha}, \ref{fig:thorngren_DTG}, and \ref{fig:thorngren_CO}. The different bands of the formed planets in the a-M space are caused by the initial spacing of the embryos with $a_{p,0}$ and $t_0$, as described in Table~\ref{tab:thorngren_param}.}
        \label{fig:thorngren_ma}
\end{figure*}

\citet{Thorngren2016} derived the heavy element content for giant exoplanets by comparing the observed planetary masses and radii with interior models. In this section we explore how the effects of pebble evaporation at evaporation lines influences the heavy element content of growing gas giants. We performed simulations on a grid of parameters given in Table \ref{tab:thorngren_param} by simulating every possible combination (in total 1620). 

We follow \citet{Thorngren2016} and only include planets in our analysis with final masses in the range of $20 M_\oplus$ and $20 M_{J}$ and stellar insulations less than $F_\star < \SI{2e8}{\erg\per\s\per\cm\squared}$. Stellar insulations are converted to final positions using the relation
 \begin{equation}
        r < \sqrt{\frac{L_\star}{4\pi F_\star}},
\end{equation}
where $L_\star=1 L_\odot$ is the stellar luminosity. 
We also select only those planets that have final orbits with less than \SI{1}{\AU} because the planetary radii, needed for the model of \citet{Thorngren2016} become increasingly difficult to determine for planets farther away.

We show in Figs.~\ref{fig:thorngren_alpha}, \ref{fig:thorngren_DTG}, and \ref{fig:thorngren_CO} the results of our simulations, where we color code different properties of the planets formed in our simulations. In Fig. \ref{fig:thorngren_alpha} we color code the disk's viscosity, in Fig. \ref{fig:thorngren_DTG} we color code the dust-to-gas ratio, and in Fig. \ref{fig:thorngren_CO} we show the atmospheric C/O ratio.

These results clearly show that the inclusion of evaporation and condensation at evaporation lines enhances the total heavy element content of the formed giant planets (right panels). The heavy element content of the giant planets is dominated by the accretion of evaporated volatiles residing in the gas.

While the value of $\alpha$ seems to determine the final mass of planets (horizontal shifting of planets in Fig.~\ref{fig:thorngren_alpha}) we can directly link the solid to gas ratio to the magnitude of the heavy element content (vertical shifting in Fig.~\ref{fig:thorngren_DTG}). This can best be shown if the fit from \citet{Thorngren2016} is reproduced for the different solid to gas ratios. We can match these quantities using a power law relation for the heavy element content and final mass \citep{Thorngren2016} of
\begin{equation}
        M_Z = \gamma_a \cdot \left(\frac{M}{M_\mathrm{J}}\right)^{ \gamma_b},
\end{equation}
where $\gamma_a$ and $\gamma_b$ are fit constants (results see Table \ref{tab:thorngren_fit}). 

We note that the slope of the heavy element content of the giant planets depends on the exact dust-to-gas ratio in the case of pebble evaporation. For $\epsilon_0 = 2.5\%$, the slope is slightly steeper than inferred from \citet{Thorngren2016}, while lower $\epsilon_0$ values result in flatter slopes. For low planetary masses, the heavy element content is very similar for the different initial dust-to-gas ratios because the heavy element content is dominated by the planetary core mass, which is set by the pebble isolation mass. The pebble isolation mass, however, does not depend on the dust-to-gas ratio, resulting in the similar heavy element contents of the small mass planets. More massive planets are dominated by gas accretion. Thus the heavy element content is dominated by the vapor content in the gas phase of the disk, which increases for increasing initial dust-to-gas ratios (see also \citealt{Garate2020} for the water content). Consequently the slope of the heavy element content as a function of planetary mass increases for increasing $\epsilon_0$.

The dependence on the solid to gas ratio is only seen if the evaporation of pebbles is taken into account. This means that the solid to gas ratio mainly influences how many heavy elements can be evaporated into the gas and consequently accreted onto the planet. The heavy element content of the low mass giant planets (below 0.2 Jupiter masses) is very similar regardless of whether pebble evaporation is taken into account or not because the heavy element content for these planets is dominated by the core, which is independent of pebble evaporation at ice lines.

It seems that $\epsilon_0 = 2.5\%$ allows the data from \citet{Thorngren2016} to be reproduced best. Under the assumption that the solid to gas ratio can be linked to the solar solid to gas ratio of $\approx 1.5\%$ \citep{Lodders2003, Asplund2009} and the value for the iron abundance [Fe/H]$=0$, this would relate to a value for [Fe/H]$=\log_{10}(2.5/1.5) \approx 0.22$.\footnote{We note that we do not change the chemical composition when increasing the dust-to-gas ratio, as observations indicate \citep[e.g.,][]{Buder2018, Bitsch2020chem}} 
The sample of planets that has been used in \citet{Thorngren2016} has a wide range of metallicities with a mean value of [Fe/H]$=0.1\pm0.2$. \citet{Buchhave2018} have shown that hot Jupiters are more likely to be found among metal-rich stars with a mean metallically of [Fe/H]$=0.25\pm0.03$. This indicates that a solid to gas ratio of 2.5\% might be a realistic proposal for the formation of these hot gas giants in line with our model.

The atmospheric C/O ratio of the planets with very large heavy element contents is mostly below 1 (Fig.~\ref{fig:thorngren_CO}), due to the efficient accretion of water vapor for these close in planets. In contrast, observations of hot Jupiters show that the C/O ratio could be super-stellar \citep{Brewer2017}. In Figs.~\ref{fig:thorngren_alpha}, \ref{fig:thorngren_DTG}, and~\ref{fig:thorngren_CO}, we show the results from planets that migrated via planet-disk interactions into the inner regions of the protoplanetary disk. However, planets can also finish their formation farther away from the central star and then scatter inward to form hot Jupiters {(for a recent review of hot Jupiter formation see \citet{Dawson2018})}. 

In Fig.~\ref{fig:thorngren_ma} we show the final orbital positions and masses of all the planets formed in our simulations with the initial conditions stated in Table~\ref{tab:thorngren_param}. With the gray band we mark the planets that fulfill the stellar insulation criterion and are within 1 AU, corresponding to the planets shown in Figs.~\ref{fig:thorngren_alpha}, ~\ref{fig:thorngren_DTG}, and~\ref{fig:thorngren_CO}. The orbital evolution of the planets ends at 0.2 AU, resulting in the vertical band at that position.

The planetary distributions show also specific bands in the mass-orbital distance plane. This is caused by the initial conditions ($a_{p,0}$ and $t_0$) of the planets, which are, for the sake of simplicity, not randomized as in population synthesis simulations, because we want to investigate general trends and not match the exact exoplanet populations. Furthermore, our simulations show a large fraction of giant planets, which is also caused by the setup of our simulations. Placing planetary seeds in the outer regions of the disk allows a more efficient growth of gas giants compared to growth in the inner disk (e.g., \citealt{Ndugu2018, Bitsch2017}). 

Due to this artificial setup, we do not compare the final positions and masses of our planets to observations as in population synthesis simulations that are specifically designed for this task (e.g., \citealt{IdaLinI, Mordasini2016, Ndugu2018, Cridland2019}). Furthermore, the planet population synthesis simulations probe different parameters (e.g., disk mass) that we did not change within our simulations (see Sect.~\ref{sec:discussion}). Our focus here is on how the evaporation of pebbles influences the heavy element content and the C/O ratio of giant planets, which is independent of the exact comparison of our synthetic planet populations to exoplanet occurrence rates.

The atmospheric C/O ratios in Fig.~\ref{fig:thorngren_ma} follow the trends already outlined before (Fig.~\ref{fig:growth}), namely that planets accreting their gaseous envelope farther away from the central star have a larger C/O ratio. This result is independent if evaporation is taken into account or not. However, as stated above, the absolute C/O ratio in the outer planets is more extreme if evaporation is taken into account. Namely, the C/O ratio is enhanced for planets forming in the outer disk due to the evaporation of carbon-rich pebbles, while the C/O ratio of planets forming in the inner disk region is very low due to the evaporation of water-rich pebbles (Fig.~\ref{fig:sigma_evap}). 

The heavy element content of the planets in the outer regions of the protoplanetary disk is smaller compared to planets in the inner regions of the protoplanetary disk (bottom in Fig.~\ref{fig:thorngren_ma}) in the case evaporation is taken into account. This is caused by the larger enrichment of heavy elements in the gas phase in the inner disk compared to the outer disk (Fig.~\ref{fig:heavy}). If no evaporation is taken into account, the heavy element content is larger for planets in the outer regions because the heavy element content is mostly set by the core mass, which is determined by the pebble isolation mass, which is increasing with orbital distance due to its dependence on the disk's aspect ratio \citep{Lambrechts2014_iso, Bitsch2018_peb_iso}.

If hot Jupiters indeed have super-stellar C/O ratios \citep{Brewer2017}, our model predicts that these planets form in the outer disk and are then scattered inward. Furthermore, our model then predict that these planets have a lower heavy element content compared to Jupiter type planets with low C/O ratios (Fig.~\ref{fig:thorngren_ma}), giving testable predictions to observations of giant planets.

\section{Discussion}\label{sec:discussion}

\subsection{Dependence on disk parameters}
As shown in Sect.~\ref{sec:results}, the results strongly depend on the disk parameters. In this section we qualitatively discuss some of the parameters that we have not investigated in detail in this work but will be investigated in future works.

\subsubsection{Fragmentation velocities}

The fragmentation velocity sets how fast pebbles can become before a collision will lead to fragmentation instead of coagulation. Fragmentation velocities are determined by laboratory experiments \citep[e.g.,][]{Blum2008,Gundlach2015,Musiolik2019} because the collision properties of dust aggregates are otherwise difficult to quantify. Typical laboratory experiments give fragmentation velocities in the range of \SIrange{1}{10}{\m\per\s}.

High values of fragmentation velocity lead to large Stokes numbers and high radial pebble velocities. This means that pebble accretion will generally be more efficient but planets need to form their core on shorter timescales, otherwise pebbles will have drifted inward and pebble accretion will not be efficient. This implies that the formation of gas giants requires early planetary embryo formation, in line with evidence from the Solar System, where an early formation of Jupiter could explain the separation between carbonaceous  and non-carbonaceous chondrites \citep{Kruijer2017}.

Lower fragmentation velocities will reduce the maximal grain size and, as such, reduce the radial drift of pebbles. As pebble accretion is size dependent (Eq. \ref{eq:peb_acc_radius}), lower fragmentation velocities might hinder pebble accretion \citep[e.g.,][]{Venturini2020a}. Changes in the fragmentation velocity at the water ice line might also lead to a pressure bump in this region \citep{Mueller2021}.

In our simulations we use a constant fragmentation velocity throughout the disk, also motivated by recent laboratory experiments that find no difference in the fragmentation velocity between silicates and water \citep{Musiolik2019}. Other simulations \citep[e.g.,][]{Izidoro2019, Guliera2020, Venturini2020a} have used different pebble sizes at ice lines (motivated by water ice evaporation) or implemented viscosity transitions (e.g., at the dead zone edge) resulting in different pebble sizes. Smaller pebbles in the inner regions of the disk can slow down the formation of super-Earths via pebble accretion \citep{Izidoro2019, Venturini2020a}. Within our model, a change in pebble sizes at the water ice line would mainly influence the inner regions of the disk, but the heavy element content in the inner disk region (Fig.~\ref{fig:heavy}) is mainly determined by the water content, which is unaffected by a change of pebble sizes at the water ice line. Regarding planetary compositions, we expect that higher fragmentation velocities will increase the heavy element content of early forming planets because the gas is polluted faster by the evaporation of the larger pebble flux.

\subsubsection{Disk mass and radius}

The disk mass obviously determines how much mass is available for planet formation. High disk masses lead to high dust masses (if the solid to gas ratio stays the same). Accretion rates of pebbles and gas linearly depend on the amount of matter available. However, type I migration rates also linearly depend on the gas surface density. Planet formation will therefore be accelerated for high disk masses. Since pebble accretion fights the decay of the pebble surface density (i.e., planets need to form before pebbles are gone) this can indirectly influences the growth of planets. 

Similarly does the disk radius determine the lifetime of the pebble surface density. Large disk radii on the one hand lead to a longer supply of pebbles from the outer disk, while smaller disks will be depleted of pebbles early on. This can lead to the growth of multiple small mass planets instead of a few large bodies \citep{Kretke2014, Levison2015}. 

Furthermore, a change of the available pebble mass and how long the pebble supply lasts, influences the heavy element content of forming giant planets. This can be inferred from Fig.~\ref{fig:thorngren_DTG}, which shows that planets forming in disks with more pebbles (large $\epsilon_0$) have a larger heavy element content, potentially similar to more massive disks.

Observations of protoplanetary disks indicate a wide spread in disk mass and radii \citep[e.g.,][]{Andrews2013}. We will investigate the effects of this in future studies.

\subsection{Model extensions}

In this subsection, we discuss how our model can be extended in the future for various aspects. While all these listed additions would improve our model, the general message that pebble evaporation at ice lines increases the heavy element content in the gas phase of the disk and thus also the heavy element content of gas accretion planets, remains.

\subsubsection{Planetesimals}

In our model, we only include the contributions of pebble and gas accretion, while we do not model the formation of planetesimals \citep{Drazkowska2016, Drazkowska2017, Lenz2019, Voelkel2020} and the accretion of those onto planetary embryos. The accretion efficiency of planetesimals in itself depends crucially on the planetesimal size \citep{Fortier2009, Guilera2011, Fortier2013, Johansen2019pla}, where the planetesimal population in the Solar System was probably dominated by large (100km) planetesimals \citep{Bottke2005, Morbidelli2009} in the inner regions, with decreasing sizes (1-10 km) toward the Kuiper belt \citep{Kenyon2012}. The accretion of planetesimals into the planetary atmosphere could prolong the envelope contraction phase \citep{Alibert2018, Venturini_2020_pla_chem, Guliera2020}, delaying gas accretion and thus resulting in large-scale inward migration. Furthermore, growing gas giants can accrete planetesimals into their envelope, which can increase the heavy element mass of the giant planet as well as the atmospheric C/O ratio. 

While we do not model planetesimal formation and accretion in our work, our results clearly show that the evaporation of inward drifting pebbles have a profound impact on the heavy element content of the gas phase and thus of gas accreting planets. Future simulations aimed to study the heavy element content of giant planets should thus include the contributions of planetesimals as well as of pebble evaporation at ice lines.

\subsubsection{Evaporation of solids}
We modeled the evaporation of dust and pebbles by assuming that the evaporation line crossing solid flux of a molecular species is converted to gas within \SI{0.001}{AU} (see Eq. \ref{eq:sink_ice_evap}). In reality the evaporation of molecular species from solids depends on desorption rates \citep{Hollenbach2009}. Molecular species are to a different extend volatile. For a molecular species to desorb it has to overcome the binding energy that keeps it on the surface of a pebble. A proper treatment of the desorption rates will not change the amount of heavy elements that is evaporated to the gas. It will rather change the location and spatial interval in which the drifting solids sublimate. However, since planets migrate, we do not expect large differences in the results, since the amount of heavy elements in the gas is the deciding factor (see Fig.~\ref{fig:heavy}).

\subsubsection{Opacities}
The midplane temperature and migration rate of the planets depends on the opacity of the protoplanetary disk. We used here for simplicity a subset of the opacities from \citet{Bitsch2021}, which are only valid for micrometer grains. However, full grain size distributions feature opacities that are not mimicked by a single grain size \citep{Savvidou2020}. In addition, local pile-ups of dust could change the local opacities and thus influence the cooling rates of the disk. As most of the material is in the form of pebbles, which are millimeters to centimeters in size, these piles might only minimally influence the opacity because the opacities are determined by the small grains rather than the large grains \citep{Savvidou2020, Savvidou2021}. We expect that a fully self-consistent treatment will mainly vary the position of the evaporation lines, but our results regarding the heavy element content or the C/O ratio will qualitatively remain valid.

The envelope contraction phase during planet formation depends on the envelope opacity. The duration of the envelope contraction phase is direct proportional to the value of the envelope opacity. We used here a value that is consistent with the findings of \citet{Movshovitz2008}. Planets migrate very fast during the envelope contraction phase. A low envelope opacity will thus allow fast gas accretion and an earlier transition into the slow type-II migration phase, compared to high envelope opacities \citep{Bitsch2021}. Future simulations should thus include a more self-consistent treatment of the envelope opacity, eventually also accounting for the bombardment of planetesimals \citep[e.g.,][]{Alibert2018}.

\subsubsection{Multiple planets}
From our Solar System and other extra solar planetary systems we know that planetary systems contain on average more than one planet. An N-body integrator can be used to solve the planet-planet interactions and evolution of multiple planetary seeds in a protoplanetary disk that grow by accretion of planetesimals \citep{Alibert2013} or pebbles \citep{Levison2015,Matsumura2017,Chambers2018,Lambrechts2019,Izidoro2019,Bitsch2019}. 

The existence of multiple protoplanets will influence the accretion of pebbles, since planets that are large enough to carve a gap in the gas surface density will stop pebbles from drifting toward possible other planets that grow further in \citep{Morbidelli2015, Morbidelli2016, Bitsch+2021, Izidoro2021}. \citet{Morbidelli2015} used this process to explain the dichotomy between the terrestrial planets and gas giants in our Solar System, while \citep{Morbidelli2016} speculated that this effect is also responsible to prevent the accretion of water-rich material onto the Earth.

In addition, a giant planet blocking pebbles exterior to its orbit influences the composition of interior planets by preventing them from accreting the material engulfed in the pebbles. \citet{Bitsch+2021} used this effect to show how the water content of inner sub-Neptunes could be used to constrain the time and formation location of giant planets relative to the water ice line.

In addition, collisions between the planets could increase the heavy element content of the formed giant planets \citep{Ginzburg2020, Ogihara2021}, even though collisions between giant planets might be rare \citep[e.g.,][]{Bitsch2020_scat}. These effects clearly show the importance of multi-body simulations in the future.

\subsubsection{Chemistry}
Including grain surface chemistry in disk models can have a large influence on the chemical composition of the dust grains and of the gas \citep{Semenov2010, Eistrup2016, Cridland2019, Krijt2020,Notsu2020}. However, the surface per mass is much higher for small dust grains than for large dust grains. \citet{Booth2019} argue that chemical reactions will be outperformed by evaporation at evaporation lines due to the fast radial transport of large dust grains. We expect that surface reactions on pebbles will therefore be less dominant than surface reactions on micrometer-sized dust grains. The effect of grain surface reactions onto the chemical composition thus depends on the growth and drift of the pebbles. Models combining both processes in detail are thus needed (e.g., \citealt{Krijt2020}).

The chemical composition of the underlying disk model is solar (Table~\ref{tab:solar_values} within our models. However, in reality stars have different composition that are not necessarily solar (e.g., \citealt{Buder2018}). Past studies have related the solar and stellar abundances to the refractory \citep{Marboeuf2014, Thiabaud2014, Thiabaud2015b, Bitsch2020chem} and volatile \citep{Marboeuf2014b, Thiabaud2015} contents of planetary building blocks and used this to predict the composition of formed planets. These simulations, however, did not take pebble drift and evaporation into account and found that the heavy element content in the gas phase is dominated by the accretion of planetesimals. \citet{Adibekyan2021} recently directly linked the stellar abundances to the metal content of well characterized rocky super-Earths, confirming indeed that the stellar abundances directly influence planetary compositions. While these studies differ in their exact approach, they all clearly show that the underlying chemical composition of the protoplanetary disk has an enormous influence on the final planetary composition in both, refractories and solids and should be investigated in more detail in the future.

Our model results clearly depend on the underlying chemical model. Different compositions result in different significance of individual evaporation lines. We have shown within this paper that the evaporation of methane is important for a large heavy element content and a large C/O ratio. Models that do not include methane or only small fractions of methane might thus show different C/O fractions within the planetary atmospheres (see also Appendix~\ref{sec:with_C}). Figures \ref{fig:growth_C} and \ref{fig:thorngren_C} demonstrate that this influences mostly the C/O ratio rather than the total heavy element content, showing the robustness of the contribution of pebble evaporation for the heavy element content of giant planets.

\subsubsection{Interior structure}
We have assumed that during pebble accretion 90\% of the accreted pebbles are attributed to the core and 10\% are attributed to the heavy-element-rich atmosphere during core buildup. However, more detailed models have revealed that less than 50\% of the accreted solids contribute to the core, while the remaining pebbles contribute to the high metallicity envelope (e.g., \citealt{Brouwers2020, Ormel2021}). Taking a larger fraction of heavy elements to be accreted into the early atmosphere rather than the core will not influence the total heavy element content of the planet, because the accreted mass is the same. It will though influence the C/O ratio of the planet. We discuss this implication in Appendix~\ref{sec:50:50}.

Interior models, however, also determine how the accreted material is distributed inside the planet and its atmosphere \citep[e.g.,][]{Vazan2018, Brouwers2020}. This could strongly influence any quantitative proposal about atmospheric contents like the discussed C/O ratios, where we always assumed a perfect mixing. Therefore, future models need to include consistent interior models, especially during the planetary buildup phase.

\subsubsection{Gas accretion} 
 
Gas accretion onto planetary cores is an active area of research \citep{Ayliffe2009, Szulagyi2016, Schulik2019, Lambrechts2019gas, Bitsch2021}. While most gas accretion recipes are derived from 1D models (e.g., \citealt{Ikoma2000}), only very few accretion recipes from 3D simulations (e.g., \citealt{Machida2010}) exist. Within our model, the gas accretion rates in our simulations (Eq.~\ref{eq:gasaccrete}) are derived for gas of H-He composition, but the gas also contains significant amount of heavy elements in vapor form (e.g., Fig.~\ref{fig:heavy}).

Previous studies show that gas enriched with heavy elements can significantly influence the gas accretion rates \citep{Hori2011, Venturini2015, Venturini2016}. Gas enriched in heavy elements seems to allow a more efficient gas contraction rate, even on cores of below a few Earth masses, which in classical simulations are not allowed to accrete gas efficiently. \citet{Venturini2015} shows that already for $Z_{\rm env}>$0.45 a significant reduction of the critical core mass for gas accretion can be achieved. However, in the models of \citet{Brouwers2020} the pebbles evaporating in the planetary atmosphere do not allow such a fast transition into runaway gas accretion, while the study of \citet{Ormel2021} shows that envelope pollution significantly reduces the time at which the planet reaches the cross over mass for runaway gas accretion in line with \citet{Venturini2015}. Furthermore, \citet{Johansen2021} shows that water-rich pebbles entering the planetary Hill sphere can evaporate high up in the planetary atmosphere, where recycling flows \citep{Lambrechts2017, Cimerman2017} could transport the water vapor away from the planet, preventing the buildup of a high Z envelope.

When the core then transitions into a rapid gas accretion mode, gas accretion rates are mostly determined via 1D simulations (e.g., \citealt{Ikoma2000}). However, 3D simulations are clearly needed to constrain gas accretion rates \citep{Szulagyi2016, Schulik2019, Lambrechts2019gas}, where complicated flow patterns around the planets can arise, complicating the picture. These different studies clearly show that gas accretion needs to be investigated in much more detail in the future.

The final mass of the gas giants in our model is determined by the lifetime of the protoplanetary disk and by the disk's viscosity, where larger viscosities allow a larger gas flux through the disk and thus result in a larger accretion rate (Eq.~\ref{eq:gasaccrete}). Longer disk lifetimes results in longer phases of gas accretion and thus larger planetary masses, while shorter disk lifetimes will result in lower planetary masses. In our model, the disk evolves viscously without photoevaporation (see below), keeping the gas disk mass large also at later times, resulting in larger planetary masses.

Within our model, the exact gas accretion rates should have only little influence on the main message of our work. The inward drifting and evaporating pebbles enrich the gas phase of the disk, which allows the accretion of high metallicity gas, resulting in high heavy element contents of planets (Figs.~\ref{fig:thorngren_alpha} to~\ref{fig:thorngren_CO}). Different implementations of gas accretion rates, would of course influence the growth tracks of planets (Fig.~\ref{fig:growth}) and the resulting C/O ratios in the planetary atmospheres, but the general trend that the C/O ratio increases with larger orbital distances would remain.
 
\subsubsection{Photoevaporation}

The disk dispersal only via viscous evolution can take a very long time, if the viscosity is low. We model the end of the disk lifetime by an exponential decay of the disk's gas surface density within the last \SI{100}{kyr} of its fixed lifetime of \SI{3}{Myr}. In reality, photoevaporation is thought to disperse the disk (for a review see \citealt{Alexander2014}).

Photoevaporation does not only disperse the disk toward the end of the disk's lifetime (after about \SI{3}{Myr}), it also alters the disk structure significantly \citep{Owen2012, Owen2013}. In fact, photoevaporation can carve a hole in the disk beyond \SI{1}{AU}, dividing the disk in two reservoirs. This behavior could have important implications for the growth and composition of planets forming in the inner regions of the disk. If the pebble supply to the inner disk is cut, the enrichment with vapor from inward drifting pebbles stops, reducing the heavy element content in the gas phase in the inner regions. In fact this process is similar to a giant planet opening a deep gap in the protoplanetary disk and blocking the pebbles in the outer disk (Fig.~\ref{fig:heavy}). We thus expect that photoevaporation has a similar effect as growing giant planets on the composition of inner growing planets (see also \citealt{Bitsch+2021}).

Furthermore, photoevaporation can significantly reduce the disk's lifetime, which has large consequences on the growth and migration of giant planets \citep{Alexander2012, Monsch2019}. Furthermore, shorter disk lifetimes would give the vapor less time to diffuse inward, similar to lower viscosities (Fig.~\ref{fig:heavy}), potentially reducing the C/O ratio of inner growing planets. Nevertheless, the inclusion of photoevaporation in our model would not influence our main results significantly, because if photoevaporation were to carve a hole in the disk, planets in the inner regions would not only be deprived of inward diffusing vapor, but of gas in general, stopping their accretion, leaving still heavily vapor-enriched planets behind.

\subsection{Heavy element content of giant planets}
 
In this work we have found that a significant contribution to the heavy element content of gas giants originates from the vapor-enriched gas phase. Past simulations have focused on the contribution to the heavy element content via solid accretion, either via planetesimals \citep{Shibata2019, Shibata2020, Venturini_2020_pla_chem} or even via giant impacts with embryos and other giants \citep{Ginzburg2020, Ogihara2021}.

The bombardment of the planetary envelope with planetesimals might allow an enrichment compatible to Jupiter's heavy element content \citep{Shibata2019, Shibata2020, Venturini_2020_pla_chem}, depending on the exact planetesimal surface density, planetesimals size as well as on the migration speed of the planet \citep{TanakaIda1999}. Giant impacts between super-Earths and giant planets or also between giant planets themselves, mostly occurring towards or after the end of the gas disk's lifetime can additionally enrich the heavy element content of giant planets \citep{Ginzburg2020, Ogihara2021}. Furthermore, giant impacts could additionally explain the structure of the core of Jupiter \citep{Liu2019}.

The main differences of these works to our here presented work is the composition of the heavy elements within the planet. Our work implies that most of the heavy elements are in volatile form, while a bombardment with planetesimals would imply a significant refractory content. We discuss this implication in much more detail in an accompanying paper.

\section{Summary and conclusion}\label{sec:conclusion}

We have performed 1D semi-analytical simulations of the formation of planets in protoplanetary disks. These simulations traced the chemical composition of the accreted pebbles and gas. Pebbles grow from small dust grains by coagulation and drift inward due to gas drag. We compare two main model approaches, where we either include the evaporation and condensation of pebbles at ice lines or not.

Planets build their core from a planetary seed by accreting pebbles while migrating through the disk. Core accretion stops when the planet has grown large enough to create a pressure bump in the surrounding gas, which will trap pebbles and hinder them from reaching the planet. The planet then starts to accrete gas and becomes a gas giant.

Our simulations show that the evaporation of pebbles at evaporation lines largely pollutes the gas with heavy elements (Fig.~\ref{fig:heavy}), in line with observations of protoplanetary disks \citep{Banzatti2020, Zhang2020}. Gas giants can therefore accrete large amounts of heavy elements by the accretion of volatile-enriched gas, with the heavy element fraction increasing as the disk viscosity decreases (Fig.~\ref{fig:thorngren_alpha}). However, larger viscosities allow the growth of more massive planets due to the more efficient gas delivery from the disk to the planet. Furthermore, our simulations indicate that the heavy element content of giant planets is lower for planets forming farther away from the host star because the gas is less enriched in heavy elements in the outer disk compared to the inner disk (Fig.~\ref{fig:thorngren_ma}).

Our simulations also indicate that the atmospheric C/O ratio of giant planets increases for planets formed farther away from the host star, especially if the planets form exterior to the water ice line, thus avoiding the accretion of water vapor, which ultimately decreases the atmospheric C/O ratio (Fig.~\ref{fig:thorngren_ma}). Our simulations show that the planetary C/O ratio increases with the formation distance of the giant planet, but at the same time the total heavy element content of the giant planet decreases. Our simulations thus predict that the C/O ratios of giant planets with large heavy element contents should be low, while the C/O ratios of giant planets with low heavy element content should be high. These predictions will be testable with large observation programs such as JWST and ARIEL.

Furthermore, our simulations clearly indicate that the heavy element content of giant planets is largely influenced by the enrichment of gas by pebble evaporation. Future simulations that aim to study the heavy element content and composition of planets should take these effects into account.

\begin{acknowledgements}
A.D.S and B.B. \ thank the European Research Council (ERC Starting Grant 757448-PAMDORA) for their financial support. A.D.S. acknowledges funding from the European Union H2020-MSCA-ITN-2019 under Grant no. 860470(CHAMELEON) and from the Novo Nordisk Foundation Interdisciplinary Synergy Program grant no. NNF19OC0057374. We thank Cornelis Dullemond for discussions about pebble condensation. We also thank the anonymous referees for their comments that helped to improve the manuscript.
\end{acknowledgements}

\begingroup
\bibliographystyle{aa}
\bibliography{ms}

\begin{thebibliography}{178}
\expandafter\ifx\csname natexlab\endcsname\relax\def\natexlab#1{#1}\fi

\bibitem[{{Adibekyan} {et~al.}(2021){Adibekyan}, {Dorn}, {Sousa}, {Santos},
  {Bitsch}, {Israelian}, {Mordasini}, {Barros}, {Delgado Mena}, {Demangeon},
  {Faria}, {Figueira}, {Hakobyan}, {Oshagh}, {Kunitomo}, {Takeda}, {Jofr{\'e}},
  {Petrucci}, \& {Martioli}}]{Adibekyan2021}
{Adibekyan}, V., {Dorn}, C., {Sousa}, S.~G., {et~al.} 2021, arXiv e-prints,
  arXiv:2102.12444

\bibitem[{{Aguichine} {et~al.}(2020){Aguichine}, {Mousis}, {Devouard}, \&
  {Ronnet}}]{Aguichine2020}
{Aguichine}, A., {Mousis}, O., {Devouard}, B., \& {Ronnet}, T. 2020, \apj, 901,
  97

\bibitem[{{Akeson} {et~al.}(2013){Akeson}, {Chen}, {Ciardi}, {Crane}, {Good},
  {Harbut}, {Jackson}, {Kane}, {Laity}, {Leifer}, {Lynn}, {McElroy}, {Papin},
  {Plavchan}, {Ram{\'\i}rez}, {Rey}, {von Braun}, {Wittman}, {Abajian}, {Ali},
  {Beichman}, {Beekley}, {Berriman}, {Berukoff}, {Bryden}, {Chan}, {Groom},
  {Lau}, {Payne}, {Regelson}, {Saucedo}, {Schmitz}, {Stauffer}, {Wyatt}, \&
  {Zhang}}]{NASA_exoplanet_archive}
{Akeson}, R.~L., {Chen}, X., {Ciardi}, D., {et~al.} 2013, \pasp, 125, 989

\bibitem[{{Alexander} {et~al.}(2014){Alexander}, {Pascucci}, {Andrews},
  {Armitage}, \& {Cieza}}]{Alexander2014}
{Alexander}, R., {Pascucci}, I., {Andrews}, S., {Armitage}, P., \& {Cieza}, L.
  2014, in Protostars and Planets VI, ed. H.~{Beuther}, R.~S. {Klessen}, C.~P.
  {Dullemond}, \& T.~{Henning}, 475

\bibitem[{{Alexander} \& {Pascucci}(2012)}]{Alexander2012}
{Alexander}, R.~D. \& {Pascucci}, I. 2012, \mnras, 422, L82

\bibitem[{{Ali-Dib}(2017)}]{Ali-Dib2017}
{Ali-Dib}, M. 2017, \mnras, 464, 4282

\bibitem[{{Alibert} {et~al.}(2013){Alibert}, {Carron}, {Fortier}, {Pfyffer},
  {Benz}, {Mordasini}, \& {Swoboda}}]{Alibert2013}
{Alibert}, Y., {Carron}, F., {Fortier}, A., {et~al.} 2013, \aap, 558, A109

\bibitem[{{Alibert} {et~al.}(2005){Alibert}, {Mordasini}, {Benz}, \&
  {Winisdoerffer}}]{Alibert2005}
{Alibert}, Y., {Mordasini}, C., {Benz}, W., \& {Winisdoerffer}, C. 2005, \aap,
  434, 343

\bibitem[{{Alibert} {et~al.}(2018){Alibert}, {Venturini}, {Helled}, {Ataiee},
  {Burn}, {Senecal}, {Benz}, {Mayer}, {Mordasini}, {Quanz}, \&
  {Sch{\"o}nb{\"a}chler}}]{Alibert2018}
{Alibert}, Y., {Venturini}, J., {Helled}, R., {et~al.} 2018, Nature Astronomy,
  2, 873

\bibitem[{{Andrews} {et~al.}(2013){Andrews}, {Rosenfeld}, {Kraus}, \&
  {Wilner}}]{Andrews2013}
{Andrews}, S.~M., {Rosenfeld}, K.~A., {Kraus}, A.~L., \& {Wilner}, D.~J. 2013,
  \apj, 771, 129

\bibitem[{{Armitage}(2013)}]{Armitage}
{Armitage}, P.~J. 2013, {Astrophysics of Planet Formation}

\bibitem[{{Asplund} {et~al.}(2009){Asplund}, {Grevesse}, {Sauval}, \&
  {Scott}}]{Asplund2009}
{Asplund}, M., {Grevesse}, N., {Sauval}, A.~J., \& {Scott}, P. 2009, \araa, 47,
  481

\bibitem[{{Ataiee} {et~al.}(2018){Ataiee}, {Baruteau}, {Alibert}, \&
  {Benz}}]{Ataiee2018}
{Ataiee}, S., {Baruteau}, C., {Alibert}, Y., \& {Benz}, W. 2018, \aap, 615,
  A110

\bibitem[{{Ayliffe} \& {Bate}(2009)}]{Ayliffe2009}
{Ayliffe}, B.~A. \& {Bate}, M.~R. 2009, \mnras, 393, 49

\bibitem[{{Banzatti} {et~al.}(2020){Banzatti}, {Pascucci}, {Bosman}, {Pinilla},
  {Salyk}, {Herczeg}, {Pontoppidan}, {Vazquez}, {Watkins}, {Krijt}, {Hendler},
  \& {Long}}]{Banzatti2020}
{Banzatti}, A., {Pascucci}, I., {Bosman}, A.~D., {et~al.} 2020, \apj, 903, 124

\bibitem[{{Baruteau} {et~al.}(2014){Baruteau}, {Crida}, {Paardekooper},
  {Masset}, {Guilet}, {Bitsch}, {Nelson}, {Kley}, \&
  {Papaloizou}}]{Baruteau2014}
{Baruteau}, C., {Crida}, A., {Paardekooper}, S.~J., {et~al.} 2014, in
  Protostars and Planets VI, ed. H.~{Beuther}, R.~S. {Klessen}, C.~P.
  {Dullemond}, \& T.~{Henning}, 667

\bibitem[{{Baumann} \& {Bitsch}(2020)}]{Baumann2020}
{Baumann}, T. \& {Bitsch}, B. 2020, \aap, 637, A11

\bibitem[{{Bell} {et~al.}(1997){Bell}, {Cassen}, {Klahr}, \&
  {Henning}}]{Bell1997}
{Bell}, K.~R., {Cassen}, P.~M., {Klahr}, H.~H., \& {Henning}, T. 1997, \apj,
  486, 372

\bibitem[{{Ben{\'\i}tez-Llambay} {et~al.}(2015){Ben{\'\i}tez-Llambay},
  {Masset}, {Koenigsberger}, \& {Szul{\'a}gyi}}]{Benitez-Llambay2015}
{Ben{\'\i}tez-Llambay}, P., {Masset}, F., {Koenigsberger}, G., \&
  {Szul{\'a}gyi}, J. 2015, \nat, 520, 63

\bibitem[{{Bergez-Casalou} {et~al.}(2020){Bergez-Casalou}, {Bitsch}, {Pierens},
  {Crida}, \& {Raymond}}]{Bergez2020}
{Bergez-Casalou}, C., {Bitsch}, B., {Pierens}, A., {Crida}, A., \& {Raymond},
  S.~N. 2020, \aap, 643, A133

\bibitem[{{Birnstiel} {et~al.}(2015){Birnstiel}, {Andrews}, {Pinilla}, \&
  {Kama}}]{Birnstiel2015}
{Birnstiel}, T., {Andrews}, S.~M., {Pinilla}, P., \& {Kama}, M. 2015, \apjl,
  813, L14

\bibitem[{{Birnstiel} {et~al.}(2009){Birnstiel}, {Dullemond}, \&
  {Brauer}}]{Birnstiel2009}
{Birnstiel}, T., {Dullemond}, C.~P., \& {Brauer}, F. 2009, \aap, 503, L5

\bibitem[{{Birnstiel} {et~al.}(2010){Birnstiel}, {Dullemond}, \&
  {Brauer}}]{Birnstiel2010}
{Birnstiel}, T., {Dullemond}, C.~P., \& {Brauer}, F. 2010, \aap, 513, A79

\bibitem[{{Birnstiel} {et~al.}(2012){Birnstiel}, {Klahr}, \&
  {Ercolano}}]{Birnstiel2012}
{Birnstiel}, T., {Klahr}, H., \& {Ercolano}, B. 2012, \aap, 539, A148

\bibitem[{{Bitsch} \& {Battistini}(2020)}]{Bitsch2020chem}
{Bitsch}, B. \& {Battistini}, C. 2020, \aap, 633, A10

\bibitem[{{Bitsch} {et~al.}(2019){Bitsch}, {Izidoro}, {Johansen}, {Raymond},
  {Morbidelli}, {Lambrechts}, \& {Jacobson}}]{Bitsch2019}
{Bitsch}, B., {Izidoro}, A., {Johansen}, A., {et~al.} 2019, \aap, 623, A88

\bibitem[{{Bitsch} \& {Johansen}(2017)}]{Bitsch2017}
{Bitsch}, B. \& {Johansen}, A. 2017, {Planet Population Synthesis via Pebble
  Accretion}, ed. M.~{Pessah} \& O.~{Gressel}, Vol. 445, 339

\bibitem[{{Bitsch} {et~al.}(2015){Bitsch}, {Lambrechts}, \&
  {Johansen}}]{Bitsch2015}
{Bitsch}, B., {Lambrechts}, M., \& {Johansen}, A. 2015, \aap, 582, A112

\bibitem[{{Bitsch} {et~al.}(2018){Bitsch}, {Morbidelli}, {Johansen}, {Lega},
  {Lambrechts}, \& {Crida}}]{Bitsch2018_peb_iso}
{Bitsch}, B., {Morbidelli}, A., {Johansen}, A., {et~al.} 2018, \aap, 612, A30

\bibitem[{{Bitsch} {et~al.}(2021){Bitsch}, {Raymond}, {Buchhave},
  {Bello-Arufe}, {Rathcke}, \& {Schneider}}]{Bitsch+2021}
{Bitsch}, B., {Raymond}, S.~N., {Buchhave}, L.~A., {et~al.} 2021, \aap, 649, L5

\bibitem[{{Bitsch} \& {Savvidou}(2021)}]{Bitsch2021}
{Bitsch}, B. \& {Savvidou}, S. 2021, \aap, 647, A96

\bibitem[{{Bitsch} {et~al.}(2020){Bitsch}, {Trifonov}, \&
  {Izidoro}}]{Bitsch2020_scat}
{Bitsch}, B., {Trifonov}, T., \& {Izidoro}, A. 2020, \aap, 643, A66

\bibitem[{{Blum} \& {Wurm}(2008)}]{Blum2008}
{Blum}, J. \& {Wurm}, G. 2008, \araa, 46, 21

\bibitem[{{Booth} {et~al.}(2017){Booth}, {Clarke}, {Madhusudhan}, \&
  {Ilee}}]{Booth2017_pebble_evap}
{Booth}, R.~A., {Clarke}, C.~J., {Madhusudhan}, N., \& {Ilee}, J.~D. 2017,
  \mnras, 469, 3994

\bibitem[{{Booth} \& {Ilee}(2019)}]{Booth2019}
{Booth}, R.~A. \& {Ilee}, J.~D. 2019, \mnras, 487, 3998

\bibitem[{{Bosman} {et~al.}(2019){Bosman}, {Cridland}, \&
  {Miguel}}]{Bosman2019}
{Bosman}, A.~D., {Cridland}, A.~J., \& {Miguel}, Y. 2019, \aap, 632, L11

\bibitem[{{Bottke} {et~al.}(2005){Bottke}, {Durda}, {Nesvorn{\'y}}, {Jedicke},
  {Morbidelli}, {Vokrouhlick{\'y}}, \& {Levison}}]{Bottke2005}
{Bottke}, W.~F., {Durda}, D.~D., {Nesvorn{\'y}}, D., {et~al.} 2005, \icarus,
  175, 111

\bibitem[{{Brauer} {et~al.}(2008){Brauer}, {Dullemond}, \&
  {Henning}}]{Brauer2008}
{Brauer}, F., {Dullemond}, C.~P., \& {Henning}, T. 2008, \aap, 480, 859

\bibitem[{Brent(1973)}]{brent_algorithms}
Brent, R. 1973, Algorithms for minimization without derivatives (Prentice-Hall)

\bibitem[{{Brewer} {et~al.}(2017){Brewer}, {Fischer}, \&
  {Madhusudhan}}]{Brewer2017}
{Brewer}, J.~M., {Fischer}, D.~A., \& {Madhusudhan}, N. 2017, \aj, 153, 83

\bibitem[{{Brouwers} \& {Ormel}(2020)}]{Brouwers2020}
{Brouwers}, M.~G. \& {Ormel}, C.~W. 2020, \aap, 634, A15

\bibitem[{{Br{\"u}gger} {et~al.}(2018){Br{\"u}gger}, {Alibert}, {Ataiee}, \&
  {Benz}}]{Bruegger2018}
{Br{\"u}gger}, N., {Alibert}, Y., {Ataiee}, S., \& {Benz}, W. 2018, \aap, 619,
  A174

\bibitem[{{Buchhave} {et~al.}(2018){Buchhave}, {Bitsch}, {Johansen}, {Latham},
  {Bizzarro}, {Bieryla}, \& {Kipping}}]{Buchhave2018}
{Buchhave}, L.~A., {Bitsch}, B., {Johansen}, A., {et~al.} 2018, \apj, 856, 37

\bibitem[{{Buder} {et~al.}(2018){Buder}, {Asplund}, {Duong}, {Kos}, {Lind},
  {Ness}, {Sharma}, {Bland-Hawthorn}, {Casey}, {de Silva}, {D'Orazi},
  {Freeman}, {Lewis}, {Lin}, {Martell}, {Schlesinger}, {Simpson}, {Zucker},
  {Zwitter}, {Amarsi}, {Anguiano}, {Carollo}, {Casagrande}, {{\v{C}}otar},
  {Cottrell}, {da Costa}, {Gao}, {Hayden}, {Horner}, {Ireland}, {Kafle},
  {Munari}, {Nataf}, {Nordlander}, {Stello}, {Ting}, {Traven}, {Watson},
  {Wittenmyer}, {Wyse}, {Yong}, {Zinn}, {{\v{Z}}erjal}, \& {Galah
  Collaboration}}]{Buder2018}
{Buder}, S., {Asplund}, M., {Duong}, L., {et~al.} 2018, \mnras, 478, 4513

\bibitem[{{Chambers}(2018)}]{Chambers2018}
{Chambers}, J. 2018, \apj, 865, 30

\bibitem[{{Chrenko} {et~al.}(2017){Chrenko}, {Bro{\v{z}}}, \&
  {Lambrechts}}]{Chrenko2017}
{Chrenko}, O., {Bro{\v{z}}}, M., \& {Lambrechts}, M. 2017, \aap, 606, A114

\bibitem[{{Cimerman} {et~al.}(2017){Cimerman}, {Kuiper}, \&
  {Ormel}}]{Cimerman2017}
{Cimerman}, N.~P., {Kuiper}, R., \& {Ormel}, C.~W. 2017, \mnras, 471, 4662

\bibitem[{{Crida} \& {Bitsch}(2017)}]{CridaBitsch2017}
{Crida}, A. \& {Bitsch}, B. 2017, \icarus, 285, 145

\bibitem[{{Crida} \& {Morbidelli}(2007)}]{CridaMorbidelli2007}
{Crida}, A. \& {Morbidelli}, A. 2007, \mnras, 377, 1324

\bibitem[{{Crida} {et~al.}(2006){Crida}, {Morbidelli}, \& {Masset}}]{Crida2006}
{Crida}, A., {Morbidelli}, A., \& {Masset}, F. 2006, \icarus, 181, 587

\bibitem[{{Cridland} {et~al.}(2019){Cridland}, {Eistrup}, \& {van
  Dishoeck}}]{Cridland2019}
{Cridland}, A.~J., {Eistrup}, C., \& {van Dishoeck}, E.~F. 2019, \aap, 627,
  A127

\bibitem[{{Dawson} \& {Johnson}(2018)}]{Dawson2018}
{Dawson}, R.~I. \& {Johnson}, J.~A. 2018, \araa, 56, 175

\bibitem[{{Dr{\k{a}}{\.z}kowska} \& {Alibert}(2017)}]{Drazkowska2017}
{Dr{\k{a}}{\.z}kowska}, J. \& {Alibert}, Y. 2017, \aap, 608, A92

\bibitem[{{Dr{\k{a}}{\.z}kowska} {et~al.}(2016){Dr{\k{a}}{\.z}kowska},
  {Alibert}, \& {Moore}}]{Drazkowska2016}
{Dr{\k{a}}{\.z}kowska}, J., {Alibert}, Y., \& {Moore}, B. 2016, \aap, 594, A105

\bibitem[{{Dr{\k{a}}{\.z}kowska} {et~al.}(2021){Dr{\k{a}}{\.z}kowska},
  {Stammler}, \& {Birnstiel}}]{Drazkowska2021}
{Dr{\k{a}}{\.z}kowska}, J., {Stammler}, S.~M., \& {Birnstiel}, T. 2021, \aap,
  647, A15

\bibitem[{Dullemond(2013)}]{LesHouches2013}
Dullemond, C.~P. 2013, Theoretical Models of the Structure of Protoplanetary
  Disks - Les Houches 2013

\bibitem[{{Dullemond} \& {Birnstiel}(2018)}]{disklab}
{Dullemond}, C.~P. \& {Birnstiel}, T. 2018, DISKLAB - A protoplanetary disk
  modeling package in Python

\bibitem[{{Dullemond} {et~al.}(2018){Dullemond}, {Birnstiel}, {Huang},
  {Kurtovic}, {Andrews}, {Guzm{\'a}n}, {P{\'e}rez}, {Isella}, {Zhu}, {Benisty},
  {Wilner}, {Bai}, {Carpenter}, {Zhang}, \& {Ricci}}]{Dullemond2018DSHARP}
{Dullemond}, C.~P., {Birnstiel}, T., {Huang}, J., {et~al.} 2018, \apjl, 869,
  L46

\bibitem[{{Eistrup} {et~al.}(2016){Eistrup}, {Walsh}, \& {van
  Dishoeck}}]{Eistrup2016}
{Eistrup}, C., {Walsh}, C., \& {van Dishoeck}, E.~F. 2016, \aap, 595, A83

\bibitem[{{Emsenhuber} {et~al.}(2020){Emsenhuber}, {Mordasini}, {Burn},
  {Alibert}, {Benz}, \& {Asphaug}}]{Emsenhuber2020}
{Emsenhuber}, A., {Mordasini}, C., {Burn}, R., {et~al.} 2020, arXiv e-prints,
  arXiv:2007.05561

\bibitem[{{Ercolano} \& {Clarke}(2010)}]{Ercolano2010}
{Ercolano}, B. \& {Clarke}, C.~J. 2010, \mnras, 402, 2735

\bibitem[{{Ercolano} {et~al.}(2009){Ercolano}, {Clarke}, \&
  {Drake}}]{Ercolano2009}
{Ercolano}, B., {Clarke}, C.~J., \& {Drake}, J.~J. 2009, \apj, 699, 1639

\bibitem[{{Flaherty} {et~al.}(2018){Flaherty}, {Hughes}, {Teague}, {Simon},
  {Andrews}, \& {Wilner}}]{Flaherty2018}
{Flaherty}, K.~M., {Hughes}, A.~M., {Teague}, R., {et~al.} 2018, \apj, 856, 117

\bibitem[{{Flock} {et~al.}(2015){Flock}, {Ruge}, {Dzyurkevich}, {Henning},
  {Klahr}, \& {Wolf}}]{Flock2015}
{Flock}, M., {Ruge}, J.~P., {Dzyurkevich}, N., {et~al.} 2015, \aap, 574, A68

\bibitem[{{Fortier} {et~al.}(2013){Fortier}, {Alibert}, {Carron}, {Benz}, \&
  {Dittkrist}}]{Fortier2013}
{Fortier}, A., {Alibert}, Y., {Carron}, F., {Benz}, W., \& {Dittkrist}, K.~M.
  2013, \aap, 549, A44

\bibitem[{{Fortier} {et~al.}(2009){Fortier}, {Benvenuto}, \&
  {Brunini}}]{Fortier2009}
{Fortier}, A., {Benvenuto}, O.~G., \& {Brunini}, A. 2009, \aap, 500, 1249

\bibitem[{{G{\'a}rate} {et~al.}(2020){G{\'a}rate}, {Birnstiel},
  {Dr{\k{a}}{\.z}kowska}, \& {Stammler}}]{Garate2020}
{G{\'a}rate}, M., {Birnstiel}, T., {Dr{\k{a}}{\.z}kowska}, J., \& {Stammler},
  S.~M. 2020, \aap, 635, A149

\bibitem[{{Ginzburg} \& {Chiang}(2020)}]{Ginzburg2020}
{Ginzburg}, S. \& {Chiang}, E. 2020, \mnras, 498, 680

\bibitem[{{Guilera} {et~al.}(2019){Guilera}, {Cuello}, {Montesinos}, {Miller
  Bertolami}, {Ronco}, {Cuadra}, \& {Masset}}]{Guilera2019}
{Guilera}, O.~M., {Cuello}, N., {Montesinos}, M., {et~al.} 2019, \mnras, 486,
  5690

\bibitem[{{Guilera} {et~al.}(2011){Guilera}, {Fortier}, {Brunini}, \&
  {Benvenuto}}]{Guilera2011}
{Guilera}, O.~M., {Fortier}, A., {Brunini}, A., \& {Benvenuto}, O.~G. 2011,
  \aap, 532, A142

\bibitem[{{Guilera} {et~al.}(2020){Guilera}, {S{\'a}ndor}, {Ronco},
  {Venturini}, \& {Miller Bertolami}}]{Guliera2020}
{Guilera}, O.~M., {S{\'a}ndor}, Z., {Ronco}, M.~P., {Venturini}, J., \& {Miller
  Bertolami}, M.~M. 2020, \aap, 642, A140

\bibitem[{{Gundlach} \& {Blum}(2015)}]{Gundlach2015}
{Gundlach}, B. \& {Blum}, J. 2015, \apj, 798, 34

\bibitem[{{G{\"u}ttler} {et~al.}(2010){G{\"u}ttler}, {Blum}, {Zsom}, {Ormel},
  \& {Dullemond}}]{Guettler2010}
{G{\"u}ttler}, C., {Blum}, J., {Zsom}, A., {Ormel}, C.~W., \& {Dullemond},
  C.~P. 2010, \aap, 513, A56

\bibitem[{{Hollenbach} {et~al.}(2009){Hollenbach}, {Kaufman}, {Bergin}, \&
  {Melnick}}]{Hollenbach2009}
{Hollenbach}, D., {Kaufman}, M.~J., {Bergin}, E.~A., \& {Melnick}, G.~J. 2009,
  \apj, 690, 1497

\bibitem[{{Hori} \& {Ikoma}(2011)}]{Hori2011}
{Hori}, Y. \& {Ikoma}, M. 2011, \mnras, 416, 1419

\bibitem[{{Ida} \& {Lin}(2004)}]{IdaLinI}
{Ida}, S. \& {Lin}, D.~N.~C. 2004, \apj, 604, 388

\bibitem[{{Ida} \& {Lin}(2008{\natexlab{a}})}]{IdaLinIV}
{Ida}, S. \& {Lin}, D.~N.~C. 2008{\natexlab{a}}, \apj, 673, 487

\bibitem[{{Ida} \& {Lin}(2008{\natexlab{b}})}]{IdaLinV}
{Ida}, S. \& {Lin}, D.~N.~C. 2008{\natexlab{b}}, \apj, 685, 584

\bibitem[{{Ida} \& {Lin}(2010)}]{IdaLinVI}
{Ida}, S. \& {Lin}, D.~N.~C. 2010, \apj, 719, 810

\bibitem[{{Ikoma} {et~al.}(2000){Ikoma}, {Nakazawa}, \& {Emori}}]{Ikoma2000}
{Ikoma}, M., {Nakazawa}, K., \& {Emori}, H. 2000, \apj, 537, 1013

\bibitem[{{Izidoro} {et~al.}(2021{\natexlab{a}}){Izidoro}, {Bitsch}, \&
  {Dasgupta}}]{Izidoro2021}
{Izidoro}, A., {Bitsch}, B., \& {Dasgupta}, R. 2021{\natexlab{a}}, \apj, 915,
  62

\bibitem[{{Izidoro} {et~al.}(2021{\natexlab{b}}){Izidoro}, {Bitsch}, {Raymond},
  {Johansen}, {Morbidelli}, {Lambrechts}, \& {Jacobson}}]{Izidoro2019}
{Izidoro}, A., {Bitsch}, B., {Raymond}, S.~N., {et~al.} 2021{\natexlab{b}},
  \aap, 650, A152

\bibitem[{{Johansen} \& {Bitsch}(2019)}]{Johansen2019pla}
{Johansen}, A. \& {Bitsch}, B. 2019, \aap, 631, A70

\bibitem[{{Johansen} \& {Lacerda}(2010)}]{Johansen2010}
{Johansen}, A. \& {Lacerda}, P. 2010, \mnras, 404, 475

\bibitem[{{Johansen} \& {Lambrechts}(2017)}]{Johansen2017peb}
{Johansen}, A. \& {Lambrechts}, M. 2017, Annual Review of Earth and Planetary
  Sciences, 45, 359

\bibitem[{{Johansen} {et~al.}(2021){Johansen}, {Ronnet}, {Bizzarro},
  {Schiller}, {Lambrechts}, {Nordlund}, \& {Lammer}}]{Johansen2021}
{Johansen}, A., {Ronnet}, T., {Bizzarro}, M., {et~al.} 2021, Science Advances,
  7, eabc0444

\bibitem[{{Kama} {et~al.}(2019){Kama}, {Shorttle}, {Jermyn}, {Folsom},
  {Furuya}, {Bergin}, {Walsh}, \& {Keller}}]{Kama2019}
{Kama}, M., {Shorttle}, O., {Jermyn}, A.~S., {et~al.} 2019, \apj, 885, 114

\bibitem[{{Kenyon} \& {Bromley}(2012)}]{Kenyon2012}
{Kenyon}, S.~J. \& {Bromley}, B.~C. 2012, \aj, 143, 63

\bibitem[{{Kley} \& {Nelson}(2012)}]{Kley2012}
{Kley}, W. \& {Nelson}, R.~P. 2012, \araa, 50, 211

\bibitem[{{Kretke} \& {Levison}(2014)}]{Kretke2014}
{Kretke}, K.~A. \& {Levison}, H.~F. 2014, \aj, 148, 109

\bibitem[{{Krijt} {et~al.}(2020){Krijt}, {Bosman}, {Zhang}, {Schwarz},
  {Ciesla}, \& {Bergin}}]{Krijt2020}
{Krijt}, S., {Bosman}, A.~D., {Zhang}, K., {et~al.} 2020, \apj, 899, 134

\bibitem[{{Kruijer} {et~al.}(2017){Kruijer}, {Burkhardt}, {Budde}, \&
  {Kleine}}]{Kruijer2017}
{Kruijer}, T.~S., {Burkhardt}, C., {Budde}, G., \& {Kleine}, T. 2017,
  Proceedings of the National Academy of Science, 114, 6712

\bibitem[{{Lambrechts} \& {Johansen}(2012)}]{Lambrechts2012}
{Lambrechts}, M. \& {Johansen}, A. 2012, \aap, 544, A32

\bibitem[{{Lambrechts} \& {Johansen}(2014)}]{Lambrechts2014}
{Lambrechts}, M. \& {Johansen}, A. 2014, \aap, 572, A107

\bibitem[{{Lambrechts} {et~al.}(2014){Lambrechts}, {Johansen}, \&
  {Morbidelli}}]{Lambrechts2014_iso}
{Lambrechts}, M., {Johansen}, A., \& {Morbidelli}, A. 2014, \aap, 572, A35

\bibitem[{{Lambrechts} \& {Lega}(2017)}]{Lambrechts2017}
{Lambrechts}, M. \& {Lega}, E. 2017, \aap, 606, A146

\bibitem[{{Lambrechts} {et~al.}(2019{\natexlab{a}}){Lambrechts}, {Lega},
  {Nelson}, {Crida}, \& {Morbidelli}}]{Lambrechts2019gas}
{Lambrechts}, M., {Lega}, E., {Nelson}, R.~P., {Crida}, A., \& {Morbidelli}, A.
  2019{\natexlab{a}}, \aap, 630, A82

\bibitem[{{Lambrechts} {et~al.}(2019{\natexlab{b}}){Lambrechts}, {Morbidelli},
  {Jacobson}, {Johansen}, {Bitsch}, {Izidoro}, \& {Raymond}}]{Lambrechts2019}
{Lambrechts}, M., {Morbidelli}, A., {Jacobson}, S.~A., {et~al.}
  2019{\natexlab{b}}, \aap, 627, A83

\bibitem[{{Lega} {et~al.}(2014){Lega}, {Crida}, {Bitsch}, \&
  {Morbidelli}}]{Lega2014}
{Lega}, E., {Crida}, A., {Bitsch}, B., \& {Morbidelli}, A. 2014, \mnras, 440,
  683

\bibitem[{{Lenz} {et~al.}(2019){Lenz}, {Klahr}, \& {Birnstiel}}]{Lenz2019}
{Lenz}, C.~T., {Klahr}, H., \& {Birnstiel}, T. 2019, \apj, 874, 36

\bibitem[{{Levison} {et~al.}(2015){Levison}, {Kretke}, \&
  {Duncan}}]{Levison2015}
{Levison}, H.~F., {Kretke}, K.~A., \& {Duncan}, M.~J. 2015, \nat, 524, 322

\bibitem[{{Liu} {et~al.}(2019){Liu}, {Hori}, {M{\"u}ller}, {Zheng}, {Helled},
  {Lin}, \& {Isella}}]{Liu2019}
{Liu}, S.-F., {Hori}, Y., {M{\"u}ller}, S., {et~al.} 2019, \nat, 572, 355

\bibitem[{{Lodato} {et~al.}(2017){Lodato}, {Scardoni}, {Manara}, \&
  {Testi}}]{Lodato2017}
{Lodato}, G., {Scardoni}, C.~E., {Manara}, C.~F., \& {Testi}, L. 2017, \mnras,
  472, 4700

\bibitem[{{Lodders}(2003)}]{Lodders2003}
{Lodders}, K. 2003, \apj, 591, 1220

\bibitem[{{Lorek} {et~al.}(2018){Lorek}, {Lacerda}, \& {Blum}}]{Lorek2018}
{Lorek}, S., {Lacerda}, P., \& {Blum}, J. 2018, \aap, 611, A18

\bibitem[{{Lynden-Bell} \& {Pringle}(1974)}]{Lynden-Bell1974}
{Lynden-Bell}, D. \& {Pringle}, J.~E. 1974, \mnras, 168, 603

\bibitem[{{Machida} {et~al.}(2010){Machida}, {Kokubo}, {Inutsuka}, \&
  {Matsumoto}}]{Machida2010}
{Machida}, M.~N., {Kokubo}, E., {Inutsuka}, S.-I., \& {Matsumoto}, T. 2010,
  \mnras, 405, 1227

\bibitem[{{Madhusudhan} {et~al.}(2017){Madhusudhan}, {Bitsch}, {Johansen}, \&
  {Eriksson}}]{Madhusudhan2017}
{Madhusudhan}, N., {Bitsch}, B., {Johansen}, A., \& {Eriksson}, L. 2017,
  \mnras, 469, 4102

\bibitem[{{Madhusudhan} {et~al.}(2014){Madhusudhan}, {Crouzet}, {McCullough},
  {Deming}, \& {Hedges}}]{Madhusudhan2014}
{Madhusudhan}, N., {Crouzet}, N., {McCullough}, P.~R., {Deming}, D., \&
  {Hedges}, C. 2014, \apjl, 791, L9

\bibitem[{{Mamajek}(2009)}]{Mamajek2009}
{Mamajek}, E.~E. 2009, in American Institute of Physics Conference Series, Vol.
  1158, Exoplanets and Disks: Their Formation and Diversity, ed. T.~{Usuda},
  M.~{Tamura}, \& M.~{Ishii}, 3--10

\bibitem[{{Marboeuf} {et~al.}(2014{\natexlab{a}}){Marboeuf}, {Thiabaud},
  {Alibert}, {Cabral}, \& {Benz}}]{Marboeuf2014b}
{Marboeuf}, U., {Thiabaud}, A., {Alibert}, Y., {Cabral}, N., \& {Benz}, W.
  2014{\natexlab{a}}, \aap, 570, A36

\bibitem[{{Marboeuf} {et~al.}(2014{\natexlab{b}}){Marboeuf}, {Thiabaud},
  {Alibert}, {Cabral}, \& {Benz}}]{Marboeuf2014}
{Marboeuf}, U., {Thiabaud}, A., {Alibert}, Y., {Cabral}, N., \& {Benz}, W.
  2014{\natexlab{b}}, \aap, 570, A35

\bibitem[{{Masset}(2017)}]{Masset2017}
{Masset}, F.~S. 2017, \mnras, 472, 4204

\bibitem[{{Matsumura} {et~al.}(2017){Matsumura}, {Brasser}, \&
  {Ida}}]{Matsumura2017}
{Matsumura}, S., {Brasser}, R., \& {Ida}, S. 2017, \aap, 607, A67

\bibitem[{{Miller} \& {Fortney}(2011)}]{Miller2011}
{Miller}, N. \& {Fortney}, J.~J. 2011, \apjl, 736, L29

\bibitem[{{Molli{\`e}re} {et~al.}(2015){Molli{\`e}re}, {van Boekel},
  {Dullemond}, {Henning}, \& {Mordasini}}]{Molliere20151Dmodel}
{Molli{\`e}re}, P., {van Boekel}, R., {Dullemond}, C., {Henning}, T., \&
  {Mordasini}, C. 2015, \apj, 813, 47

\bibitem[{{Monsch} {et~al.}(2019){Monsch}, {Ercolano}, {Picogna}, {Preibisch},
  \& {Rau}}]{Monsch2019}
{Monsch}, K., {Ercolano}, B., {Picogna}, G., {Preibisch}, T., \& {Rau}, M.~M.
  2019, \mnras, 483, 3448

\bibitem[{{Morbidelli} {et~al.}(2016){Morbidelli}, {Bitsch}, {Crida},
  {Gounelle}, {Guillot}, {Jacobson}, {Johansen}, {Lambrechts}, \&
  {Lega}}]{Morbidelli2016}
{Morbidelli}, A., {Bitsch}, B., {Crida}, A., {et~al.} 2016, \icarus, 267, 368

\bibitem[{{Morbidelli} {et~al.}(2009){Morbidelli}, {Bottke}, {Nesvorn{\'y}}, \&
  {Levison}}]{Morbidelli2009}
{Morbidelli}, A., {Bottke}, W.~F., {Nesvorn{\'y}}, D., \& {Levison}, H.~F.
  2009, \icarus, 204, 558

\bibitem[{{Morbidelli} {et~al.}(2015){Morbidelli}, {Lambrechts}, {Jacobson}, \&
  {Bitsch}}]{Morbidelli2015}
{Morbidelli}, A., {Lambrechts}, M., {Jacobson}, S., \& {Bitsch}, B. 2015,
  \icarus, 258, 418

\bibitem[{{Morbidelli} \& {Nesvorny}(2012)}]{Morbidelli2012}
{Morbidelli}, A. \& {Nesvorny}, D. 2012, \aap, 546, A18

\bibitem[{{Mordasini} {et~al.}(2012){Mordasini}, {Alibert}, {Georgy},
  {Dittkrist}, {Klahr}, \& {Henning}}]{Mordasini2012}
{Mordasini}, C., {Alibert}, Y., {Georgy}, C., {et~al.} 2012, \aap, 547, A112

\bibitem[{{Mordasini} {et~al.}(2016){Mordasini}, {van Boekel}, {Molli{\`e}re},
  {Henning}, \& {Benneke}}]{Mordasini2016}
{Mordasini}, C., {van Boekel}, R., {Molli{\`e}re}, P., {Henning}, T., \&
  {Benneke}, B. 2016, \apj, 832, 41

\bibitem[{{Mousis} {et~al.}(2021){Mousis}, {Aguichine}, {Bouquet}, {Lunine},
  {Danger}, {Mandt}, \& {Luspay-Kuti}}]{Mousis2021}
{Mousis}, O., {Aguichine}, A., {Bouquet}, A., {et~al.} 2021, arXiv e-prints,
  arXiv:2103.01793

\bibitem[{{Movshovitz} \& {Podolak}(2008)}]{Movshovitz2008}
{Movshovitz}, N. \& {Podolak}, M. 2008, \icarus, 194, 368

\bibitem[{{M{\"u}ller} {et~al.}(2021){M{\"u}ller}, {Savvidou}, \&
  {Bitsch}}]{Mueller2021}
{M{\"u}ller}, J., {Savvidou}, S., \& {Bitsch}, B. 2021, \aap, 650, A185

\bibitem[{{M{\"u}ller} {et~al.}(2020){M{\"u}ller}, {Ben-Yami}, \&
  {Helled}}]{Mueller2020}
{M{\"u}ller}, S., {Ben-Yami}, M., \& {Helled}, R. 2020, \apj, 903, 147

\bibitem[{{Musiolik} \& {Wurm}(2019)}]{Musiolik2019}
{Musiolik}, G. \& {Wurm}, G. 2019, \apj, 873, 58

\bibitem[{{Nakagawa} {et~al.}(1986){Nakagawa}, {Sekiya}, \&
  {Hayashi}}]{Nakagawa1986}
{Nakagawa}, Y., {Sekiya}, M., \& {Hayashi}, C. 1986, \icarus, 67, 375

\bibitem[{{Ndugu} {et~al.}(2018){Ndugu}, {Bitsch}, \& {Jurua}}]{Ndugu2018}
{Ndugu}, N., {Bitsch}, B., \& {Jurua}, E. 2018, \mnras, 474, 886

\bibitem[{{Ndugu} {et~al.}(2021){Ndugu}, {Bitsch}, {Morbidelli}, {Crida}, \&
  {Jurua}}]{Ndugu2021}
{Ndugu}, N., {Bitsch}, B., {Morbidelli}, A., {Crida}, A., \& {Jurua}, E. 2021,
  \mnras, 501, 2017

\bibitem[{{Nelson} {et~al.}(2013){Nelson}, {Gressel}, \&
  {Umurhan}}]{Nelson2013}
{Nelson}, R.~P., {Gressel}, O., \& {Umurhan}, O.~M. 2013, \mnras, 435, 2610

\bibitem[{{Notsu} {et~al.}(2020){Notsu}, {Eistrup}, {Walsh}, \&
  {Nomura}}]{Notsu2020}
{Notsu}, S., {Eistrup}, C., {Walsh}, C., \& {Nomura}, H. 2020, \mnras, 499,
  2229

\bibitem[{{{\"O}berg} {et~al.}(2011){{\"O}berg}, {Murray-Clay}, \&
  {Bergin}}]{Oeberg2011}
{{\"O}berg}, K.~I., {Murray-Clay}, R., \& {Bergin}, E.~A. 2011, \apjl, 743, L16

\bibitem[{{Ogihara} {et~al.}(2021){Ogihara}, {Hori}, {Kunitomo}, \&
  {Kurosaki}}]{Ogihara2021}
{Ogihara}, M., {Hori}, Y., {Kunitomo}, M., \& {Kurosaki}, K. 2021, \aap, 648,
  L1

\bibitem[{{Okuzumi}(2009)}]{Okuzumi2009}
{Okuzumi}, S. 2009, \apj, 698, 1122

\bibitem[{{Ormel} \& {Klahr}(2010)}]{Ormel2010}
{Ormel}, C.~W. \& {Klahr}, H.~H. 2010, \aap, 520, A43

\bibitem[{{Ormel} {et~al.}(2015){Ormel}, {Shi}, \& {Kuiper}}]{Ormel2015}
{Ormel}, C.~W., {Shi}, J.-M., \& {Kuiper}, R. 2015, \mnras, 447, 3512

\bibitem[{{Ormel} {et~al.}(2021){Ormel}, {Vazan}, \& {Brouwers}}]{Ormel2021}
{Ormel}, C.~W., {Vazan}, A., \& {Brouwers}, M.~G. 2021, \aap, 647, A175

\bibitem[{{Owen} {et~al.}(2012){Owen}, {Clarke}, \& {Ercolano}}]{Owen2012}
{Owen}, J.~E., {Clarke}, C.~J., \& {Ercolano}, B. 2012, \mnras, 422, 1880

\bibitem[{{Owen} {et~al.}(2013){Owen}, {Scaife}, \& {Ercolano}}]{Owen2013}
{Owen}, J.~E., {Scaife}, A. M.~M., \& {Ercolano}, B. 2013, \mnras, 434, 3378

\bibitem[{{Paardekooper}(2014)}]{Paardekooper2014}
{Paardekooper}, S.~J. 2014, \mnras, 444, 2031

\bibitem[{{Paardekooper} {et~al.}(2011){Paardekooper}, {Baruteau}, \&
  {Kley}}]{Paardekooper2011}
{Paardekooper}, S.~J., {Baruteau}, C., \& {Kley}, W. 2011, \mnras, 410, 293

\bibitem[{{Pascucci} \& {Sterzik}(2009)}]{Pascucci2009}
{Pascucci}, I. \& {Sterzik}, M. 2009, \apj, 702, 724

\bibitem[{{Pierens}(2015)}]{Pierens2015}
{Pierens}, A. 2015, \mnras, 454, 2003

\bibitem[{{Pinilla} {et~al.}(2021){Pinilla}, {Lenz}, \&
  {Stammler}}]{Pinilla2021}
{Pinilla}, P., {Lenz}, C.~T., \& {Stammler}, S.~M. 2021, \aap, 645, A70

\bibitem[{{Piso} {et~al.}(2015){Piso}, {{\"O}berg}, {Birnstiel}, \&
  {Murray-Clay}}]{Piso2015_snowline}
{Piso}, A.-M.~A., {{\"O}berg}, K.~I., {Birnstiel}, T., \& {Murray-Clay}, R.~A.
  2015, \apj, 815, 109

\bibitem[{{Press} {et~al.}(1992){Press}, {Teukolsky}, {Vetterling}, \&
  {Flannery}}]{Brentq_impl}
{Press}, W.~H., {Teukolsky}, S.~A., {Vetterling}, W.~T., \& {Flannery}, B.~P.
  1992, {Numerical recipes in FORTRAN. The art of scientific computing}

\bibitem[{{Pringle}(1981)}]{Pringle1981}
{Pringle}, J.~E. 1981, \araa, 19, 137

\bibitem[{{Ram{\'\i}rez} {et~al.}(2020){Ram{\'\i}rez}, {Cridland}, \&
  {Molli{\`e}re}}]{Ramirez2020}
{Ram{\'\i}rez}, V., {Cridland}, A.~J., \& {Molli{\`e}re}, P. 2020, \aap, 641,
  A87

\bibitem[{{Raymond} {et~al.}(2009){Raymond}, {O'Brien}, {Morbidelli}, \&
  {Kaib}}]{Raymond2009}
{Raymond}, S.~N., {O'Brien}, D.~P., {Morbidelli}, A., \& {Kaib}, N.~A. 2009,
  \icarus, 203, 644

\bibitem[{{Savitzky} \& {Golay}(1964)}]{savgol_filter}
{Savitzky}, A. \& {Golay}, M.~J.~E. 1964, Analytical Chemistry, 36, 1627

\bibitem[{{Savvidou} \& {Bitsch}(2021)}]{Savvidou2021}
{Savvidou}, S. \& {Bitsch}, B. 2021, \aap, 650, A132

\bibitem[{{Savvidou} {et~al.}(2020){Savvidou}, {Bitsch}, \&
  {Lambrechts}}]{Savvidou2020}
{Savvidou}, S., {Bitsch}, B., \& {Lambrechts}, M. 2020, \aap, 640, A63

\bibitem[{{Schulik} {et~al.}(2019){Schulik}, {Johansen}, {Bitsch}, \&
  {Lega}}]{Schulik2019}
{Schulik}, M., {Johansen}, A., {Bitsch}, B., \& {Lega}, E. 2019, \aap, 632,
  A118

\bibitem[{{Semenov} {et~al.}(2010){Semenov}, {Hersant}, {Wakelam}, {Dutrey},
  {Chapillon}, {Guilloteau}, {Henning}, {Launhardt}, {Pi{\'e}tu}, \&
  {Schreyer}}]{Semenov2010}
{Semenov}, D., {Hersant}, F., {Wakelam}, V., {et~al.} 2010, \aap, 522, A42

\bibitem[{{Shakura} \& {Sunyaev}(1973)}]{Shakura1973}
{Shakura}, N.~I. \& {Sunyaev}, R.~A. 1973, \aap, 500, 33

\bibitem[{{Shibata} {et~al.}(2020){Shibata}, {Helled}, \&
  {Ikoma}}]{Shibata2020}
{Shibata}, S., {Helled}, R., \& {Ikoma}, M. 2020, \aap, 633, A33

\bibitem[{{Shibata} \& {Ikoma}(2019)}]{Shibata2019}
{Shibata}, S. \& {Ikoma}, M. 2019, \mnras, 487, 4510

\bibitem[{{Sotiriadis} {et~al.}(2017){Sotiriadis}, {Libert}, {Bitsch}, \&
  {Crida}}]{Sotiriadis2017}
{Sotiriadis}, S., {Libert}, A.-S., {Bitsch}, B., \& {Crida}, A. 2017, \aap,
  598, A70

\bibitem[{{Szul{\'a}gyi} {et~al.}(2016){Szul{\'a}gyi}, {Masset}, {Lega},
  {Crida}, {Morbidelli}, \& {Guillot}}]{Szulagyi2016}
{Szul{\'a}gyi}, J., {Masset}, F., {Lega}, E., {et~al.} 2016, \mnras, 460, 2853

\bibitem[{{Takeuchi} \& {Lin}(2002)}]{Takeuchi2002}
{Takeuchi}, T. \& {Lin}, D.~N.~C. 2002, \apj, 581, 1344

\bibitem[{{Tanaka} \& {Ida}(1999)}]{TanakaIda1999}
{Tanaka}, H. \& {Ida}, S. 1999, \icarus, 139, 350

\bibitem[{{Thiabaud} {et~al.}(2014){Thiabaud}, {Marboeuf}, {Alibert}, {Cabral},
  {Leya}, \& {Mezger}}]{Thiabaud2014}
{Thiabaud}, A., {Marboeuf}, U., {Alibert}, Y., {et~al.} 2014, \aap, 562, A27

\bibitem[{{Thiabaud} {et~al.}(2015{\natexlab{a}}){Thiabaud}, {Marboeuf},
  {Alibert}, {Leya}, \& {Mezger}}]{Thiabaud2015b}
{Thiabaud}, A., {Marboeuf}, U., {Alibert}, Y., {Leya}, I., \& {Mezger}, K.
  2015{\natexlab{a}}, \aap, 580, A30

\bibitem[{{Thiabaud} {et~al.}(2015{\natexlab{b}}){Thiabaud}, {Marboeuf},
  {Alibert}, {Leya}, \& {Mezger}}]{Thiabaud2015}
{Thiabaud}, A., {Marboeuf}, U., {Alibert}, Y., {Leya}, I., \& {Mezger}, K.
  2015{\natexlab{b}}, \aap, 574, A138

\bibitem[{{Thorngren} {et~al.}(2016){Thorngren}, {Fortney}, {Murray-Clay}, \&
  {Lopez}}]{Thorngren2016}
{Thorngren}, D.~P., {Fortney}, J.~J., {Murray-Clay}, R.~A., \& {Lopez}, E.~D.
  2016, \apj, 831, 64

\bibitem[{{Turner} {et~al.}(2014){Turner}, {Fromang}, {Gammie}, {Klahr},
  {Lesur}, {Wardle}, \& {Bai}}]{Turner2014}
{Turner}, N.~J., {Fromang}, S., {Gammie}, C., {et~al.} 2014, in Protostars and
  Planets VI, ed. H.~{Beuther}, R.~S. {Klessen}, C.~P. {Dullemond}, \&
  T.~{Henning}, 411

\bibitem[{{Valletta} \& {Helled}(2020)}]{Valletta2020}
{Valletta}, C. \& {Helled}, R. 2020, \apj, 900, 133

\bibitem[{{Vazan} {et~al.}(2018){Vazan}, {Helled}, \& {Guillot}}]{Vazan2018}
{Vazan}, A., {Helled}, R., \& {Guillot}, T. 2018, \aap, 610, L14

\bibitem[{{Venturini} {et~al.}(2016){Venturini}, {Alibert}, \&
  {Benz}}]{Venturini2016}
{Venturini}, J., {Alibert}, Y., \& {Benz}, W. 2016, \aap, 596, A90

\bibitem[{{Venturini} {et~al.}(2015){Venturini}, {Alibert}, {Benz}, \&
  {Ikoma}}]{Venturini2015}
{Venturini}, J., {Alibert}, Y., {Benz}, W., \& {Ikoma}, M. 2015, \aap, 576,
  A114

\bibitem[{{Venturini} {et~al.}(2020){Venturini}, {Guilera}, {Ronco}, \&
  {Mordasini}}]{Venturini2020a}
{Venturini}, J., {Guilera}, O.~M., {Ronco}, M.~P., \& {Mordasini}, C. 2020,
  \aap, 644, A174

\bibitem[{{Venturini} \& {Helled}(2020)}]{Venturini_2020_pla_chem}
{Venturini}, J. \& {Helled}, R. 2020, \aap, 634, A31

\bibitem[{{Voelkel} {et~al.}(2020){Voelkel}, {Klahr}, {Mordasini},
  {Emsenhuber}, \& {Lenz}}]{Voelkel2020}
{Voelkel}, O., {Klahr}, H., {Mordasini}, C., {Emsenhuber}, A., \& {Lenz}, C.
  2020, \aap, 642, A75

\bibitem[{{Weber} {et~al.}(2018){Weber}, {Ben{\'\i}tez-Llambay}, {Gressel},
  {Krapp}, \& {Pessah}}]{Weber2018}
{Weber}, P., {Ben{\'\i}tez-Llambay}, P., {Gressel}, O., {Krapp}, L., \&
  {Pessah}, M.~E. 2018, \apj, 854, 153

\bibitem[{{Weidenschilling}(1977)}]{Weidenschilling1977dyn}
{Weidenschilling}, S.~J. 1977, \mnras, 180, 57

\bibitem[{{Zhang} {et~al.}(2020){Zhang}, {Bosman}, \& {Bergin}}]{Zhang2020}
{Zhang}, K., {Bosman}, A.~D., \& {Bergin}, E.~A. 2020, \apjl, 891, L16

\end{thebibliography}
\endgroup

\newpage
\begin{appendix}

\section{Parameters in our model}

We list in this subsection all the variables and their meaning used in this work.

\begin{table*}
        \centering
        \begin{tabular}{c | c}
                \hline\hline
                Parameter & Explanation\\
                \hline
                $\nu$ & alpha viscosity parameter\\
                $c_s$ & sound speed\\
                $\Omega_\mathrm{K}$ & Kepler angular velocity\\
                $G$ & Gravitational constant\\
                $M_\star$ & Mass of central star\\
                $r$ & radius coordinate of the disk\\
                $k_\mathrm{B}$ & Boltzmann constant\\
                $T_\mathrm{mid}$ & Temperature of the midplane\\
                $\mu$ & mean molecular weight\\
                $m_\mathrm{p}$ & proton mass\\
                $\Delta v$ & sub-Keplerian azimuthal speed\\
                $v_\mathrm{K}$ & Keplerian velocity\\
                $v_\varphi$ & azimuthal velocity of the gas\\
                $P$ & Gas pressure\\
                $H_\mathrm{gas}$ & scale height of the disk\\
                St & Stokes number of a particle\\
                $\tau_f$ & stopping time of a particle\\
                $a$ & particle size\\
                $\rho_\bullet$ & density of a dust and pebble particle\\
                $\Sigma_\mathrm{gas}$ & Gas surface density\\
                $u_\mathrm{Z}$ & general pebble/dust drift velocity\\
                $u_\mathrm{gas}$ & gas diffusion velocity\\
                $\epsilon$ & solid to gas ratio\\
                $\Sigma_\mathrm{Z}$ & total dust surface density (pebbles and dust)\\
                $f_m$ & mass fraction of pebbles\\
                $\Sigma_\mathrm{peb}$ & surface density of pebbles\\
                $\Sigma_\mathrm{dust}$ & surface density of dust\\
                $\hat u_\mathrm{Z}$ & mass averaged dust drift velocity (dust and pebbles)\\
                $u_\mathrm{peb}$ & pebble drift velocity\\
                $u_\mathrm{dust}$ & dust drift velocity\\       
                Y & molecular species\\
                $t$ & time\\
                $\dot \Sigma$ & source/sink term for viscous evolution\\
                $M_0$ & initial disk mass\\
                $R_0$ & scaling radius of the disk\\
                $\psi$ & logarithmic gradient of the viscosity at the inner edge\\
                $r_\mathrm{in}$ & radius at the inner edge of the disk\\
                $\xi$ & normed time\\
                $t_\nu$ & viscous time\\
                $\dot\Sigma^\mathrm{evap}_\mathrm{Y}$ & evaporation source term\\
                $\dot\Sigma^\mathrm{cond}_\mathrm{Y}$ & condensation source term\\
                $\dot\Sigma^\mathrm{acc,peb}_\mathrm{Y}$ & source term due to the discount of accreted pebbles\\
                $r_\mathrm{ice,Y}$ & position of the evaporation line of species Y\\
                $a_\mathrm{dust}$ & dust size\\
                $a_\mathrm{peb}$ & pebble size\\
                $\epsilon_p$ & pebble sticking efficiency\\
                $\mu_\mathrm{Y}$ & mass of species Y (in proton masses)\\
                $\dot\Sigma^\mathrm{W}_\mathrm{gas,Y}$ & photoevaporation source term\\
                $t_\mathrm{evap}$ & start of photoevaporation\\
                $\tau_\mathrm{decay}$ & decay timescale of the disk\\
                $\dot M_{\rm disk}$ & disk accretion rate\\
                \hline
        \end{tabular}
 \caption{Explanations of variables related to the disk used in our model.}
 \label{tab:variables1} 
\end{table*}
\begin{table*}
        \centering
        \begin{tabular}{c | c}
                \hline\hline
                Parameter & Explanation\\
                \hline
                $J_p$ & angular momentum of the planet\\
                $M$ & Mass of the planet\\
                $a_p$ & distance of planet to the star\\
                $\tau_M$ & Migration timescale\\
                $\Gamma$ & Torque that acts on the planet\\
                $L$ & accretion luminosity of the planet\\
                $\dot M_\mathrm{peb}$ & accretion rates of pebbles onto the planet\\
                $\Theta$ & numerical parameter used for the dynamical torque\\
                $\mathcal{P}$ & numerical gap parameter\\
                $\mathcal{R}$ & Reynolds number\\
                $R_\mathrm{H}$ & Hill radius\\
                $q$ & mass ratio of planet to host star\\
                $f_\mathrm{gap}$ & relative Depth of surface density gap caused by planet\\
                $f(\mathcal{P})$ & gravitational gap depth \\
                $f_\mathrm{A}$ & gap depth caused by accretion \\
        $\tau_{\mathrm{II}}$ & Migration timescale for type II migration \\
                $M_t$ & transition mass, where pebble accretion becomes efficient\\
                $M_\mathrm{iso}$ & pebble isolation mass\\
                $M_a$ & mass of planetary envelope\\
                $R_\mathrm{acc}$ & impact radius of pebble accretion\\
                $\rho_\mathrm{peb}$ & density of pebbles in the disk\\
                $\delta v$ & approach speed of pebbles\\
                $H_\mathrm{peb}$ & pebble scale hight\\
                $\mathrm{\alpha_z}$ & vertical mixing of pebbles\\
                $\dot{M}_{\rm gas, Ikoma}$ & Ikoma gas accretion rate\\
                $\dot{\tau}_{\rm KH}$ & Kelvin-Helmholtz contraction rate\\
                $M_c$ & Core mass of the planet\\
                $\kappa_{\rm env}$ & envelope opacity\\
                $\dot M_{\rm gas, low}$ & low branch of the Machida gas accretion rate\\
                $\dot M_{\rm gas, high}$ & high branch of the Machida gas accretion rate\\
                $\dot M_{\rm HS}$ & horseshoe depletion rate \\
                $T_{\rm HS}$ & synodic period at the border of the horseshoe region\\
                $\Omega_{\rm HS}$ & synodic orbital frequency at the border of the horseshoe region\\
                $r_{\rm HS}$ & half width of the horseshoe region\\
                $x_{s}$ & half width of the horseshoe region normed to planetary position\\
                $h$ & aspect ratio at the planetary position\\
                $\Delta t$ & numerical time step\\
                $\Sigma_{\rm HS}$ & surface density of the horseshoe region\\
                $\aleph$ & gap profile\\
                $\sigma$ & std. deviation of gap profile\\
                $F_\star$ & stellar insulation at planetary position\\
                $\gamma_a,\gamma_b$ & Fit parameter for heavy element content\\
                $M_Z$ & mass of heavy elements in planet\\
                \hline
        \end{tabular}
 \caption{Explanations of variables related to the planet used in our model.}
 \label{tab:variables2}         
\end{table*}

\begin{table*}
        \centering
        \begin{tabular}{c | c}
                \hline\hline
                Parameter & Explanation\\
                \hline
                $L_\star$ & stellar luminosity\\
                $Q_\mathrm{irr}$ & heat flux on the disk surface\\
                $F$ & stellar Flux\\
                $\varphi$ & fraction of the stellar light that heats the disk\\
                $T_\mathrm{eff}$ & effective temperature of the disk\\
                $Q_+$ & viscous heat flux due to accretion\\
                $\tau_d$ & optical depth\\
                $\kappa_d$ & dust opacity\\
                $T_\mathrm{mid}$ & midplane temperature\\
                $T_\mathrm{visc}$ & temperature caused by viscous heating\\
                $\alpha_\Sigma$ & slope of the radial gas surface density profile\\
                $\Sigma_\mathrm{tot}$ & total surface density (gas and solids)\\
                $\Sigma_\mathrm{bg}$ & surface density of the H+He background gas\\
                $\Sigma_\mathrm{v}$ & surface density of the heavy molecular species in the gas phase\\
                $\epsilon_{0, \mathrm{chem}}$ & intrinsic heavy to gas ratio of the chemical model\\
                $m$ & mass of a dust grain\\
                $n_Y$ & number density of gaseous molecules of species Y\\
                $v_\mathrm{th, Y}$ & mean thermal velocity projected onto a surface \\
                \hline
        \end{tabular}
 \caption{Explanations of variables used in our model related to the appendix.}
 \label{tab:variables3} 
\end{table*}

\section{Temperature}\label{sec:num_T_it}
\begin{figure}
        \centering
        \includegraphics[width=.45\textwidth]{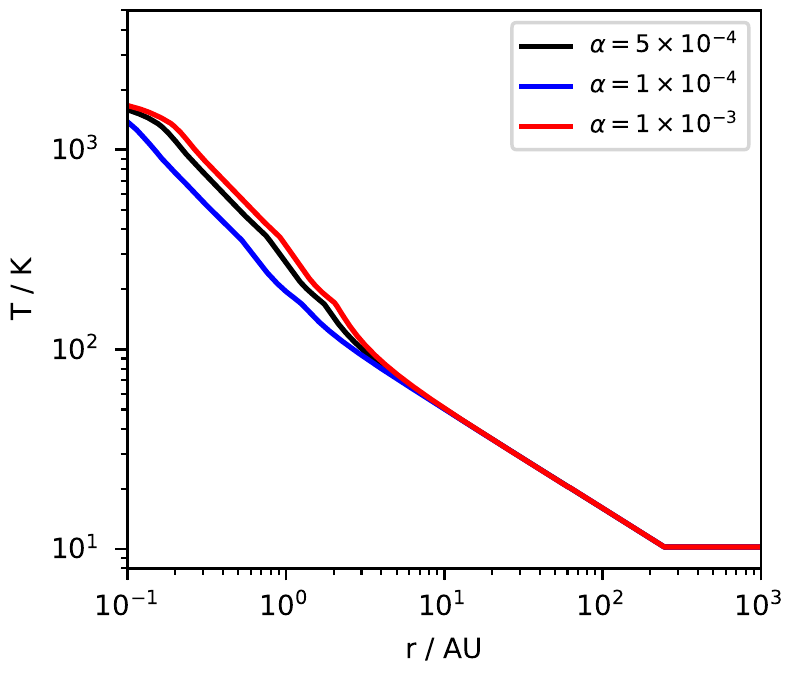}
        \caption{Midplane temperature profile (black line) of the protoplanetary disk model. The temperature depends on two constituents: viscous heating (dominant in the inner part) and irradiation from the central star (dominant in the outer part). The different colors show the disk's temperature for different viscosities. Larger viscosities result in more viscous heating and thus higher disk temperatures. The other disk parameters can be found in Table \ref{tab:parameters}.}
        \label{fig:T-profil}
\end{figure}

The temperature of the disk (see Fig.~\ref{fig:T-profil}) is mainly dependent on two physical processes: Heating through viscous accretion and irradiation by the central star. Irradiation can be described by the simplified assumption that a fraction $\varphi$ (e.g., a certain solid angle) of the flux $F = \frac{L_\star}{4\pi r^2}$ from the central star heats the surface of the protoplanetary disk \citep{Armitage,LesHouches2013}. Here $L_\star$ denotes the Luminosity of the host star. If we assume that the irradiation accounts for the effective temperature $T_\mathrm{eff}$ of the disk (the emission temperature), we get the heat flux on the disk surface
\begin{equation}\label{eq:evo_heat_irr}
        Q_{\textrm{irr}}=\frac{L_\star}{4\pi r^2}\varphi \hspace{1cm} \Rightarrow T_{\textrm{eff}}=\left(\frac{Q_{\textrm{irr}}}{2\sigma_{\textrm{SB}}}\right)^{1/4},
\end{equation}
where $\sigma_{\textrm{SB}}$ stands for the Stefan–Boltzmann constant. We use a constant value of $\varphi=0.05$ and $L_\star=L_\odot$ throughout this paper.
The heat flux due to viscous accretion heats the midplane. This heat flux can be described by \citep{Pringle1981}
\begin{equation}
        Q_+ = \frac{9}{4}{\Sigma_{\mathrm{gas}}\nu\Omega_{\textrm{K}}}^2.
\end{equation}
Using the definition for the optical depth $\tau_d$
\begin{equation}
        \tau_d = \frac12\Sigma_\mathrm{gas}\epsilon_0\kappa_d
,\end{equation}
we get the midplane temperature
\begin{equation}\label{eq:evo_heat_vischeat}
        T_{\textrm{mid}}^4 = T_{\textrm{visc}}^4 + T_{\textrm{eff}}^4 = \frac{3}{8}\frac{\tau_{\mathrm{d}}Q_+}{\sigma_{\textrm{SB}}}+\frac{Q_{\textrm{irr}}}{2\sigma_{\textrm{SB}}}
.\end{equation}

In order to find the midplane temperature, we applied the Brent method \citep{brent_algorithms, Brentq_impl} that uses the sign change in an interval in order to determine the root of an equation. We applied the Brent root finding method to solve the equation
\begin{equation}
        0 = \frac{3}{8}\frac{\tau_{\mathrm{d}}Q_+}{\sigma_{\textrm{SB}}}+\frac{Q_{\textrm{irr}}}{2\sigma_{\textrm{SB}}} - T_{\textrm{mid}}^4
\end{equation}
for every grid cell individually. We used a sign change interval of $\left[T_\mathrm{eff},1.5\times 10^{5}\SI{}{K}\right]$.
Protoplanetary disks likely have a background temperature due to the effects of irradiation from heavy stars that form nearby. We therefore used a minimum value of \SI{10}{K} for the effective temperature. 

When a good solution to Eq. $\ref{eq:evo_heat_vischeat}$ has been found we interpolate the temperature to a linear spaced grid by increasing the resolution drastically, apply a Savitzky–Golay filter \citep{savgol_filter} that smoothes the radial temperature profile and then interpolate back. 
As the disk evolves in time, the temperature in the inner regions decreases, due to the reduced gas surface density. However, the evolution of the gas surface density is quite minimal, especially for low viscosities (Fig.~\ref{fig:sigma_evap}). Therefore, we do not evolve the temperature profile of our disk in time for simplicity. 

\section{Comparison of the dynamical core of \texttt{TwoPopPy} with \texttt{chemcomp}}
\label{sec:codecomp}

We show in this section the comparison of \texttt{chemcomp} with \texttt{TwoPopPy}\footnote{The version of \texttt{TwoPopPy} used for the comparison has the git-hash: \href{https://github.com/birnstiel/two-pop-py/tree/6ac432718bffc3cf197a9e3d78fca492847c36f4}{\texttt{6ac432718bffc3cf197a9e3d78fca492847c36f4}}} \citep{Birnstiel2012} regarding the evolution of the disk's gas surface density and the pebble surface density for our standard disk model with different viscosities (Table~\ref{tab:parameters}). However, for this test we use the same temperature profile as in \cite{Birnstiel2012}, which is just a power-law compared to our temperature profile corresponding to viscous and stellar heating (Appendix~\ref{sec:num_T_it}). Furthermore, we only include one solid and one gas species (hydrogen-helium) for the code comparison, in contrast to the several species we included in our main work (Table~\ref{tab:comp}).

\begin{figure*}
    \centering
    \includegraphics[width=\textwidth]{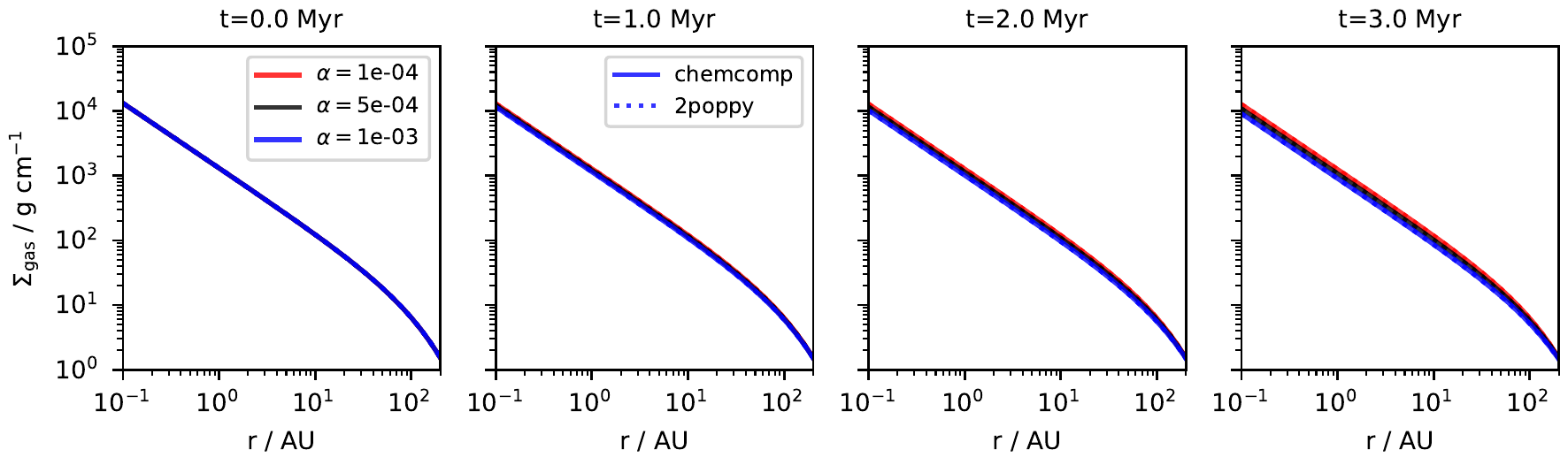}
    \caption{Comparison of the gas surface density evolution for different values of $\alpha$ for \texttt{TwoPopPy} with \texttt{chemcomp} for different times.}
    \label{fig:gascompare}
\end{figure*}

\begin{figure*}
    \centering
    \includegraphics[width=\textwidth]{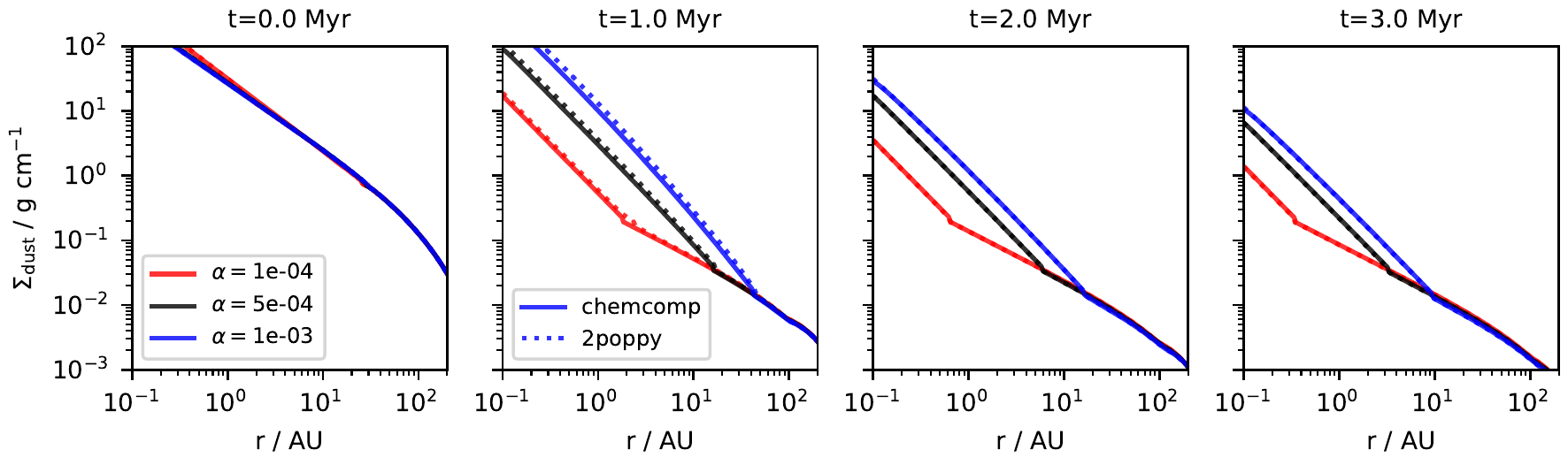}
    \caption{Comparison of the dust surface density evolution for different values of $\alpha$ for \texttt{TwoPopPy} with \texttt{chemcomp} for different times.}
    \label{fig:dustcompare}
\end{figure*}

We show in Fig.~\ref{fig:gascompare} the evolution of the gas surface density in time for the two codes for different viscosities. The evolution of the gas surface density is practically identical. The comparison of the evolution of the dust surface density in time (Fig.~\ref{fig:dustcompare}) also reveals a very similar evolution. At around 1 Myr, the dust seems to evolve slightly faster in \texttt{chemcomp}, especially for higher viscosities. However, after \SI{2}{Myr}, the differences in the dust surface density between the two codes is minimal, verifying our approach.


\section{Model with carbon}\label{sec:with_C}
\begin{figure}
        \centering
        \includegraphics[width=.45\textwidth]{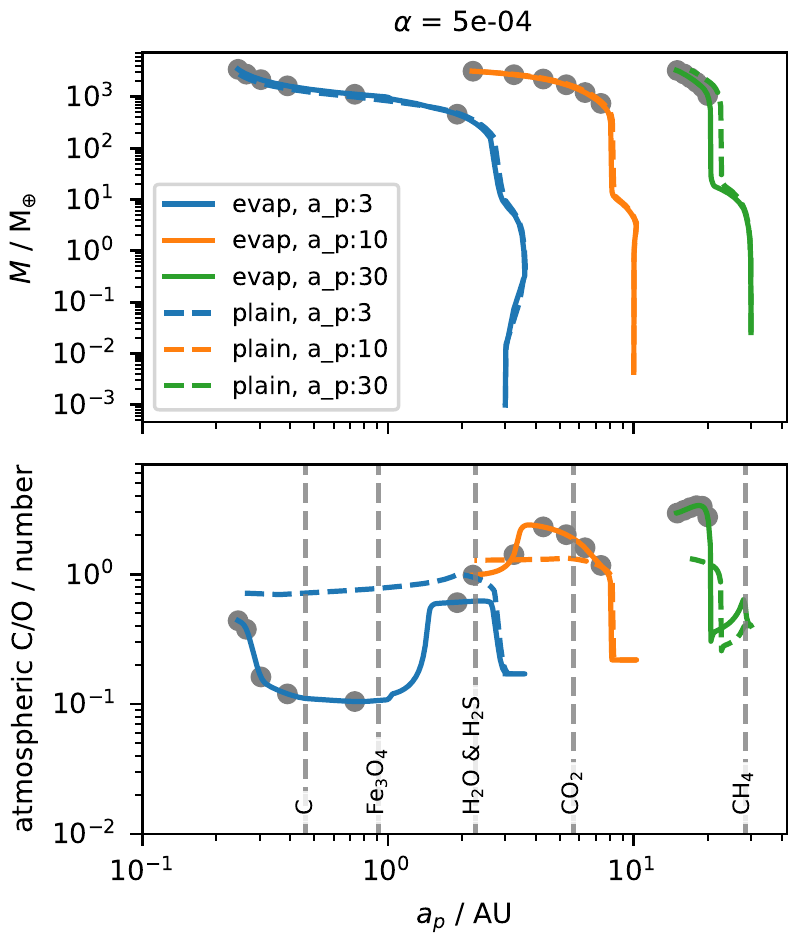}
        \caption{Like Fig. \ref{fig:growth}, using the same model parameters, but using the carbon grain model. Model parameters can be found in Table \ref{tab:parameters}.}
        \label{fig:growth_C}
\end{figure}
\begin{figure}
        \centering
        \includegraphics[width=.45\textwidth]{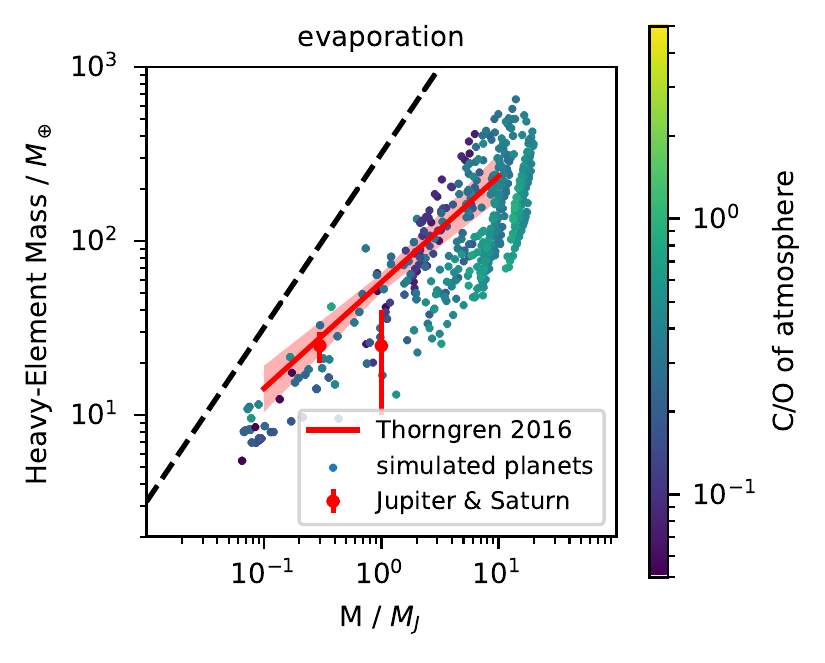}
        \caption{Like Fig. \ref{fig:thorngren_CO} but using the carbon grain model. The plot shows the model with evaporation.}
        \label{fig:thorngren_C}
\end{figure}

The results shown in the main part of this work use chemical compositions that do not include carbon grains (see Table \ref{tab:comp}, $v_\mathrm{Y, no C}$). The inclusion of carbon grains will mainly influence the C/O ratio, since it shifts the sublimation of carbon containing material from the methane evaporation line to the carbon evaporation front in the inner disk regions. Figure \ref{fig:growth_C} shows how this influences the atmospheric C/O content of planets that grow in a protoplanetary disk with carbon grains, where the same parameters as in Fig. \ref{fig:growth} have been used. Clearly, the planets forming in the outer regions of the disk now harbor a smaller C/O ratio due to the lack of methane. The inner planet (blue line), though, is mostly unaffected by the inclusion of carbon grains because once the planet crosses the carbon grain evaporation front, water vapor has already diffused inward and diluted the effects of the evaporating carbon grains on the C/O ratio. At late times, the inward diffusing carbon-rich gas from the outer disk results in an increase in the planetary C/O ratio, as for the model without carbon grains (Fig.~\ref{fig:growth}).

Figure \ref{fig:thorngren_C} shows that the C/O ratio of inner gas giants is slightly smaller in the model without carbon grains. Planets migrating across the carbon grain evaporation front early on can increase their carbon content slightly compared to planets in the model without carbon grains. However, the C/O ratio is only slightly affected because the carbon grain abundance is low compared to the water abundance, which dominates the C/O ratio of the planets accreting gas in the inner disk. However, the total heavy element contents of the planets remain unaffected because the planets accrete most of the gas at a few AU, where most of the volatiles have already evaporated, thus enriching the gas phase.

\section{Composition}
\label{sec:mucompo}
\subsection{Surface densities}
The total surface density of gas and dust is given as
\begin{equation}\label{eq:surface_tot}
        \Sigma_\mathrm{tot}(r) = \Sigma_\mathrm{gas}(r) + \Sigma_\mathrm{Z}(r),
\end{equation}
where $\Sigma_\mathrm{gas}$ is initialized from the disk mass via Eq. \ref{eq:gas_density_analytically}. We initialize the dust surface density by forcing a constant (in radius) initial heavy molecular species content ($\epsilon_0$) on the disk. 

We can write 
\begin{equation}
        \Sigma_\mathrm{gas}(r) = \Sigma_\mathrm{bg}(r) + \Sigma_\mathrm{v}(r),
\end{equation}
where $\Sigma_\mathrm{bg}(r)$ is the contribution of the background gas (consisting out of Hydrogen and Helium) and $\Sigma_\mathrm{v}(r)$ is the contribution from molecular species (see Table \ref{tab:comp}).
We can thus write
\begin{equation}
        \Sigma_\mathrm{tot}(r) = \Sigma_\mathrm{bg}(r) + \Sigma_\mathrm{v}(r) + \Sigma_\mathrm{Z}(r) \ .
\end{equation}

The intrinsic fraction of heavy molecular species $\epsilon_{0,\mathrm{chem}}$ relative to the hydrogen abundance in our chemical model is given by the sum over all molecular volume mixing ratios 
\begin{equation}
        \epsilon_{0,\mathrm{chem}}=\sum_Y{\mu_Y\times(\mathrm{Y/H})}=0.0179.
\end{equation}
It should be noted that we want to rescale our chemical model to a heavy molecular species content of $\epsilon_0$ (e.g., $\epsilon_0 = 2\%$). This heavy molecular species content can be thought of as the dust-to-gas ratio, when all molecular species are frozen out.

The dust-to-gas ratio $\epsilon(T)$ is the fraction of heavy molecular species in solids given by the rescaling of the chemical model as
\begin{equation}\label{eq:dtg}
        \mathrm{\epsilon}(T) = \frac{\epsilon_0}{\epsilon_{0,\mathrm{chem}}}\times\sum_{Y\in\{\mathrm{dust}\}}{\mu_Y\times(\mathrm{Y/H})}.
\end{equation}

Using the surface density of the background gas $\Sigma_\mathrm{bg}(r)$ we can now reformulate Eq. \ref{eq:surface_tot} as
\begin{equation}\label{eq:surface_tot_re}
\begin{split}
    \Sigma_\mathrm{tot}(r) & = \Sigma_\mathrm{bg}(r) + (\epsilon_0-\epsilon(r))\Sigma_\mathrm{bg}(r) + \epsilon(r)\Sigma_\mathrm{bg}(r)\\
        & = \Sigma_\mathrm{bg}(r)(1 +\epsilon_0),
\end{split}
\end{equation}
where $\epsilon_0$ is the mass fraction of heavy molecular species in the disk (i.e., also given by Eq. \ref{eq:dtg} when all species are part of the dust). 
The background gas surface density can therefore be calculated as
\begin{equation}\label{eq:bg_gas}
        \Sigma_\mathrm{bg}(r) = \Sigma_\mathrm{gas} \times (1+(\epsilon_0-\epsilon))^{-1} \ .
\end{equation}

The dust is then initialized as
\begin{equation}
        \Sigma_\mathrm{Z} = \epsilon(T) \times \Sigma_\mathrm{bg} .
\end{equation}

\subsection{Dust composition}

For every molecular species available in dust we used
\begin{equation}
        \frac{\Sigma_\mathrm{Z,Y}}{\Sigma_\mathrm{Z}} = \frac{\mu_Y\times \mathrm{(Y/H)}}{\sum_{i\in\{\mathrm{dust}\}}\mu_i}.
\end{equation}

\subsection{Gas composition}
The molecular weight of the hydrogen-helium mixture in a protoplanetary disk is given by \citep{Garate2020}
\begin{equation}
        \mu_\mathrm{H+He} = 2.3.
\end{equation}

For the heavy molecular species available in gas we can use
\begin{equation}
        \frac{\Sigma_\mathrm{gas,Y}}{\Sigma_\mathrm{v}} = \frac{\mu_Y\times \mathrm{(Y/H)}}{\sum_{i\in\{\mathrm{gas}\}}\mu_i},
\end{equation}
which yields the individual mass fractions of the heavy molecular species Y in the vapor. Using this equation we can now calculate
\begin{equation}\label{eq:gas_parts}
        \frac{\Sigma_\mathrm{gas,Y}}{\Sigma_\mathrm{gas}} = \frac{\Sigma_\mathrm{gas,Y}}{\Sigma_\mathrm{v}} \times \frac{\Sigma_\mathrm{v}}{\Sigma_\mathrm{gas}} = \frac{\mu_Y\times \mathrm{(Y/H)}}{\sum_{i\in\{\mathrm{gas}\}}\mu_i}\times \frac{\epsilon_0 - \epsilon}{1 + (\epsilon_0 - \epsilon)},
\end{equation}
while the background gas species (H+He) is simply given by Eq. \ref{eq:bg_gas}.
We note that the sum of the individual gas surface densities of the different molecular species must be the total gas surface density (the same also applies for the dust surface density).

We can now calculate the mean molecular weight from the sum over all molecular species (including the background gas):
\begin{equation}\label{eq:eq_mu}
        \mu = \Sigma_\mathrm{gas} \times \left(\sum_{Y\in \{\mathrm{gas}\}}\frac{\Sigma_{gas,Y}}{\mu_Y}\right)^{-1}.
\end{equation}

\section{Gas, dust, and planetary velocities}
\label{sec:velocities}

\begin{figure*}
    \includegraphics[width=\textwidth]{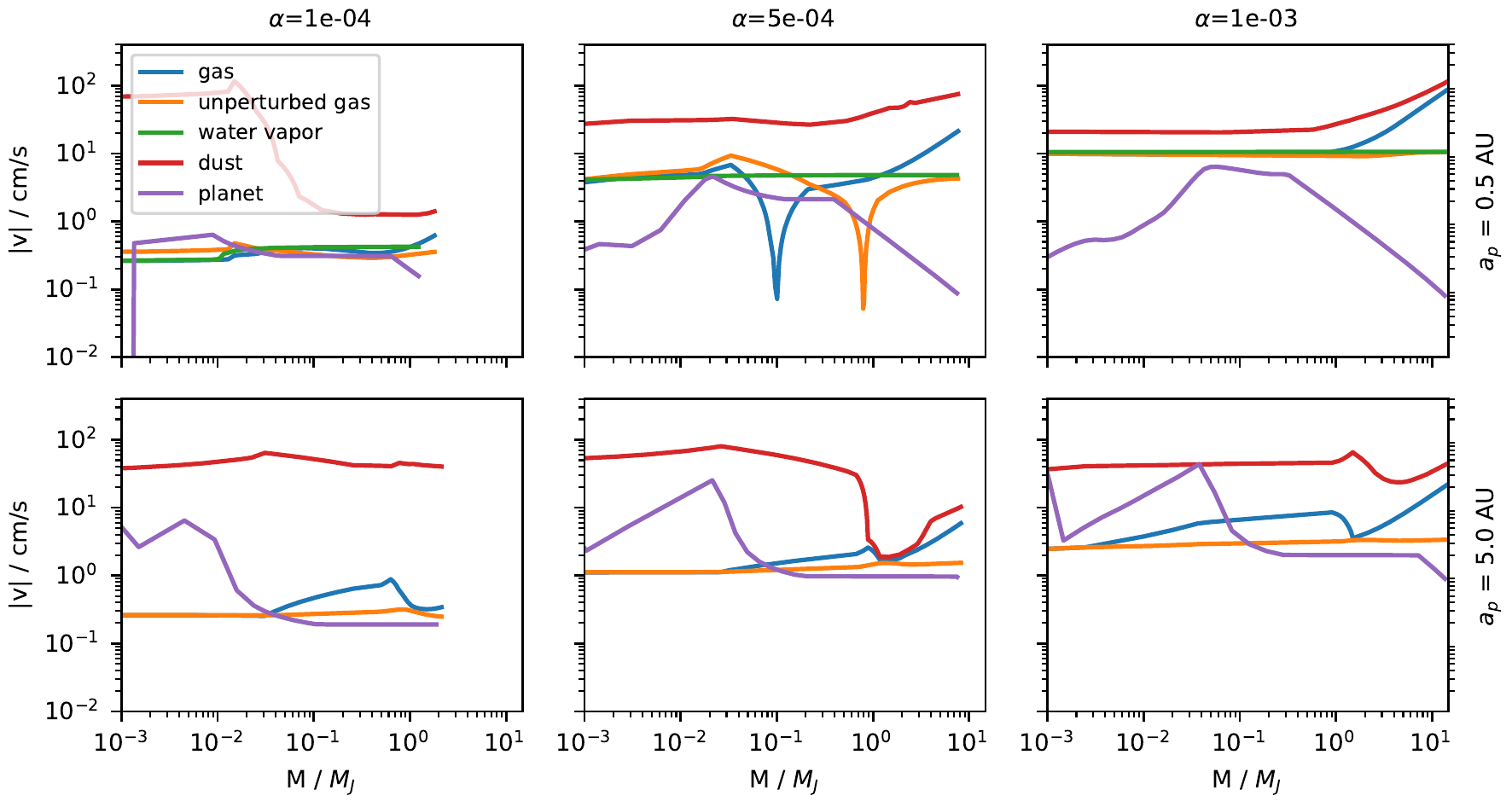}
    \caption{Radial velocities of dust, water vapor, perturbed and unperturbed gas, and the planet as a function of the planetary mass for non-migrating planets at 0.5 AU (top) and 5.0 AU (bottom). From left to right the disk's viscosity is increasing. The curves stop once the disk's lifetime reaches 3 Myr.}
    \label{fig:velocities}
\end{figure*}

The accretion of vapor-enriched gas onto planetary atmospheres depends on the relative speed of the gas to the planet. As the planet accretes gas that is provided by viscous evolution, the gas only reach the planet if the gas moves faster than the planet.

In Fig.~\ref{fig:velocities} we show the radial velocities of the dust, the perturbed and unperturbed gas and planets as a function of the planetary mass.  We note that the change of the planetary mass corresponds to the growth of the planet in the simulations, showing effectively a time evolution. The curves stop once the disk has reached its lifetime of \SI{3}{Myr}. The different velocities are extracted from the same simulations used in Figs.~\ref{fig:watertimenoplanet}, \ref{fig:watertime05}, and \ref{fig:watertime50}. The unperturbed velocities are extracted from simulations without planets.

The dust velocities originate from Eq.~\ref{eq:vel_dust} and clearly show that dust moves fastest, allowing the enrichment of vapor in the inner disk by volatile transporting pebbles. The dust velocities change in time, due to the evolution of the dust and gas profile of the protoplanetary disk.

The unperturbed gas velocities (where there is no planet embedded in the disk) show some slight variations in time. This effect is caused by slight changes of the gas surface density in time due to evaporation, which influence the gas velocities (Eq.~\ref{eq_the:radial_velocity_general}).

The velocities of the water vapor (only displayed at 0.5 AU because there is no water vapor at 5.0 AU) of simulations without planets are larger than the velocities of the unperturbed gas. This is caused by the lower vapor surface density compared to the total gas surface density, resulting in faster velocities (Eq.~\ref{eq_the:radial_velocity_general}).

The perturbed gas velocities (blue line in Fig.~\ref{fig:velocities}) correspond to simulations with embedded planets. The main changes in the perturbed gas velocities are caused by the gap opening of the growing planet, which we mimic by changing the disk's viscosity at the planetary position. As the planet growth, the gap becomes deeper, mimicked by increasing the disk's viscosity at the planetary position. An increase in the disk's viscosity automatically increases the velocities of the perturbed gas profile. This approach is designed to keep the radial mass flux across the gap constant. The additional variations in the perturbed gas profile are caused by the further changes of the gas surface density profile due to evaporation of the inward drifting pebbles.

The velocities of the planet reflect the different migration prescriptions. Initially the planet is in type-I migration, which increases with planetary mass. As soon as the planet becomes massive and opens a partial gap, the migration speed decreases toward the initially constant type-II migration rate. Once the planet becomes very massive, its migration speed further decreases due to the inertia, which scales linearly with planetary mass, resulting in a further decrease in the planet's migration rate. This effect is clearly more important for higher viscosities, where the planet can grow faster.

The planetary migration rate of the planet, once it starts to open a partial gap, is always lower than the gas velocity. This effect can also be seen by comparing the viscous type-II migration rate ($\tau_{\rm visc} = a_P^2 /\nu$, e.g., \citealt{Baruteau2014}) with the gas velocities in a viscously evolving disk 
\begin{equation}
    v_{\rm r,gas} = - 3 \alpha \frac{c_s^2}{v_{\rm K}} \left( \frac{3}{2} - \alpha_{\Sigma} \right)
\end{equation}
analytically \citep{Takeuchi2002}. Here $\alpha_{\Sigma}$ denotes the slope of the radial gas surface density profile. This allows the planet to accrete volatile-enriched material brought by inward drifting pebbles in gas form. 

\section{Ice condensation onto dust grains}\label{sec:cond} 
The mass increase per grain (with mass $m=\frac43\pi\rho_\bullet a^3$) per second due to the condensation (with sticking efficiency $\epsilon_p=0.5$) of gaseous molecules of species Y with number density
\begin{equation}
        n_\mathrm{Y} = \frac{\Sigma_\mathrm{gas,Y}}{\sqrt{2\pi}H_\mathrm{gas} m_\mathrm{Y}}   
\end{equation}
and mass $m_\mathrm{Y}=\mu_\mathrm{Y}m_p$ (e.g., CO: $m_\mathrm{Y}=(12+16)m_p$)
is given by
\begin{equation}\label{eq:per_grain_increase}
        \frac{\mathrm{d}m}{\mathrm{d}t}=4\pi a^2 n_\mathrm{Y} v_\mathrm{th,Y} \epsilon_p m_\mathrm{Y},
\end{equation}
where $v_\mathrm{th,Y}$ is the mean thermal velocity projected onto a surface given by
\begin{equation}
        v_\mathrm{th,Y} = \sqrt{\frac{k_B T}{2\pi m_\mathrm{Y}}}.
\end{equation}
This per grain increase translates to an increase in the solid surface density by
\begin{equation}\label{eq:surface_density_increase}
                \dot\Sigma^{\mathrm{cond}}_\mathrm{Y} = \Sigma_\mathrm{peb} \frac{1}{m_\mathrm{peb}}\frac{\mathrm{d}m_\mathrm{peb}}{\mathrm{d}t}+\Sigma_\mathrm{dust} \frac{1}{m_\mathrm{dust}}\frac{\mathrm{d}m_\mathrm{dust}}{\mathrm{d}t}.
\end{equation}
Inserting Eq. \ref{eq:per_grain_increase} into Eq. \ref{eq:surface_density_increase} yields:
\begin{equation}
                \dot\Sigma^{\mathrm{cond}}_\mathrm{Y} = \frac{3}{\rho_\bullet}\left(\frac{\Sigma_\mathrm{peb}}{a_\mathrm{peb}}+\frac{\Sigma_\mathrm{dust}}{a_\mathrm{dust}}\right)n_\mathrm{Y}v_\mathrm{th,Y}\epsilon_p m_\mathrm{Y}.
\end{equation}
If we eliminate $v_\mathrm{th,Y}$ and $n_\mathrm{Y}$ we get
\begin{equation}
        \dot\Sigma^{\mathrm{cond}}_\mathrm{Y} = 
\frac{3\epsilon_p}{2\pi\rho_\bullet}\Sigma_{\mathrm{gas,Y}}\left(\frac{\Sigma_\mathrm{dust}}{a_\mathrm{dust}}+\frac{\Sigma_\mathrm{peb}}{a_\mathrm{peb}}\right)\Omega_\mathrm{k}\sqrt{\frac{\mu}{\mu_\mathrm{Y}}}.
\end{equation}

\section{Water ice content in the pebbles}
\label{sec:watercontent}

The efficiency of evaporation of inward drifting pebbles depends crucially on the pebble size and their velocities (e.g., \citealt{Piso2015_snowline, Drazkowska2017}), where larger pebbles drift inward faster (e.g., \citealt{Brauer2008}). In our model the size of the pebbles is determined by a coagulation/fragmentation equilibrium, where the exact pebble size depends on the viscosity of the disk \citep{Birnstiel2012}. Lower viscosities result in larger pebble sizes, allowing them to drift inward further compared to smaller pebbles before they evaporate.

\begin{figure*}
    \centering
    \includegraphics[width = \textwidth]{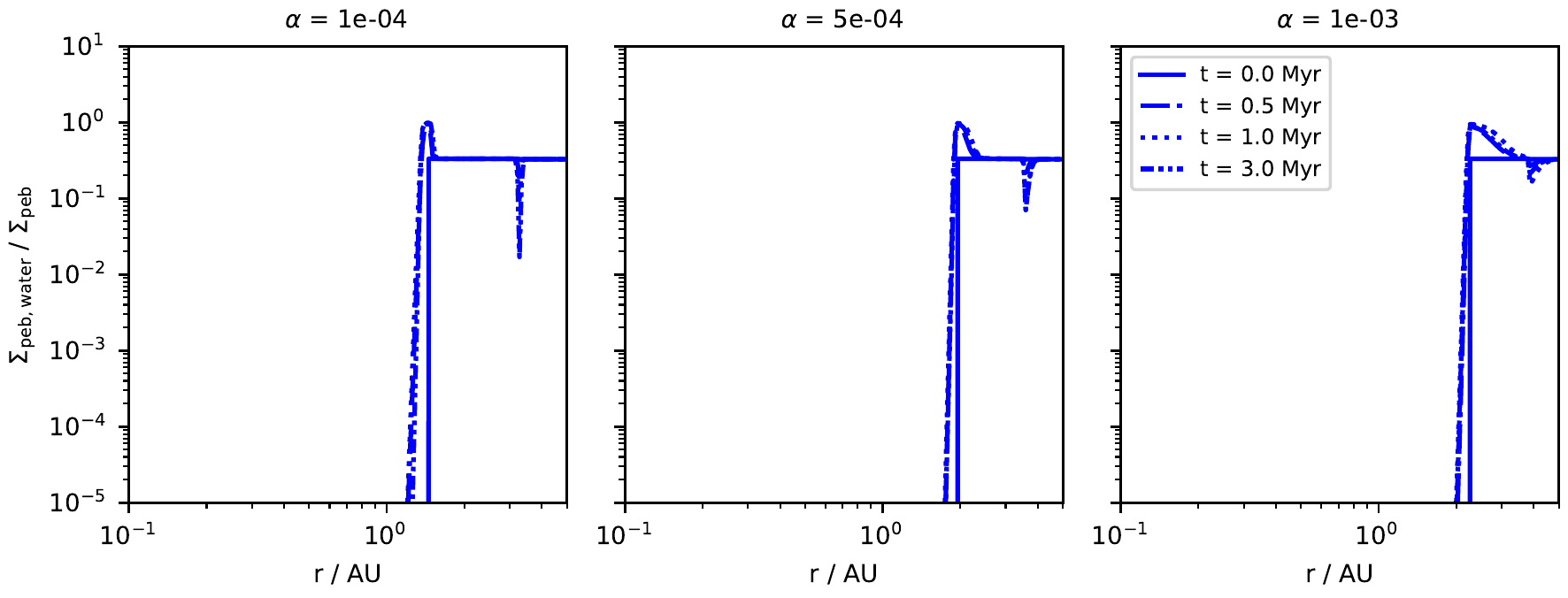}
    \caption{Ratio of the water ice surface density in pebbles compared to the total pebble surface density for different viscosities (left to right) as a function of time. The peak around the water ice line is caused by the condensation of water vapor, while the dip around 3-4 AU is caused by the condensation of CO$_2$, which consequently decreases the fraction of all solids at this position, including water ice.}
    \label{fig:iceline_water}
\end{figure*}

We show in Fig.~\ref{fig:iceline_water} the ratio of the water ice surface density to the total pebble surface density for different disk viscosities and times. As soon as the pebbles cross the water ice line, they start to evaporate and the water ice fraction decreases. In the cases of lower viscosities, water-ice-rich particles can penetrate farther into the inner disk compared to higher viscosities because of the increased particles size and thus increased particle speed (Eq.~\ref{eq:sink_ice_evap}). This result is in line with the simulations of \citet{Piso2015_snowline}, who also showed that larger particles can penetrate deeper into the disk before they evaporate compared to smaller particles.

Figure~\ref{fig:iceline_water} also reveals an increase in the water ice content in the solids close to the water ice line, which is caused by condensation of outward diffusing water vapor. Furthermore, we also see a dip in the water ice content at around 3-4 AU (depending on the disk's viscosity). This dip is caused by the condensation of CO$_2$ vapor, which increases the CO$_2$ content in the solids and consequently decreases the fraction of all other species, including water ice. 

The pebbles close to the water ice line have Stokes numbers around 10$^{-2}$ to 0.1 (low viscosity) or 10$^{-3}$ to 10$^{-2}$ (high viscosities), corresponding to particles sizes of $\sim$10 cm (low viscosities) or $\sim$1 cm (high viscosities). The resulting inward drift velocities are in the range of meters per second. Consequently pebbles evaporate within a close distance to the water ice line, in line with the simulations of \citet{Drazkowska2017}.

\section{Core mixing in the atmosphere}
\label{sec:50:50}

In our main paper, we have shown the atmospheric C/O ratios, where our model initially contributes 10\% of the accreted solids during the core buildup phase into the early planetary atmosphere until pebble isolation mass is reached. However, more detailed simulations indicate that the planetary atmosphere buildup might already start before the pebble isolation mass is reached and that a larger fraction of the solids might be accreted into the planetary atmosphere \citep{Brouwers2020, Valletta2020}. Nevertheless, a change in the amount of solids that can be accreted in the early atmosphere buildup will not change the total heavy element content of the planet, but would rather change the atmospheric C/O ratio. 

\begin{figure*}
        \centering
        \includegraphics[width=\textwidth]{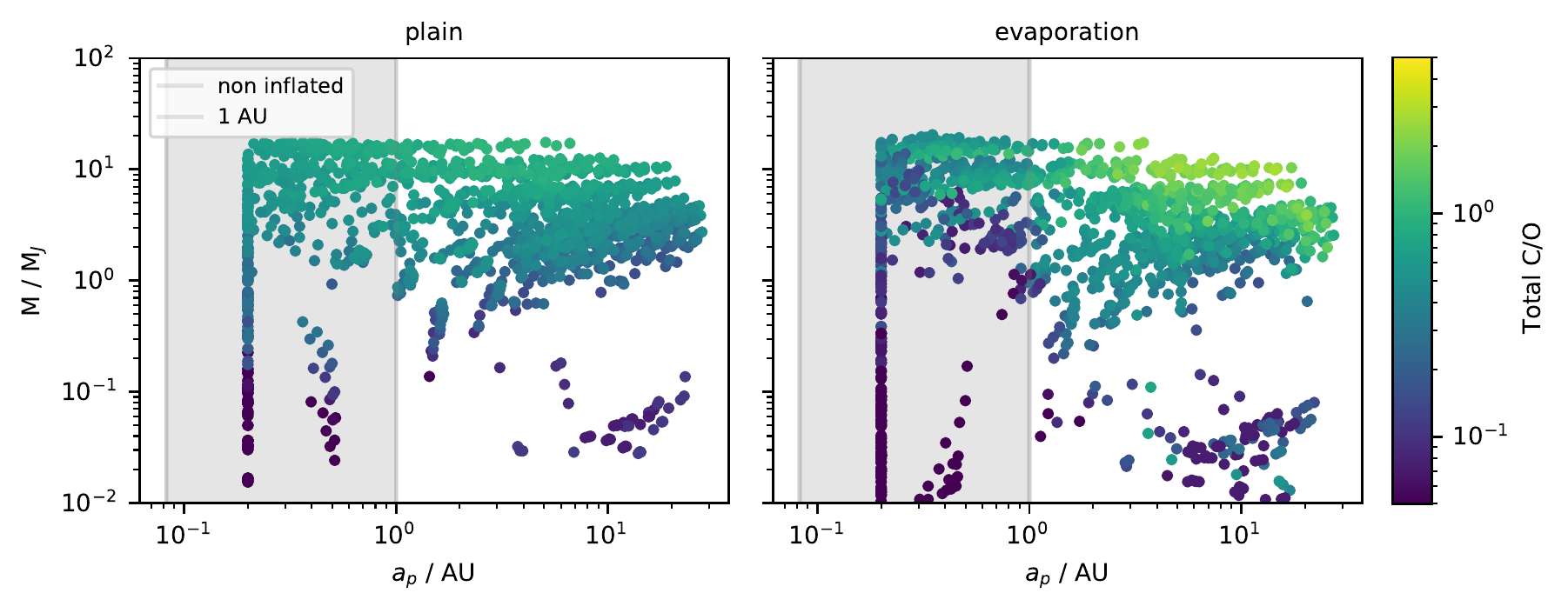}
        \caption{Like Fig.~\ref{fig:thorngren_ma}, but the color coding shows the total C/O ratio of core and atmosphere mixed together.}
        \label{fig:thorngren_ma_tot}
\end{figure*}

We show in Fig.~\ref{fig:thorngren_ma_tot} the total C/O ratio under the assumption that the whole planetary core is mixed evenly into the planetary atmosphere. The situation shown here is clearly an extreme assumption; however, the total C/O ratio of the planetary atmosphere in reality might thus be in between a complete mixture of the core in the atmosphere and the situation shown in Fig.~\ref{fig:thorngren_ma}, where core and atmosphere are completely separated. While the exact C/O ratio is reduced for the mixing scenario, the general trend that planets forming farther away from the central star should harbor a larger C/O remains intact for both models with and without evaporation of pebbles.

\section{Fit parameter} 
The solid to gas ratio fit parameters from Fig. \ref{fig:thorngren_DTG} can be found in Table \ref{tab:thorngren_fit}. \citet{Thorngren2016} found $\gamma_a=57.9\pm7.03$ and $\gamma_b=0.61\pm0.08$ for his fit on observed exoplanets. This is best matched with our simulations by $\epsilon_0=2.5\%$ in the evaporation model. 

\begin{table*}
        \begin{subtable}{\textwidth}
                \centering
                \begin{tabular}{c | c  c  c  c}
                \hline\hline
                simulation & $\epsilon_0=\num{1e-2}$ & $\epsilon_0=\num{1.5e-2}$ & $\epsilon_0=\num{2e-2}$ & $\epsilon_0=\num{2.5e-2}$ \\
                \hline
                plain & $6.4\pm0.3$ & $7.5\pm0.3$ & $9.7\pm 0.3$ & $11.5\pm 0.4$\\

                evap & $28.6\pm1.6$ & $47.3\pm3.6$ & $65.1\pm 7.7$ & $82.1\pm 11.4$\\
                \hline
        \end{tabular}
        \caption{Fit parameter $\gamma_a$}
        \end{subtable}
        \begin{subtable}{\textwidth}
                \centering
                \begin{tabular}{c | c  c  c  c}
                \hline\hline
                simulation & $\epsilon_0=\num{1e-2}$ & $\epsilon_0=\num{1.5e-2}$ & $\epsilon_0=\num{2e-2}$ & $\epsilon_0=\num{2.5e-2}$ \\
                \hline
                plain & $0.65\pm 0.02$ & $0.70\pm 0.02$ & $0.69\pm 0.01$ & $0.70\pm 0.01$\\
                evap & $0.53\pm0.02$ & $0.49\pm0.03$ & $0.52\pm 0.05$ & $0.54\pm 0.06$\\
                \hline
        \end{tabular}
        \caption{Fit parameter $\gamma_b$}
        \end{subtable}
        \caption{Fit results for Fig.~\ref{fig:thorngren_DTG}.}
        \begin{tablenotes}
        	\item \textbf{Notes:} The total heavy element content is related to the final mass by the power law $M_Z = \gamma_a \cdot M ^{ \gamma_b}$.
        \end{tablenotes}
        \label{tab:thorngren_fit}
\end{table*}

\end{appendix}
\end{document}